%% file: ceers_xi_ion_fesc.tex
\documentclass[twocolumn]{aastex7}
\usepackage{apjfonts} 

\newif\preprint 

\usepackage{amsmath,amstext}
\usepackage[T1]{fontenc}
\usepackage{comment}
\usepackage{natbib}
\usepackage{multirow}
\usepackage{verbatim}
\usepackage{hyperref}
\hypersetup{
    colorlinks=true,
    linkcolor=blue,
    filecolor=magenta,      
    urlcolor=cyan,
  }
  \usepackage{footmisc}
\graphicspath{{./}{figures/}{spec_plots/}}

\newcommand{\editone}[1]{{#1}}

\begin{document}
\include{defs}

\newcommand{\todo}[1]{\textcolor{aggiemaroon}{\tt #1}}

\title{\large \bf Galaxies in the Epoch of Reionization Are All Bark and No Bite --- Plenty of Ionizing Photons,\\ Low Escape Fractions}

\correspondingauthor{Casey Papovich}
\email{papovich@tamu.edu}

\author[0000-0001-7503-8482]{Casey Papovich}
\affiliation{Department of Physics and Astronomy, Texas A\&M University, College
Station, TX, 77843-4242 USA}
\affiliation{George P.\ and Cynthia Woods Mitchell Institute for
 Fundamental Physics and Astronomy, Texas A\&M University, College
 Station, TX, 77843-4242 USA}
 \email{papovich@tamu.edu}

\author[0000-0002-6348-1900]{Justin W. Cole}
\altaffiliation{NASA FINESST Fellow}
\affiliation{Department of Physics and Astronomy, Texas A\&M
  University, College Station, TX, 77843-4242 USA}
\affiliation{George P.\ and Cynthia Woods Mitchell Institute for
  Fundamental Physics and Astronomy, Texas A\&M University, College
  Station, TX, 77843-4242 USA}
   \email{papovich@tamu.edu}

  \author[0000-0003-3424-3230]{Weida Hu}
\affiliation{Department of Physics and Astronomy, Texas A\&M
  University, College Station, TX, 77843-4242 USA}
\affiliation{George P.\ and Cynthia Woods Mitchell Institute for
  Fundamental Physics and Astronomy, Texas A\&M University, College
  Station, TX, 77843-4242 USA}
 \email{weidahu@tamu.edu}

\author[0000-0001-8519-1130]{Steven L. Finkelstein}
\affiliation{Department of Astronomy, The University of Texas at Austin, Austin, TX, USA}
 \email{stevenf@astro.as.utexas.edu}

\author[0000-0001-9495-7759]{Lu Shen}
\affiliation{Department of Physics and Astronomy, Texas A\&M
  University, College Station, TX, 77843-4242 USA}
\affiliation{George P.\ and Cynthia Woods Mitchell Institute for
  Fundamental Physics and Astronomy, Texas A\&M University, College
  Station, TX, 77843-4242 USA}
 \email{lushen@tamu.edu}

\author[0000-0002-7959-8783]{Pablo Arrabal Haro}
\altaffiliation{NASA Postdoctoral Fellow}
\affiliation{Astrophysics Science Division, NASA Goddard Space Flight Center, 8800 Greenbelt Rd, Greenbelt, MD 20771, USA}
\email{pablo.arrabalharo@nasa.gov}

\author[0000-0001-5758-1000]{Ricardo O. Amor\'{i}n}
\affiliation{Instituto de Astrof\'{i}sica de Andaluc\'{i}a (CSIC), Apartado 3004, 18080 Granada, Spain}
 \email{amorin@iaa.es}

\author[0000-0001-8534-7502]{Bren E. Backhaus}
\affil{Department of Physics and Astronomy, University of Kansas, Lawrence, KS 66045, USA}
 \email{bren.backhaus@ku.edu}

\author[0000-0002-9921-9218]{Micaela B. Bagley}
\affiliation{Department of Astronomy, The University of Texas at Austin, Austin, TX, USA}
\affiliation{Astrophysics Science Division, NASA Goddard Space Flight Center, 8800 Greenbelt Rd, Greenbelt, MD 20771, USA}
 \email{micaela.b.bagley@nasa.gov}

\author[0000-0003-0883-2226]{Rachana Bhatawdekar}
\affiliation{European Space Agency (ESA), European Space Astronomy Centre (ESAC), Camino Bajo del Castillo s/n, 28692 Villanueva de la Cañada, Madrid, Spain}
 \email{rachanab@gmail.com}

\author[0000-0003-2536-1614]{Antonello Calabr\`o}
\affiliation{INAF Osservatorio Astronomico di Roma, Via Frascati 33, 00078 Monte Porzio Catone, Rome, Italy}
 \email{antonello.calabro@inaf.it}

\author[0000-0002-1482-5818]{Adam C. Carnall}
\affiliation{Institute for Astronomy, University of Edinburgh, Royal Observatory, Edinburgh, EH9 3HJ, UK}
 \email{adamc@roe.ac.uk}

\author[0000-0001-7151-009X]{Nikko J. Cleri}
\affiliation{Department of Astronomy and Astrophysics, The Pennsylvania State University, University Park, PA 16802, USA}
\affiliation{Institute for Computational and Data Sciences, The Pennsylvania State University, University Park, PA 16802, USA}
\affiliation{Institute for Gravitation and the Cosmos, The Pennsylvania State University, University Park, PA 16802, USA}
 \email{cleri@psu.edu}

\author[0000-0002-3331-9590]{Emanuele Daddi}
\affiliation{Universit\'e Paris-Saclay, Universit\'e Paris Cit\'e, CEA, CNRS, AIM, 91191, Gif-sur-Yvette, France}
 \email{edaddi@cea.fr}

\author[0000-0001-5414-5131]{Mark Dickinson}
\affiliation{NSF's National Optical-Infrared Astronomy Research Laboratory, 950 N. Cherry Ave., Tucson, AZ 85719, USA}
 \email{mark.dickinson@noirlab.edu}

\author[0000-0001-9440-8872]{Norman A. Grogin}
\affiliation{Space Telescope Science Institute, 3700 San Martin Drive, Baltimore, MD 21218, USA}
 \email{nagrogin@stsci.edu}

\author[0000-0002-4884-6756]{Benne W. Holwerda}
\affiliation{Department of Physics, University of Louisville,40292 KY Louisville, USA}
\email{benne.holwerda@louisville.edu}  

\author[0000-0002-6790-5125]{Anne E. Jaskot}
\affiliation{Department of Physics and Astronomy, Williams College, Williamstown, MA 01267, USA}
 \email{08aej@williams.edu}

\author[0000-0002-6610-2048]{Anton M. Koekemoer}
\affiliation{Space Telescope Science Institute, 3700 San Martin Drive, Baltimore, MD 21218, USA}
 \email{koekemoer@stsci.edu}

\author[0000-0003-1354-4296]{Mario Llerena}
\affiliation{INAF - Osservatorio Astronomico di Roma, via di Frascati 33, 00078 Monte Porzio Catone, Italy}
 \email{mario.llerena@inaf.it}

\author[0000-0003-1581-7825]{Ray A. Lucas}
\affiliation{Space Telescope Science Institute, 3700 San Martin Drive, Baltimore, MD 21218, USA}
 \email{lucas@stsci.edu}

\author[0000-0002-9572-7813]{Sara Mascia}
\affiliation{Institute of Science and Technology Austria (ISTA), Am Campus 1, A-3400 Klosterneuburg, Austria}
 \email{sara.mascia@ist.ac.at}

\author[0000-0001-9879-7780]{Fabio Pacucci}
\affiliation{Center for Astrophysics $\vert$ Harvard \& Smithsonian, 60 Garden St, Cambridge, MA 02138, USA}
\affiliation{Black Hole Initiative, Harvard University, 20 Garden St, Cambridge, MA 02138, USA}
 \email{fabio.pacucci@cfa.harvard.edu}

\author[0000-0001-8940-6768]{Laura Pentericci}
\affiliation{INAF - Osservatorio Astronomico di Roma, via di Frascati 33, 00078 Monte Porzio Catone, Italy} \email{laura.pentericci@inaf.it}

\author[0000-0003-4528-5639]{Pablo G. P\'erez-Gonz\'alez}
\affiliation{Centro de Astrobiolog\'{\i}a (CAB/CSIC-INTA), Ctra. de Ajalvir km 4, Torrej\'on de Ardoz, E-28850, Madrid, Spain}
 \email{pgperez@cab.inta-csic.es}

\author[0000-0003-3382-5941]{Nor Pirzkal}
\affiliation{ESA/AURA Space Telescope Science Institute}
 \email{npirzkal@stsci.edu}

\author[0000-0003-1405-378X]{Srinivasan Raghunathan}
\affiliation{Center for AstroPhysical Surveys, National Center for Supercomputing Applications, Urbana, IL, 61801, USA}
\email{srinirag@illinois.edu}

\author[ 0000-0001-7755-4755]{Lise-Marie Seill\'e}
\affiliation{ 
Aix Marseille Univ, CNRS, CNES, LAM, Marseille, France
}
 \email{lise-marie.seille@lam.fr}

\author[0000-0002-6748-6821]{Rachel S.~Somerville}
\affiliation{Center for Computational Astrophysics, Flatiron Institute, 162 5th Avenue, New York, NY 10010, USA}
 \email{rsomerville@flatironinstitute.org}

\author[0000-0003-3466-035X]{{L. Y. Aaron} {Yung}}
\affiliation{Space Telescope Science Institute, 3700 San Martin Drive, Baltimore, MD 21218, USA}
 \email{yung@stsci.edu}

%

\begin{abstract}
Early results from \jwst\ suggest that epoch-of-reionization (EoR) galaxies produce copious ionizing photons, which, if they escape efficiently, could cause reionization to occur too early.  We study this problem using \jwst\ imaging and prism spectroscopy for \editone{412} galaxies at $4.5 < z < 9.0$.  We fit these data simultaneously with stellar--population and nebular--emission models that include a parameter for the fraction of ionizing photons that escape the galaxy, \fesc.   We find that the ionization production efficiency, $\xiion = Q(\mathrm{H_0}) / \luv$, increases with redshift and decreasing UV luminosity, but shows significant scatter, $\sigma( \log \xiion|z,\muv) \simeq 0.3$ dex.   The inferred escape fractions averaged over the population \editone{are low, ranging  from$\langle \fesc \rangle \simeq 2.6\pm 1.4$\% at $6 < z < 9$ to $6.5\pm 2.2$\% at $4.5 < z < 6$ with weak or no indication of evolution with redshift}.  This implies that in our models \editone{most} of the ionizing photons need to be absorbed to account for the nebular emission.   We compute the impact of our results on reionization, including the distributions for \xiion\ and \fesc, and the evolution and uncertainty of the UV luminosity function.  Considering galaxies brighter than $\muv < -16$~mag, we would produce \editone{an IGM hydrogen-ionized fraction of $x_e = 0.5$ at $5.3 < z < 5.8$}, possibly too late compared to constraints from from QSO sightlines.   Including fainter galaxies, $\muv < -14$~mag, \editone{we obtain $x_e = 0.5$ at $6.0 < z < 8.1$}, fully consistent with QSO and CMB data.  This implies that EoR galaxies produce plenty of ionizing photons, but these do not efficiently escape.  This may be a result of high gas column densities combined with burstier star-formation histories, which limit the time massive stars are able to clear channels through the gas for ionizing photons to escape.  
\end{abstract}


\section{Introduction}\label{section:introduction}

  \jwst\ has transformed our ability to characterize the nature of galaxies at redshifts as high as $z\sim 14$ \citep[e.g.,][]{Bunker_2023,Arrabal_Haro_2023b,Carniani_2024,Kokorev_2025}.   From the analyses of \jwst\ data, galaxies at $z > 7$ appear to be dominated by younger stellar populations \citep[e.g.,][]{Papovich_2023}, with stronger nebular emission \citep{Endsley_2024,Calabro_2024,Mascia_2024,Pahl_2024,Simmonds_2024,Simmonds_2024b}, and evidence of more bursty star-formation histories \citep{Endsley_2024,Cole_2025,Trussler_2025}, compared to lower redshift galaxies \citep[e.g.,][]{Shen_2024,Cole_2025}.

The properties of galaxies at $z > 6$ are crucially important because this period corresponds to the epoch of reionization (EoR) of the intergalactic medium (IGM).  Constraints from the polarization of the cosmic microwave background (CMB) show that  the ionized fraction of the IGM rises from nil at the epoch of recombination, $z\approx 1100$, to $x_e = 0.5$ by $z=8.4$ (\citealt{Planck_2020}).\footnote{The
\citet{Planck_2020} analysis {also} suggests a non-zero optical depth of
CMB photons scattering off free electrons at $z \approx 15$, which
implies ionization of the IGM had begun by this epoch.}  This can be combined with observations of the optical depth of ionizing radiation from QSOs to show that reionization had effectively ``ended'' by $z \sim 5-6$ when the ionized fraction was $x_e \gtrsim 0.9$ \citep[e.g.,][]{McGreer_2015,Greig_2017,Becker_2019,Hu_2019,Mason_2019,Wang_2020,Whitler_2020,Nakane_2024}.    
%

Galaxies during the first billion years appear to be the most likely cause of this reionization \citep[see,][]{Robertson_2022}.  
Observations from \jwst\ have already shown evidence that galaxies at $z \gtrsim 7$ have sufficient ionizing radiation to complete reionization by this time \citep{Simmonds_2024,Simmonds_2024b}.   The primary measurement of interest is the so-called ionizing production efficiency, $\xiion$, defined as the ratio of production rate of photons with energy $>$13.6~eV, $Q(\mathrm{H_0})$, in units of s$^{-1}$, to the luminosity density at 1500~\AA, $L_\mathrm{UV}$, in units of erg~s$^{-1}$~Hz$^{-1}$ \citep{Leitherer_1995}.   The analysis of \jwst\ data shows that \xiion\ \editone{spans a wide range, from as low $\log \xiion \simeq 24.5$ to as high as} $\log \xiion \simeq 25-26$ \editone{\citep[e.g.,][]{Atek_2024,Prieto-Lyon_2023,Pahl_2024,Simmonds_2024,Simmonds_2024b,Begley_2025,Hayes_2025,llerena_2024},} possibly exceeding the canonical value, $\log \xiion = 25.2$ expected for young stellar populations and observed for galaxies at $z\sim 2$ \citep{Robertson_2015,Shivaei_2018}.   That is, EoR galaxies appear to produce sufficient ionizing radiation for reionization to proceed.

However, we lack a good understanding of how much of this radiation escapes into the IGM.  This is measured by the escape fraction, \fesc, which is the relative amount of photons with energy greater than the ionization energy of hydrogen (also referred to as Lyman-continuum radiation, LyC, at wavelengths $<$912~\AA).  Theoretical constraints predict that the escape fraction is low for EoR galaxies as the column density of gas that fuels star formation is very large \citep[e.g.,][]{Ferrara_2013,Xu_2016}.  The key is how efficiently feedback from star formation can clear channels through the gas through which the LyC can escape \citep{Kimm_2014}.  Current predictions for EoR galaxies, with masses like those we currently detect with \jwst, are that they have lower escape fractions, with $\fesc \sim 1-15\%$ \citep{Ma_2015,Xu_2016,Ferrara_2019,Kostyuk_2023,Yeh_2023}\editone{, although some simulations argue for an increase in \fesc\ with increasing redshift \citep[e.g.,][]{Rosdahl_2022,Trebitsch_2022}}.  As these galaxies are expected to drive reionization \citep{Lewis_2020}, it is of paramount importance that we make constraints on \fesc\ in galaxies at these redshifts. 

Measuring the escape fraction is challenging.  
At low redshifts, it must be observed from space.   Some results for starburst and extreme-emission line galaxies show high values, $\fesc = 6-35$\% \citep{Izotov_2016,Izotov_2021}, but that the value for an individual galaxy appears to depend strongly on the geometry \citep{Choi_2020}.    Work from the Low-Redshift Lyman Continuum Survey (LzLCS, \citealt{Flurry_2022a}) showed that for $z\sim 0.3$ galaxies, \fesc\ generally increases with decreasing galaxy luminosity, increasing ionization parameter (often parameterized by the \oiii/\oii\ ratio or by the Balmer emission equivalent width [EW]), decreasing UV spectral slope ($\beta_\mathrm{UV}$) and decreasing dust attenuation \citep[e.g.,][]{Flurry_2022b}.  It is not known if these correlations are causal or related to underlying processes.  \citet{Jaskot_2024} showed that \fesc\ has a smaller scatter when using a multivariate analysis, implying a more complicated relation between \fesc\ and galaxy properties.  Nevertheless, the application of the LzLCS relations to EoR galaxies predicts the latter should have relatively high escape fractions, in excess of 10\% \citep{Chisholm_2022,Cullen_2023,Mascia_2024}.  
%

 At higher redshifts the escape fraction can only be detected directly for galaxies at $z\lesssim 4$, as at higher redshifts the density of \ion{H}{1} absorbers in the IGM becomes too large for us to detect any escaping LyC photons \citep[e.g.,][]{Inoue_2014}.  Measurements of \fesc\ from galaxies \editone{at $1 < z < 3.5$ find low \textit{average} values, $\langle \fesc \rangle \leq 7$\% \citep{Siana_2010,Boutsia_2011,Nestor_2013,Grazian_2017,Matthee_2017,Steidel_2018,Pahl_2021,Pahl_2023,Saxena_2022,Begley_2022,Jung_2024,Wang_2025},} and these values agree with constraints on global escape fraction measurements of the \ion{H}{1} photoionization rate in the IGM from the \lya\ forest, which require $\fesc \simeq 1-2$\%
 %
 %
 \citep{Haardt_2012, Khaire_2015}. 
 However, \fesc\ may vary significantly, where individual reports span 20--100\% \editone{\citep[see, e.g.,][]{Izotov_2021,Saxena_2022}}.  Moreover, the average \fesc\ appears to be higher for galaxies with stronger emission lines \editone{\citep[e.g.,][]{Steidel_2018,Pahl_2021},} which may be more typical of EoR galaxies.  Therefore, it remains unclear what, if anything, we can infer about the nature of \fesc\ for EoR galaxies using these results.


Combining all of these results poses a potential problem.  Pre-\jwst\ studies \citep{Robertson_2015, Finkelstein_2019, Yung_2020b} found a wide range of \fesc\ ($\fesc \simeq 3 - 30$\%) would be capable of producing reionization.  \jwst\ observations now show that EoR galaxies produce higher quantities of ionizing photons (see above), which, when combined with high \fesc\ values would cause reionization to occur too early \citep{Munoz_2024}. These details depend crucially on the distributions of \xiion\ and \fesc\, their dependence on the UV luminosity, and on the evolution of the UVLF. For the latter, we have come to a consensus, using deep field studies from \hst\ and \jwst\,  that the UVLF at $7<z<10$ extends as a power-law for galaxies as faint as $\muv < -15$ \citep{Finkelstein_2022b,Leung_2023,PG_2023,Adams_2024,Donnan_2024}. Studies from lensed galaxies provide evidence that this extends to at least $\muv < -13$ \citep{Livermore_2017,Bhatawdekar_2019,Atek_2024}, notably fainter than the LMC and SMC \citep{Maucherat-Joubert_1980}.  
%
%
\citet{Munoz_2024} showed that given this evolution of the UVLF, if EoR galaxies have high \xiion, \textit{and} high \fesc, then the global emissivity, $\dot n$ / (s$^{-1}$~Mpc$^{-3}$), is potentially too high, creating a  ``photon budget crisis'' \citep[see also,][]{Ocvirk_2021}.   This makes it imperative that we constrain $\xiion$ and $\fesc$ in EoR galaxies. 

%

Here we study this problem using a new analysis of \jwst\ imaging and spectroscopy from the  Cosmic Evolution Early Release Science (CEERS) survey and \jwst\ Advanced Deep Extragalactic survey (JADES) of galaxies at $4.5 < z < 9.0$ that have photometry from \hst\ and \jwst, and spectroscopy from the \jwst/NIRSpec prism.  These data cover the rest-frame region from the rest-frame far-UV ($\sim$1500~\AA) through the H$\beta$ (+ \oiii) lines (or to \ha, depending on the redshift).   We model these  spectroscopic and imaging data for each galaxy simultaneously, where the strength of the nebular emission is compared to the shape of the rest-UV/optical continuum.  With this we can infer an \textit{indirect} estimate of the escape fraction, as the modeling constrains the number of ionizing photons and the nebular emission constrains the number that do \textit{not} escape. 
%
%
There is precedence for this type of analysis based on observations of \ion{H}{2} regions \cite{Oey_1997} and resolved stars in the starburst galaxy NGC~4214 \citep{Choi_2020}.   In both cases, these works use the observations of the far-UV and nebular emission to constrain the number of LyC photons, and then compared these to the emission in the nebular component.  
%
%
At high redshift, similar analyses have attempted to constrain \fesc, finding that some galaxies have \fesc\ = 50--80\% \citep{Topping_2022}.   Here, we test the ability of the SED modeling to constrain \fesc\ using both real and simulated galaxies (see Section~\ref{section:results_fesc} and Appendix~\ref{appendix:fesc_sed}).  Applying this to the real data for the $4.5 < z < 9.0$ galaxies, we find that the nebular emission is sufficiently strong to account for nearly all of the produced ionizing photons:  the escape fractions are low, less than a few percent for the majority of galaxies.    

The outline for the \textit{Paper} is as follows.   In
Section~\ref{section:data} we discuss the CEERS and JADES datasets, and how we selected our sample. 
 In Section~\ref{section:analysis} we discuss our analysis methods to model the stellar populations and nebular emission of the galaxies based on their imaging and spectroscopic data. We also discuss the measurements of emission line fluxes from the spectroscopic data.  In section~\ref{section:results} we discuss the constraints on \xiion\ and \fesc, their distributions as a function of UV luminosity and redshift, and how these constrain $\dot{n}$.  In Section~\ref{section:discussion} we discuss how \xiion\ and \fesc\ depend on galaxy properties, and we discuss the implications our results have for reionization.  In Section~\ref{section:summary} we present our conclusions.

Throughout we use a flat cosmology with $\Omega_{m,0}=0.315$, $H_0 =
67.4$~km s$^{-1}$ Mpc$^{-1}$, and $\Omega_{b,0}=0.0493$ \citep{Planck_2020}.    All magnitudes
reported here are on the \textit{Absolute Bolometric} (AB) system
\citep{Oke_1983}.   Throughout we use the \citet{Chabrier_2003} initial
mass function (IMF) for relevant quantities.  We also frequently drop the units for the ionization production efficiency,  $\xiion$, which are s$^{-1}$ (erg~s$^{-1}$~Hz$^{-1}$)$^{-1}$, throughout. 

\section{Data and Sample}\label{section:data}

For this work, we select galaxies with excellent multiwavelength coverage from imaging from \hst\ and \jwst, and spectroscopy from the \jwst\ NIRSpec/PRISM.  The prism data are important because they cover the wavelength range 0.8--5.3~\micron\ contiguously, with varying resolution of $R\sim 80$ at $\simeq$2.5~\micron\ to $R\sim 300$ at $\simeq 5$~\micron. This provides measurements of the galaxy continua and emission features that can be modeled to constrain properties of the stellar populations and nebular gas.  However, the prism spectra face additional challenges including the fact that the amount of flux received depends on the galaxy morphology, location of the galaxy within the MSA slitlet, the wavelength-dependent image quality and spatial resolution.  This is a primary reason we combine the prism data with the photometric imaging. Because the imaging provides matched aperture flux densities over a wide range of wavelength, typically 0.6--5~\micron, we can model these simultaneously with the prism data to measure any flux-calibration issues from the data itself.  

\subsection{CEERS Data}\label{section:ceers_data}

CEERS is a \jwst\ Early Release Survey that provides \jwst\ imaging and spectroscopy of the legacy CANDELS/EGS field \citep{Finkelstein_2025a}.  CEERS includes NIRCam and MIRI imaging, as described in \citep{Bagley_2023} and \citep{Yang_2023}.  We use an updated photometric catalog following a similar procedure as that in \citet{Finkelstein_2024}.  This includes matched-aperture flux density measurements from \jwst/NIRCam with the F115W, F150W, F200W, F277W, F356W, F410M, and F444W filters, combined with \hst\ imaging from ACS in the F606W and F814W filters, and from WFC3/IR in the F125W, and F160W, with partial coverage from F105W and F140W, drawn from all the archival \hst\ data on these fields, with most of these data coming from CANDELS \citep{Grogin_2011, Koekemoer_2011}. For \editone{22 of the 114} of the CEERS spectroscopic sources ($\approx 20$\%), no \jwst/NIRCam imaging exists. \editone{For these} we use the \hst--based catalog of \citet{Finkelstein_2022a}, which includes the \hst\ imaging above combined with \spitzer/IRAC imaging at 3.6 and 4.5~\micron.  \editone{We include these objects for completeness, but we have tested that removing them does not change the results.} The CEERS MIRI data include deep imaging in F560W and F770W in two pointings that overlap with NIRCam \citep{Yang_2023}, and we include those data where they are available following \citet{Papovich_2023}.  

We use the \jwst/NIRSpec prism data for CEERS, with additional NIRSpec/PRISM data obtained in the CEERS field from \citet{Arrabal_Haro_2023b}.   The data reduction steps are described in \citet{Arrabal_Haro_2023a}, and we use the most recent version available (v0.9, Arrabal Haro et al. in prep.)  We then use an internal catalog of spectroscopic redshifts from the CEERS team based using the LIne MEasuring library \citep[LIME][]{Fernandez_2024}.  

\subsection{JADES Data}\label{section:jades_data}

We include published data from JADES (\citealt{Eisenstein_2023}), which provides \jwst/NIRCam imaging and NIRSpec/PRISM data for galaxies in the GOODS-N and GOODS-S fields.    Here, we use the photometric catalog from \citet{Hainline_2024}, which includes flux-density measurements from \jwst/NIRCam, and \hst/ACS and WFC3/IR. For GOODS-N, these include imaging from NIRCam using the F090W, F115W, F150W, F182M, F200W, F210M, F277W, F335M, F356W, F410M, and F444W filters, combined with \hst\ imaging from ACS in F435W, F606W, F775W, F814W, and F850LP, and WFC3/IR in F105W, F125W, F140W, and F160W.    The GOODS-S catalog includes all of the filters used for GOODS-N combined with additional NIRCam medium-band imaging in F430M, F460M, and F480M \citep{Williams_2023}. 

The JADES NIRSpec/PRISM data provide spectroscopy for $\sim$4000 galaxies in the GOODS-N and GOODS-S fields \citep{DEugenio_2024}. 

\begin{figure}[t]
  \begin{centering}
    \includegraphics[width=0.47\textwidth]{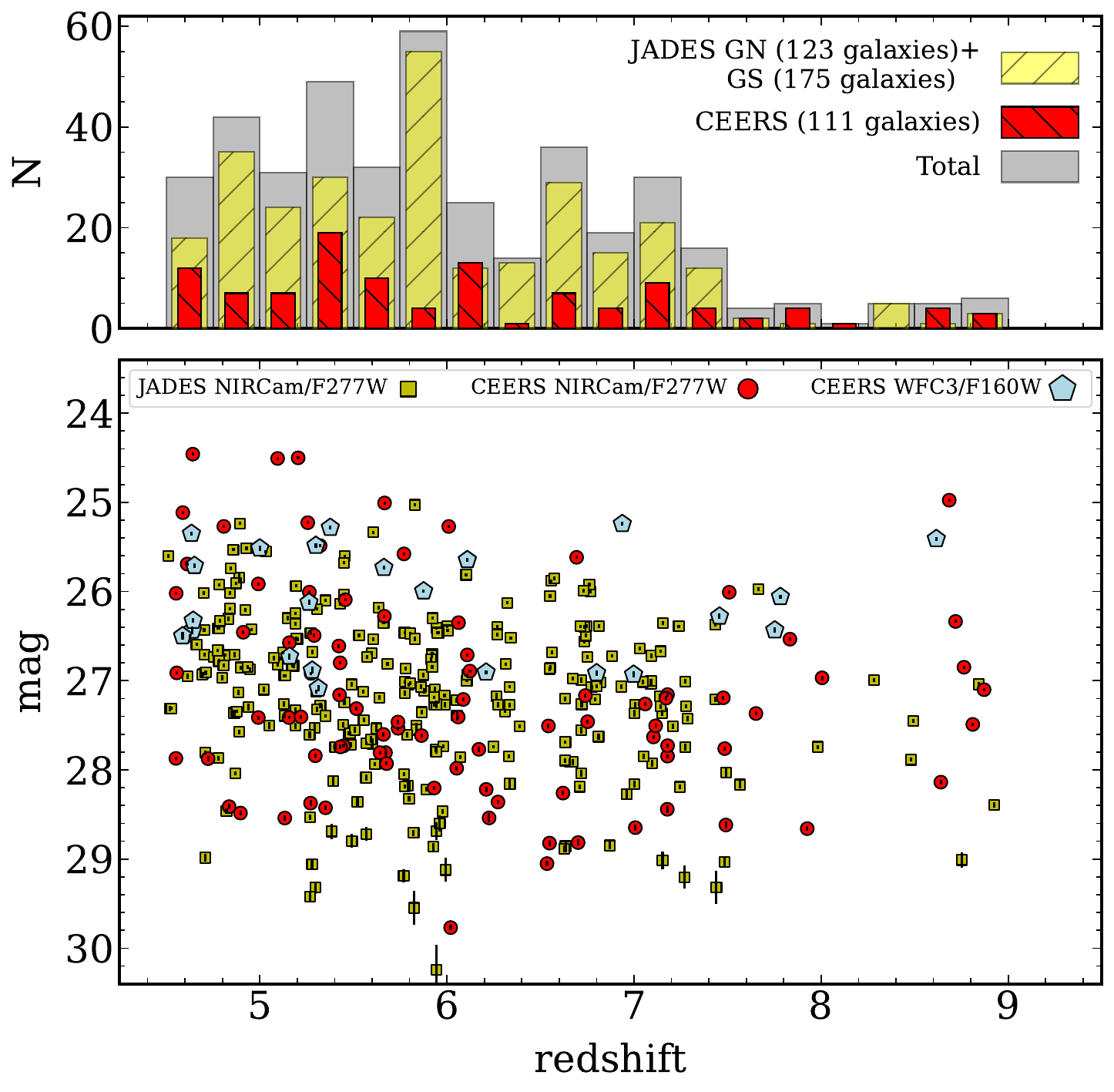}
  \end{centering}
    \vspace{-18pt}  \caption{Redshift and magnitude distributions of the spectroscopic sources at $4.5 < z < 9.0$ used in this work  The bottom panel shows the redshift versus the NIRCam F277W magnitude distribution for galaxies in JADES and CEERS. A minority of the CEERS galaxies do not include NIRCam imaging, and we show the WFC3/F160W magnitude. The top panel shows the histograms show the distribution of sources from JADES, CEERS, and the total (the sum of the two), as labeled.   JADES includes more galaxies overall, but weighted toward galaxies at the lower end of our redshift range.  The CEERS sample is more evenly spread in redshift, and provides greater coverage at the higher redshift end of the redshift range.  }\label{fig:zhist}
  \end{figure}

\subsection{Sample Selection}\label{section:samples}

Our goal is to study the ionization conditions, escape fractions, and stellar population parameters in galaxies at redshifts near and into the EoR.   To this end, we select galaxies from JADES and CEERS following conditions.   (1) The galaxies have coverage from broad-band imaging using the photometric catalogs above.  (2) The galaxies have NIRSpec/PRISM data from the datasets above.  (3) The galaxies have a confirmed spectroscopic redshift that places them at $4.5 < z < 9.0$.  For this last step, the lower redshift bound ensures that the NIRSpec/PRISM data cover the rest-frame far-UV ($\sim 1500$~\AA), while the upper redshift bound ensures the data cover the important rest-frame optical strong emission lines, \hb\ + \oiii\ $\lambda\lambda$4959, 5008.   The former will be important for constraining the rest-frame UV luminosity while the latter will be important for constraining the number of ionizing photons, and therefore the ionizing production efficiency and inferred escape fraction.

We select from the JADES catalog \citep{DEugenio_2024} all galaxies with $4.5 \leq z < 9.0$ with a spectroscopic quality flag of ``A'' or ``B''.   This results in a sample of 129 galaxies in GN and 183 galaxies in GS, after removing duplicates (keeping the spectrum with higher S/N following a visual inspection, see \citealt{DEugenio_2024}).  \editone{We then remove candidate AGN, either with X-ray detections \citep{Xu_2016,Luo_2017} or identified as possible AGN from \citet{Juodzbalis_2025} or \citet{Kocevski_2025}  This reduced the JADES sample to 123 galaxies in GN and 175 galaxies in GS.}  From CEERS, we select galaxies from the internal team release that includes data from all CEERS NIRSpec observations, with redshifts from an automatic line finder \citep{Fernandez_2024}, and visual inspection by multiple CEERS-team members. \editone{We again remove possible AGN using \citet{Kocevski_2025} and no galaxies are detected in the current X-ray coverage of the CEERS field.}  We identify 114 unique galaxies in CEERS in this redshift range.

Figure~\ref{fig:zhist} shows the redshift distribution of the samples.  While JADES provides more spectra overall, their distribution is weighted toward galaxies at the lower redshift end of our redshift range.   The CEERS data fills distribution at higher redshifts.  This redshift range covers galaxies from the period after reionization ends, $z \gtrsim 5$ \citep{Becker_2019}, through the period when it begins, $z \gtrsim 8$ \citep{Planck_2020}.   Nevertheless, selection effects will remain important as our sample requires the presence of spectroscopic feature(s) to be included in the sample (i.e., emission lines).  For all of the galaxies in our sample, this includes the presence of the \oiii\ 4959, 5008 doublet and other lines.  Future studies with deeper exposures, primarily from the NIRSpec/PRISM will be needed to determine more complete samples based on redshifts that are not necessarily dependent on the presence of emission lines.  This will be possible in the future from surveys such as CAPERS \citep{Dickinson_2024}.   We return to the importance of selection effects below in Section~\ref{section:discussion}.


\begin{deluxetable*}{c@{\hskip 25pt}lcc}
\tablecolumns{4}
\tablewidth{0pt}
\tablecaption{Parameter Settings for \bagpipes\label{table:bagpipes}}
\tablehead{\colhead{Model Component} & \colhead{Parameter} & \colhead{Prior}  & \colhead{Limits} }
  %
%
\startdata
& redshift, $z$ & $N(\mu=z_\mathrm{sp}$, $\sigma=5\sigma_z$)$^\dag$ &  $z_\mathrm{sp} \pm 0.2$  \\[2pt]\hline
\multirow{3}{*}{\shortstack{Star-Formation History: \citep{Iyer_2019} \\
    Dense-basis, Gaussian Processes  } } & instantaneous SFR / \msol\ yr$^{-1}$ & $\log_{10}$ & ($10^{-3}$, $10^3$) \\ 
& lookback times to bin $X$, $\{t_X\}$ / yr &  Dirichlet & $X=\{25,50,75\}$ \\ 
& stellar mass, $\log(M_\ast / M_\odot)$ & uniform & (6, 13)  \\[2pt] \hline
\multirow{2}{*}{Dust attenuation} 
&attenuation law & \ldots & \cite{Calzetti_2001} \\ 
&attenuation, $A(V)$ / mag & uniform & (0, 2) \\[2pt] \hline
metallicity & $Z / Z_\odot$ & uniform & (0, 1) \\[2pt] \hline 
\multirow{2}{*}{Nebular Emission} 
&\editone{ionization parameter, $\log U$} & uniform & \editone{($-4$, 0)} \\ 
& escape fraction, \fesc & $\log_{10}$ & ($10^{-4}$, 1) \\[2pt] \hline
\multirow{2}{*}{IGM damping wing} & Size of Bubble, $R / \mathrm{cMpc}$ & uniform & (0,150) \\
& average neutral fraction, $\bar x_e$ & uniform & (0.0, 1.0) \\[2pt] \hline
\multirow{3}{*}{NIRSpec specific} & line-spread factor (LSF) & uniform & (1.0, 6.6) \\
 & velocity dispersion, $\sigma_v / (\mathrm{km~s^{-1}})$ & $\log_{10}$ & (50, 1000) \\ 
 & noise scaling factor & $\log_{10}$ & (1, 10) \\[3pt]\cline{2-4}
 \multirow{3}{*}{\shortstack{ flux-calibration factor applied \\to spectra: second order polynomial,\\ $A_0 + A_1\lambda + A_2\lambda^2$}}
 & $A_0$ & $\log_{10}$ &  $ (0.1, 10)$ \\
  & $A_1$ & $N(\mu=0, \sigma=0.25)$ & $ (-0.5, 0.5)$ \\
  & $A_2$ &  $N(\mu=0, \sigma=0.25)$ & $(-0.5, 0.5)$ \\
\enddata
\tablenotetext{$\dag$}{$N(\mu, \sigma)$ corresponds to a Gaussian, ``normal'', prior described by mean, $\mu$, and variance, $\sigma^2$}
%
\end{deluxetable*}
%

\section{Analysis}\label{section:analysis}

\subsection{Spectral Energy Distribution
  Modelling}\label{section:bagpipes}

We fit the spectral energy distributions (SEDs) of the galaxies in our sample using stellar population synthesis models.   \editone{We model each galaxy using \bagpipes\ \citep[v1.3.1,][]{Carnall_2018}}, which is a Bayesian SED-fitting code that has the capability to fit simultaneously the multiband photometry and spectroscopy with stellar population synthesis models and nebular emission models generated over a wide range of parameters.  The code has flexibility in the type of stellar population models, star-formation history, dust attenuation, and nebular emission.  The code computes posterior probability densities for model parameters given the data by calculating a likelihood weighted by priors, and samples the posteriors for the \editone{parameters using \texttt{nautilus} \citep{nautilus}.  The newest version of \bagpipes\ also includes updated \texttt{Cloudy} nebular models (v25, \citealt{cloudy25}).}

We have added several features to \bagpipes\ for our analysis.  First, one of our goals is to determine the constraints on the inferred LyC escape fraction, \fesc.  By default, the current nebular emission models in \bagpipes\ generated from \texttt{Cloudy} assume 100\% of the incident radiation is absorbed and used to generate the nebular emission (i.e,. \fesc = 0).    Following \citet{Morales_2024}, we have added a parameter (\texttt{nebular:fesc}) such that the nebular emission is reduced by a factor of $1 - \fesc$.  This models the LyC escape as a fraction of sightlines that are ``clear'' (akin to a covering fraction equal to $(1-\fesc)$).     As we show, the majority of galaxies favor very low escape fractions of $< 5$\%, implying such clear channels are infrequent.   In a future work, we plan to explore other types of \fesc, including cases where the nebulae are truncated or ``density bounded'' \citep{Nakajima_2014,Plat_2019}.

Second, there are several processes that can impact the shape of the UV continuum with respect to the stars.  One process is the nebular continuum, produced from free-bound and the two-photon emission from the $2s\rightarrow1s$ transition in hydrogen.  These are already included in the \bagpipes\ \texttt{Cloudy} models.  
%
%
Another process is increased \ion{H}{1} absorption from the IGM or ISM of the host galaxy, both of which can alter the shape of the UV spectra, particularly around \lya.   Several studies have measured an increase in the strength of the \ion{H}{1} damping wing in galaxies at the EOR \citep{Heintz_2024,Umeda_2024}.   
We have added to \bagpipes\ the \ion{H}{1} damping wing to allow for this effect following the formalism of \citet{Dijkstra_2014}.  This requires two parameters, \texttt{IGM:R}, the comoving distance to the patch of IGM at the edge of the ionized ``bubble'' around a galaxy, and \texttt{IGM:x\_HI}, the average gas-neutral fraction of the patch of the IGM (see also \citealt{Heintz_2024}). This allows the models not to rely overly on the nebular continuum to reproduce the shape of the rest-UV continuum.  

Third, through testing we found that modeling of the JWST/NIRSpec data frequently under-fit the spectral resolution.  \bagpipes\ uses a model for the variable spectral resolution of the NIRSpec/PRISM available from the \textit{Space Telescope Science Institute} (STScI)\footnote{\url{https://jwst-docs.stsci.edu/files/97979440/97979447/1/1596073265467/jwst\_nirspec\_prism\_disp.fits}}, but this assumes galaxies are unresolved and are centered in the MSA slits.  The reality is that the effects of galaxy morphology and location within the slit lead to a type of morphological spectral broadening similar to that seen in slitless data \citep{Simons_2023}.  \editone{Indeed, recent updates to the NIRSpec calibration find this effect is wavelength dependent and can be as much as 50\%\footnote{\url{https://www.stsci.edu/contents/news/jwst/2025/new-pathloss-corrections-for-nirspec-multi-object-and-fixed-slit-spectroscopy-are-available}}.}  To compensate, we added a nuisance parameter that multiplies the spectral resolution by a factor with respect to the spectral file (we refer to this as a line-spread factor (LSF)).  This parameter improves the quality of the fits, particularly noticeable in the width of the nebular emission lines.   We allow the galaxy spectra to be broadened by an intrinsic velocity dispersion, $\sigma_v$, though in practice this is much less than the instrument dispersion.  We also allow the noise to be scaled by a nuisance parameter, and we allow for a multiplicative scaling factor (assumed to be a second order polynomial) to match the spectral flux to the broad-band photometry.  

Table~\ref{table:bagpipes} lists the parameters, limits, and priors we used with \bagpipes\ for the SED fitting.   We use the Binary Population and Spectral Synthesis  (BPASS, v2.2.1) models from \citet{Stanway_2018} formed with a
Chabrier IMF with an upper-stellar mass cutoff of 300~\msol, and that include the effects of stellar binaries.  This choice is motivated by the fact that high-redshift galaxies appear to require additional ionization from high-mass stars \citep[e.g.,][]{Katz_2024}, and as we show below these models reasonably reproduce both the spectral continua and emission line intensities \citep[consistent with other results, see, e.g.,][]{Seille_2024}.    
%
%
For the star-formation history, we employ the parameterization of \citet{Iyer_2019}, which uses Gaussian Processes to represent the SFR in different time bins, with a boundary condition that it produces the instantaneous SFR and stellar mass at the moment of observation. 
 We choose three time bins, representing when the star-formation history has formed, 25, 50, and 75\% of the stellar mass ($t_{25}$, $t_{50}$, and $t_{75}$).  \editone{We have also tested the use of star-formation histories with fixed bins in time \citep[e.g.,][]{Leja_2019}.  Using bins that sample the time quasi-logarithmically following \citet{Begley_2025}, we find that this has no net impact on our results.  We nevertheless adopt the star-formation histories parameterized by Gaussian Processes \citep{Iyer_2019} as they provide flexibility in the star-formation history to account for bursts and quenching events without setting a prior on the width of the star-formation time bins.}
 
 We adopt the \citet{Calzetti_2001} dust attenuation law, which is appropriate for more highly attenuated, blue star-forming galaxies (where for less attenuated galaxies the differences are small between \citealt{Calzetti_2001} and laws like the SMC, see \citealt{Salmon_2016}).  We also assume the stellar continua and nebular emission are attenuated by the same amount \citep{Reddy_2015}, which is appropriate as our galaxies generally have low dust attenuation, with young ages \citep{Lecroq_2024}, so systematic shifts in the attenuation will have minimal impact on our results \citep{Sanders_2021,Shen_2024}. \editone{We use the dust attenuation derived from the SED fitting to correct emission line fluxes and the UV continuum.  We opt not to use the Balmer lines as not all lines are available for all galaxies and the SED fitting uses all available information \citep[see also][]{Vijayan_2025}.  As we show below (Section~\ref{section:xiion}), the \xiion\ values derived from dust-corrected \ha\ and \hb\ are consistent, which further justifies these choices.}  \editone{For the escape fraction, we adopted a logarithmic prior, that is, linear in $\log \fesc$.  This choice resulted from our experiments that showed most galaxies have low \fesc, where the logarithmic prior allowed us to better sample the posterior at low values of \fesc.  We have checked that this choice has no impact on our conclusions.}
 
We then use \bagpipes\ to fit these models to the observed photometry and NIRSpec/PRISM data for each galaxy. 
 We used the Texas A\&M High Performance Research Computing Grace cluster\footnote{Named for Grace Hopper, \url{https://en.wikipedia.org/wiki/Grace_Hopper}. } to perform the calculations.  Grace is a Dell x86 HPC cluster, and has 800 48-core compute nodes, each with 384 Gb RAM.   Run times for the fit to converge for each galaxy are typically \editone{1 to 4 hrs of CPU time.} 

\begin{figure*}
       \gridline{
    \fig{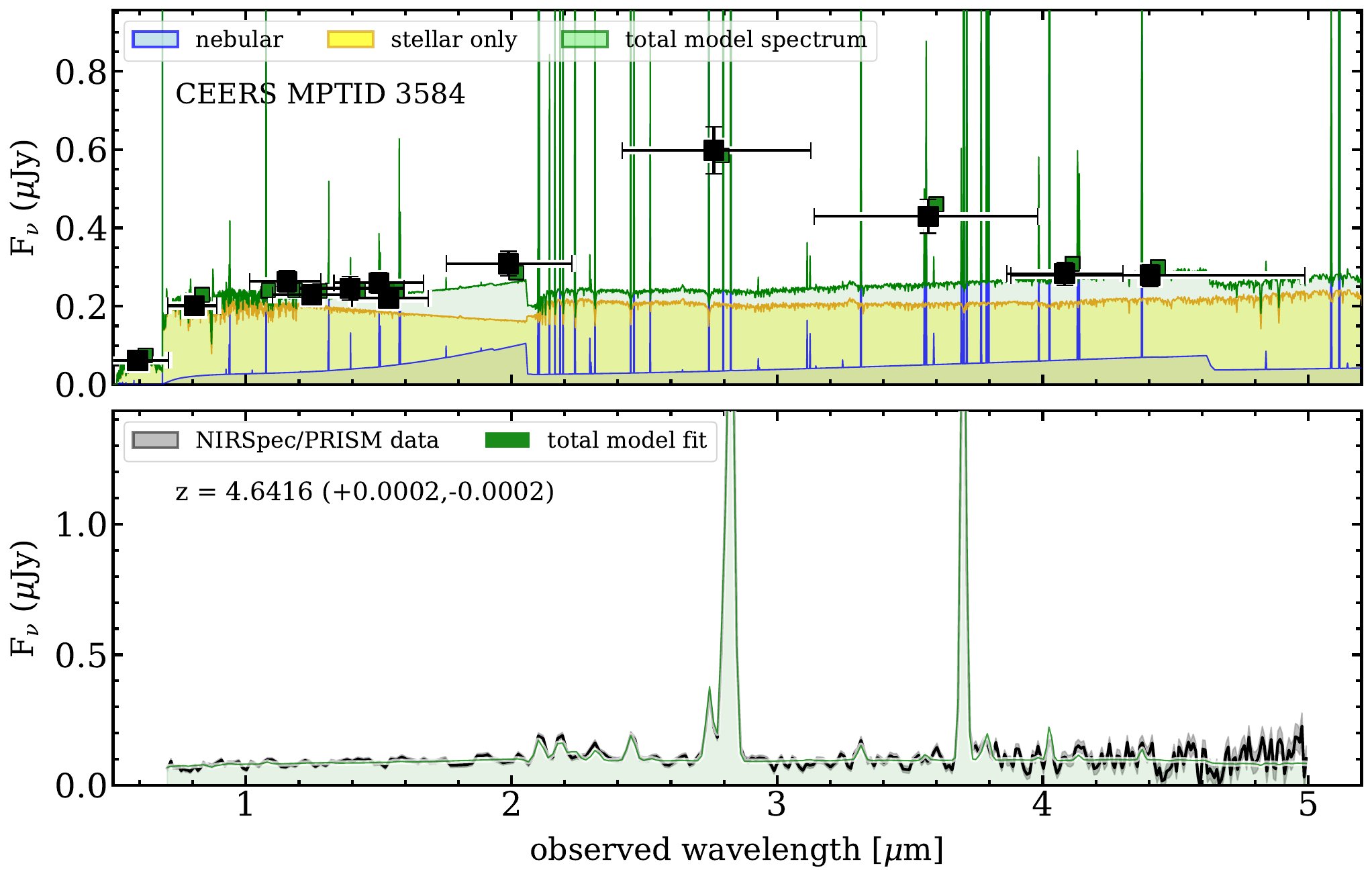}{0.5\textwidth}{}
    \fig{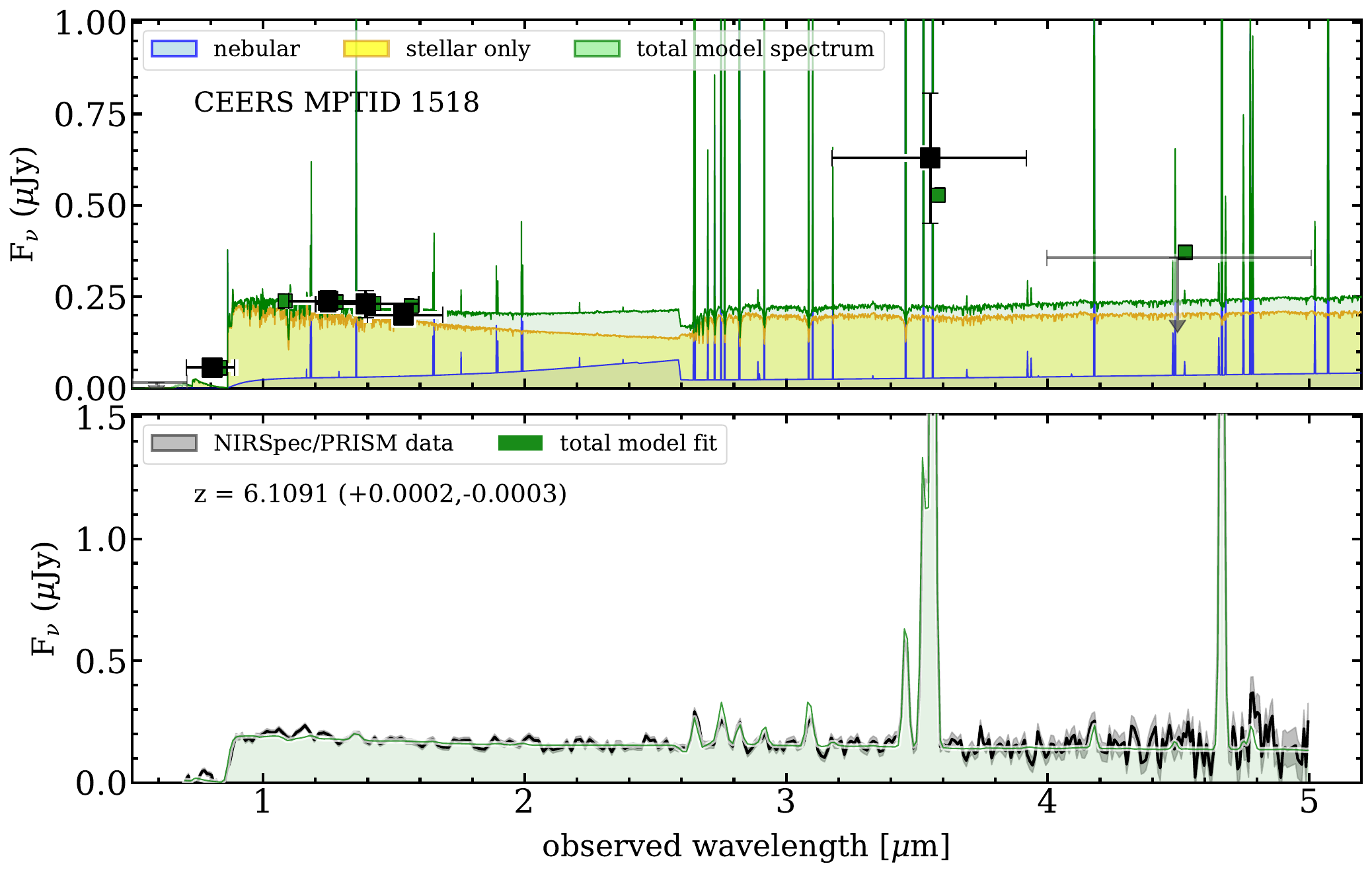}{0.5\textwidth}{}
  }\vspace{-12pt}
           \gridline{
    \fig{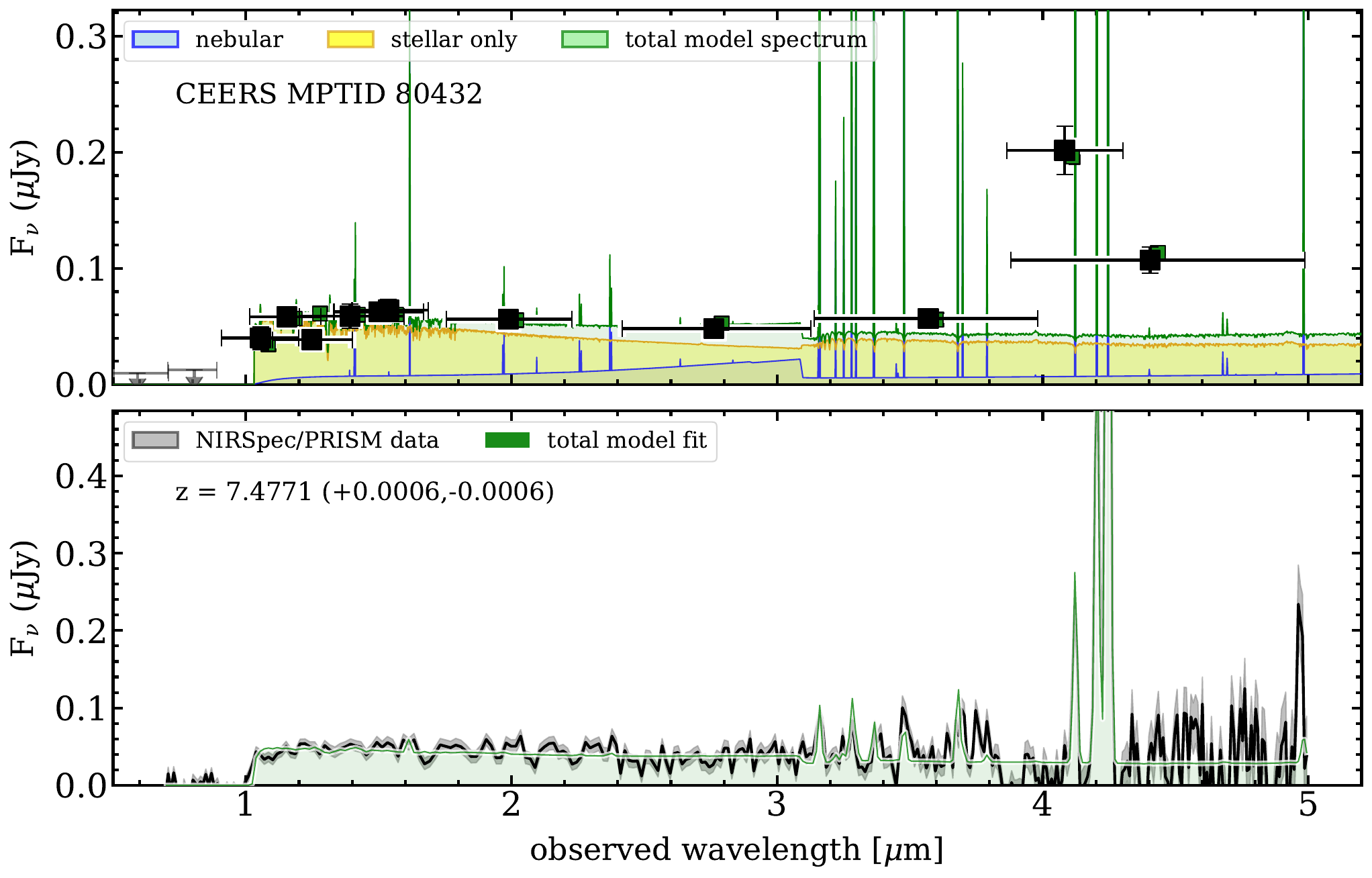}{0.5\textwidth}{}
    \fig{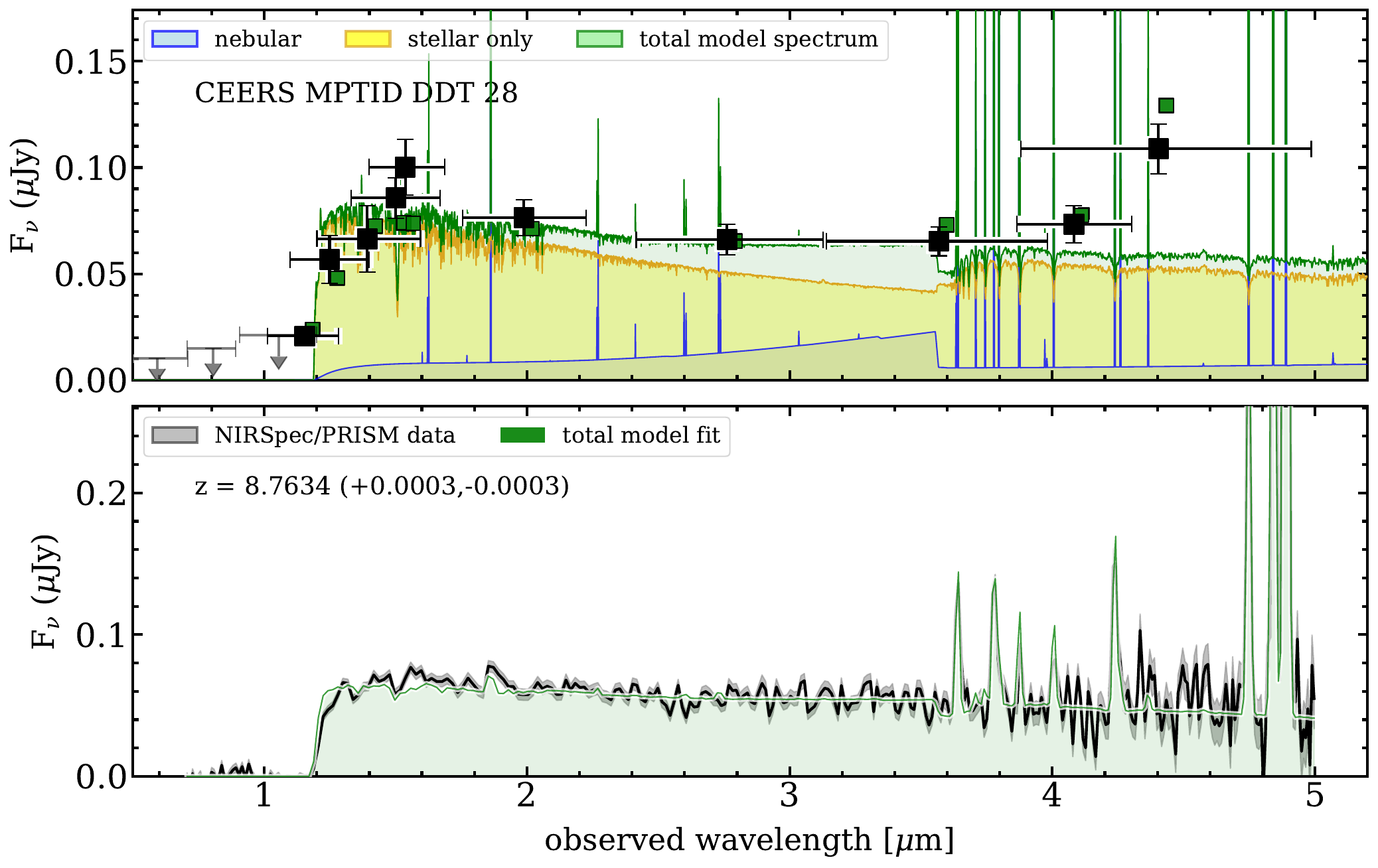}{0.5\textwidth}{}
  }
    \vspace{-22pt}
    \caption{Examples of SED fits to galaxies in the CEERS sample. Each pair of plots shows the best-fit model for an individual galaxy, labeled by the MPT ID number. In each pair of plots, the top panel shows the model fit to the broad-band photometry.  The black data points with error bars show the measured flux densities. The curves show a best-fit total model (green) and the contributions to that model from the stellar light (yellow) and nebular emission (blue).  The green squares are the model photometry.  The bottom panel of each galaxy shows the same model fit (green curve and shading) to the NIRSpec prism data (grey).  }\label{fig:sedfits_ceers}
\end{figure*}

\begin{figure*}
       \gridline{
    \fig{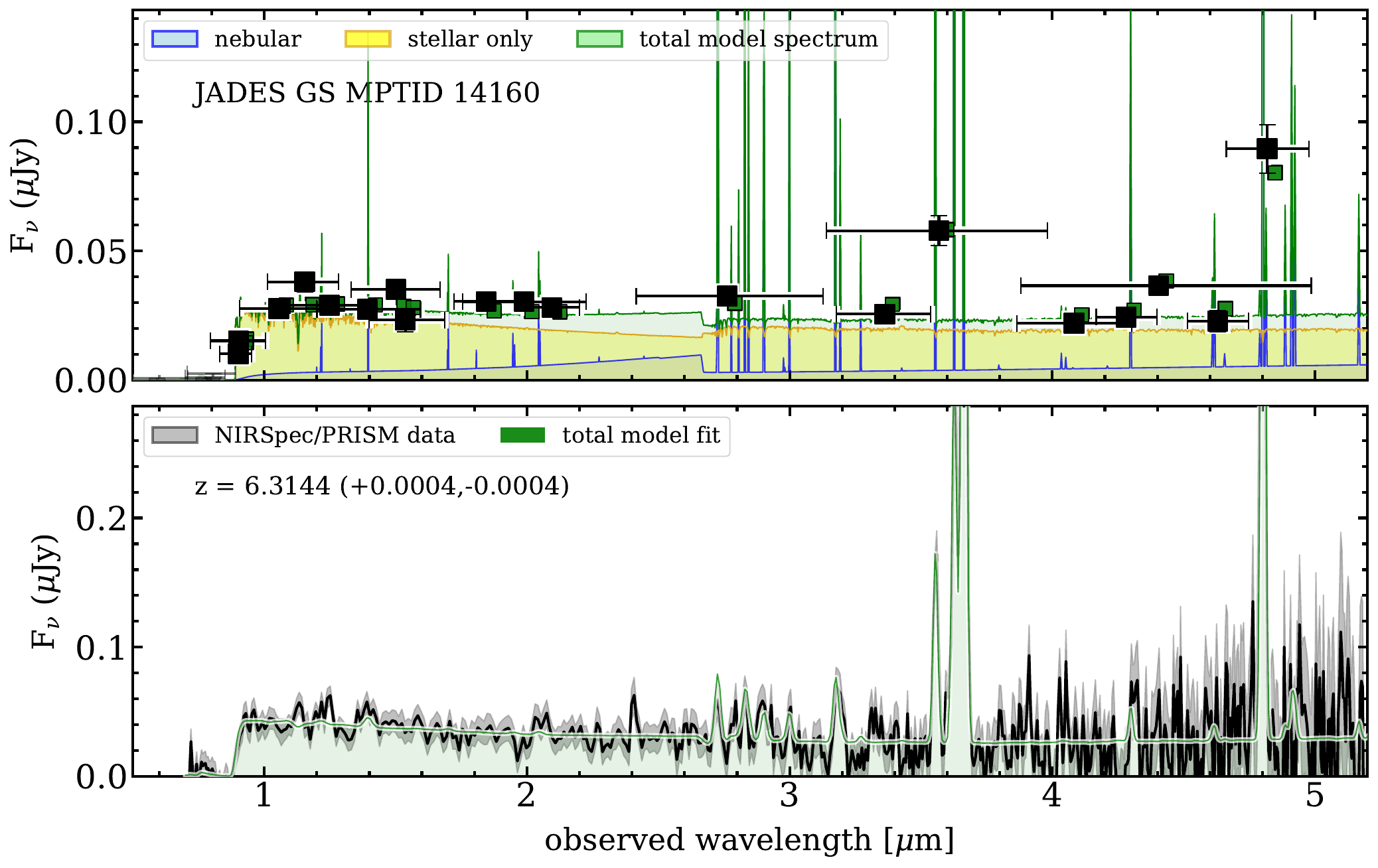}{0.5\textwidth}{}
    \fig{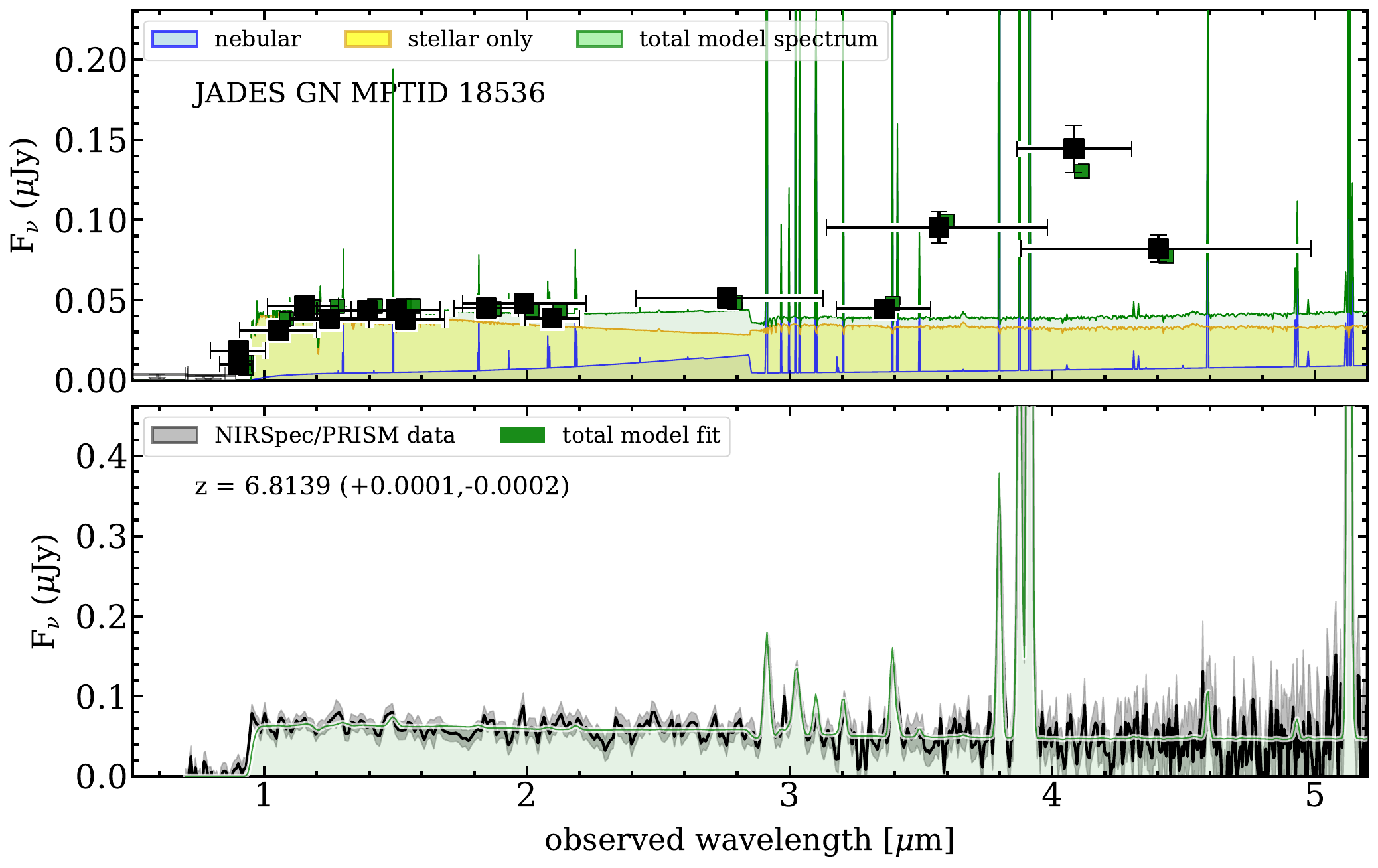}{0.5\textwidth}{}
  }\vspace{-12pt}
           \gridline{
    \fig{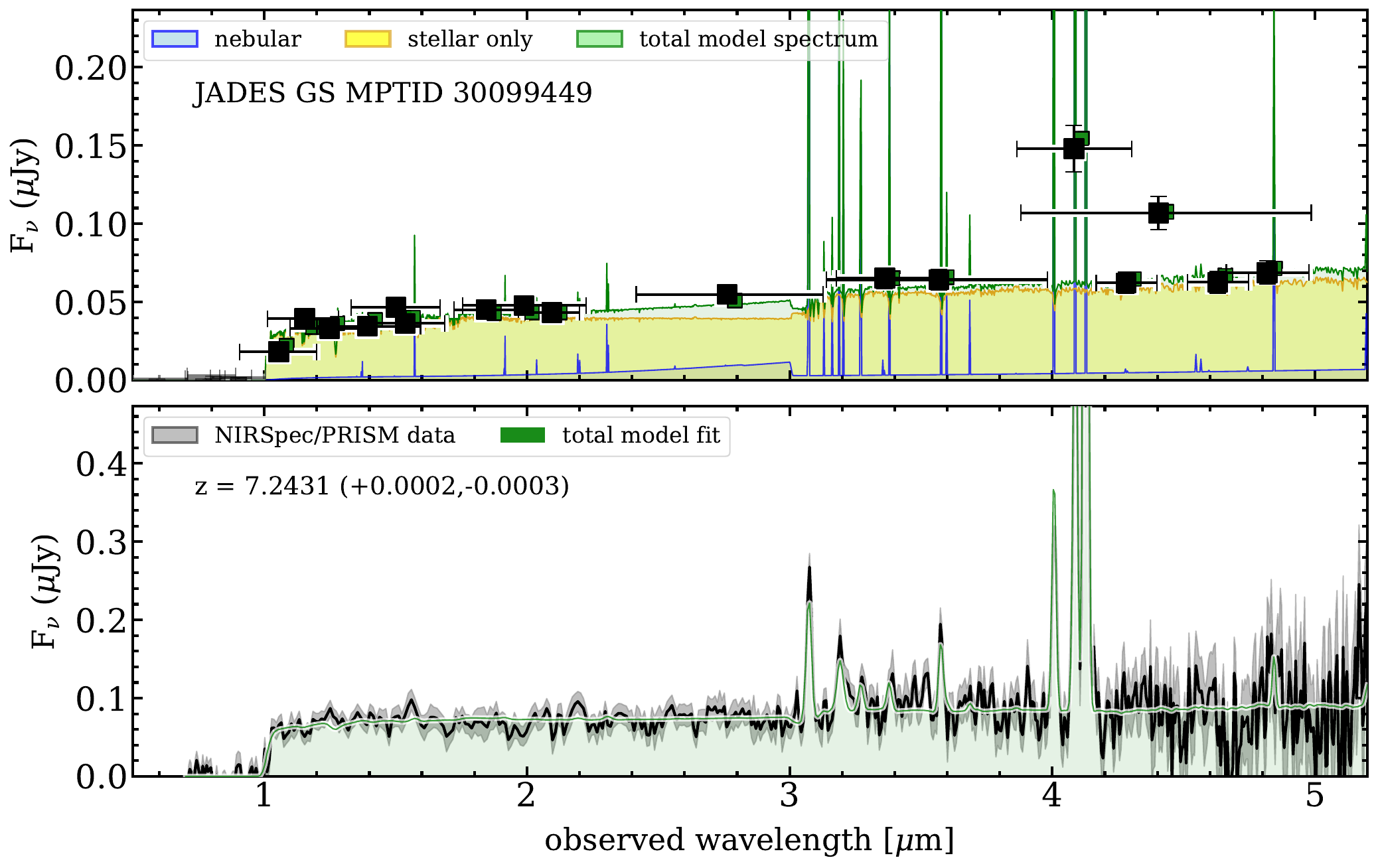}{0.5\textwidth}{}
    \fig{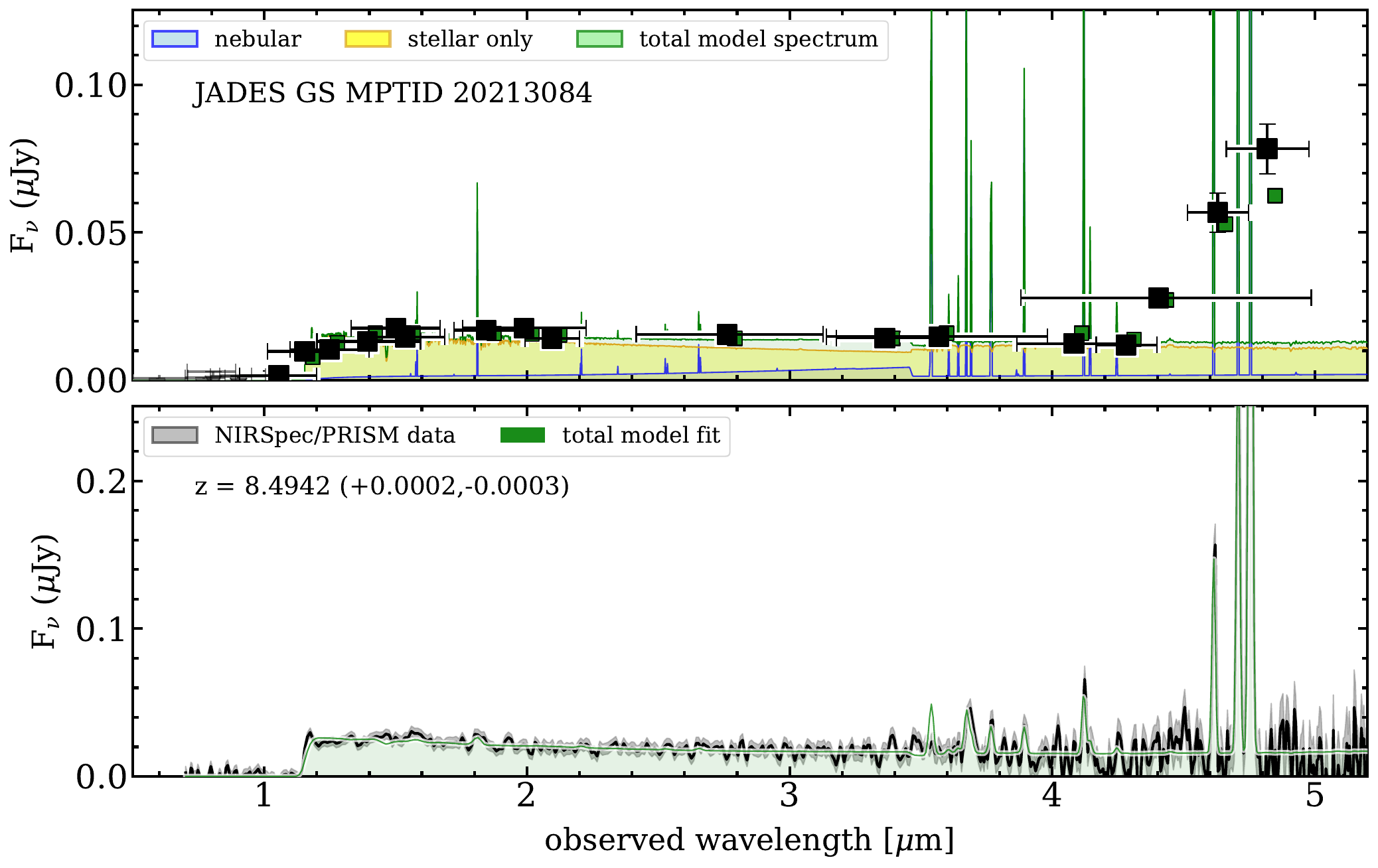}{0.5\textwidth}{}
  }
    \vspace{-22pt}
    \caption{Same as Figure~\ref{fig:sedfits_ceers} but for example galaxies in the JADES sample.  }\label{fig:sedfits_jades}
\end{figure*}

Figures~\ref{fig:sedfits_ceers} and \ref{fig:sedfits_jades} show results from the SED modeling of example galaxies from the CEERS and JADES samples, respectively.  These  galaxies illustrate the quality of the fits from which we derive results.  The galaxies all show strong rest-UV continua with a ``break'' at rest-frame \lya.  All galaxies also show evidence of rest-optical emission lines, \oiii\ 5008, and \hb\ 4861, and \ha\ 6564 for galaxies with $z < 6.7$.  These features are well-fit in both the prism data (lower panels in the figures) and for the broad-band photometry (upper panels).  In some cases, the prism data show evidence of rest-UV emission lines, e.g., \ion{C}{3}] 1909 in CEERS MPTID DDT 28 (in Figure~\ref{fig:sedfits_ceers}). 

\subsection{Measuring \muv, $\beta_\mathrm{UV}$, and \fesc}

For this analysis, we are focused on the evolution of the ionizing production efficiency, \xiion, and LyC escape fraction, \fesc, as a function of \muv.  We will also consider trends between \fesc\ and the UV spectral slope, $\beta_\mathrm{UV}$, $f_\lambda \propto \lambda^{\beta_\mathrm{UV}}$, defined using regions devoid of emission features over rest-frame $1268-2580$~\AA\ \citep{Calzetti_1994}.  For the UV spectral slope, we measure this from the NIRSpec data after correcting for the slit-loss flux calibration (see above).    For \muv\ and \fesc, we derive their constraints for each galaxy from the SED fits.   For each galaxy, we take 400 model fits drawn from the posterior.  For each model we measure the 1500~\AA\ rest-frame flux density, $F_{1500}$, using a top-hat filter from 1450--1550~\AA, and then calculate the rest-frame UV (1500~\AA) absolute magnitude by applying the distance modulus. 
%
%
In practice we compute \muv\ twice, once for the dust-corrected spectrum and once without.  We use the dust-corrected measurement to estimate \xiion, but we use the observed \muv\ with no dust correction when we consider the production of ionizing photons based on the galaxy UVLF (in Section~\ref{section:ndot_ion}).   For each galaxy we take \muv\ and \fesc\ as the median of the draws from the posterior, with an uncertainty from the 16th to 84th percentiles.  We also derive constraints on other stellar--population parameters, such as stellar mass, SFR, and dust attenuation.  The latter we use to correct derived quantities and emission line fluxes.  


\subsection{Measuring Emission Line Fluxes\\and Equivalent Widths}\label{section:lines}

We measure emission line EWs and line fluxes from the NIRSpec/PRISM spectra.  In principal, we could take the emission-line measurements from the \bagpipes\ SED fits directly, however, these models assume \textit{a priori} knowledge.  This includes especially the chemical abundances, where \bagpipes\ scales the metallicity with a Solar abundance pattern.   As this assumption is still under debate, with several studies arguing for $\alpha$-enrichment in these early galaxies \citep[e.g.,][]{Beverage_2024,Park_2024}, we opt to measure the emission fluxes directly from the data, using the SED fits as guidance.  In the Appendix, we compare the emission line fluxes measured directly to those predicted from the SED modeling.  We show that these are in good agreement for the strong optical emission lines (\oiii, \hb, and \ha), but with higher scatter for \oii. 
%

For each galaxy, we correct for dust attenuation and calibration/slit-loss corrections using the results from the SED modeling (see Section~\ref{section:bagpipes}).   This factor varies from $\sim$0.5 -- 2.0 at the wavelength of the redshifted \hb-line for each galaxy, with a median (inner-68\%-tile range) of 0.77 (0.60--0.99) for JADES (with no difference seen between GOODS-N and GOODS-N) and 1.52 (0.83--2.62) for CEERS.  Applying this factor ensures the NIRSpec spectra have been adjusted in a robust fashion to match the photometry, and it corrects for any wavelength-dependent calibration issues.  \editone{We have compared the EWs of the \hbeta\ and \oiii\ lines measured from the raw spectra and compared them to the corrected spectra and find they agree with a difference of $<10^{-4}$ dex with a scatter of $\leq 0.04$~dex.}
%
%
We then take use the SED modeling to provide an estimate of the continuum of each spectrum, excluding the contribution of the nebular emission lines.   This step is crucial for measuring accurate equivalent widths, and for measuring accurate Balmer emission line fluxes in the case that there is non-negligible stellar absorption present.  

We then model the strong nebular lines, \oii, \neiii, \oiii, \hb, and \ha, with Gaussians.  Because the \oii\ $\lambda\lambda 3727$, 2729 doublet is unresolved, we fit it with a single Gaussian component and measure the sum of the two lines.  For \oiii, in some cases the line is unresolved, especially at the lower end of our redshift distribution, $z\sim 4.5$, which places these lines at $\sim2.8-3.0$~\micron, where the NIRSpec/PRISM resolution is $R\sim 80-100$.  We therefore fit both lines simultaneously, forcing them to have the same dispersion and assuming a line ratio of \oiii\ $\lambda 5008$ / \oiii\ $\lambda 4959$ = 2.98.   We measure \ha\ with a single Gaussian, but we note that it is blended with \nii\ $\lambda\lambda$6548, 6583 at the resolution of the prism. This is reasonable as we expect the contribution of \nii\ to be \nii/\ha $<$ 10\% given the high redshifts and low metallicity of the galaxies in our sample \citep{Faisst_2018}.  Furthermore, we adopt $\xiion$ constraints based on \hbeta\ as this line is accessible with the NIRSpec/PRISM data for all galaxies in our sample.  Nevertheless, we show below that the \xiion\ values measured from \hbeta\ are consistent with those from \ha\ for the cases where both lines are available.   


In the analysis that follows, we explore trends with galaxy EW and emission line ratios.  For the emission-line ratios, we primarily study the ratio of \oiii\ to \oii, defined here as
\begin{equation}\label{equation:032}
\mathrm{O}_{32} \equiv \frac{ \oiii \lambda 4959 + \oiii \lambda 5008}{\oii\ \lambda 3726 + \oii \lambda 3729}.
\end{equation}
We will compare our results to other studies that explore trends in the ratio of these sums, available primarily from other space-based spectroscopic studies using the \hst\ and \jwst\ grisms \citep{Papovich_2022,Shen_2024}.  

\subsection{Measuring the Ionizing Production Efficiency}\label{section:xiion}

\begin{figure}[t]
\begin{center}\includegraphics[width=0.45\textwidth]{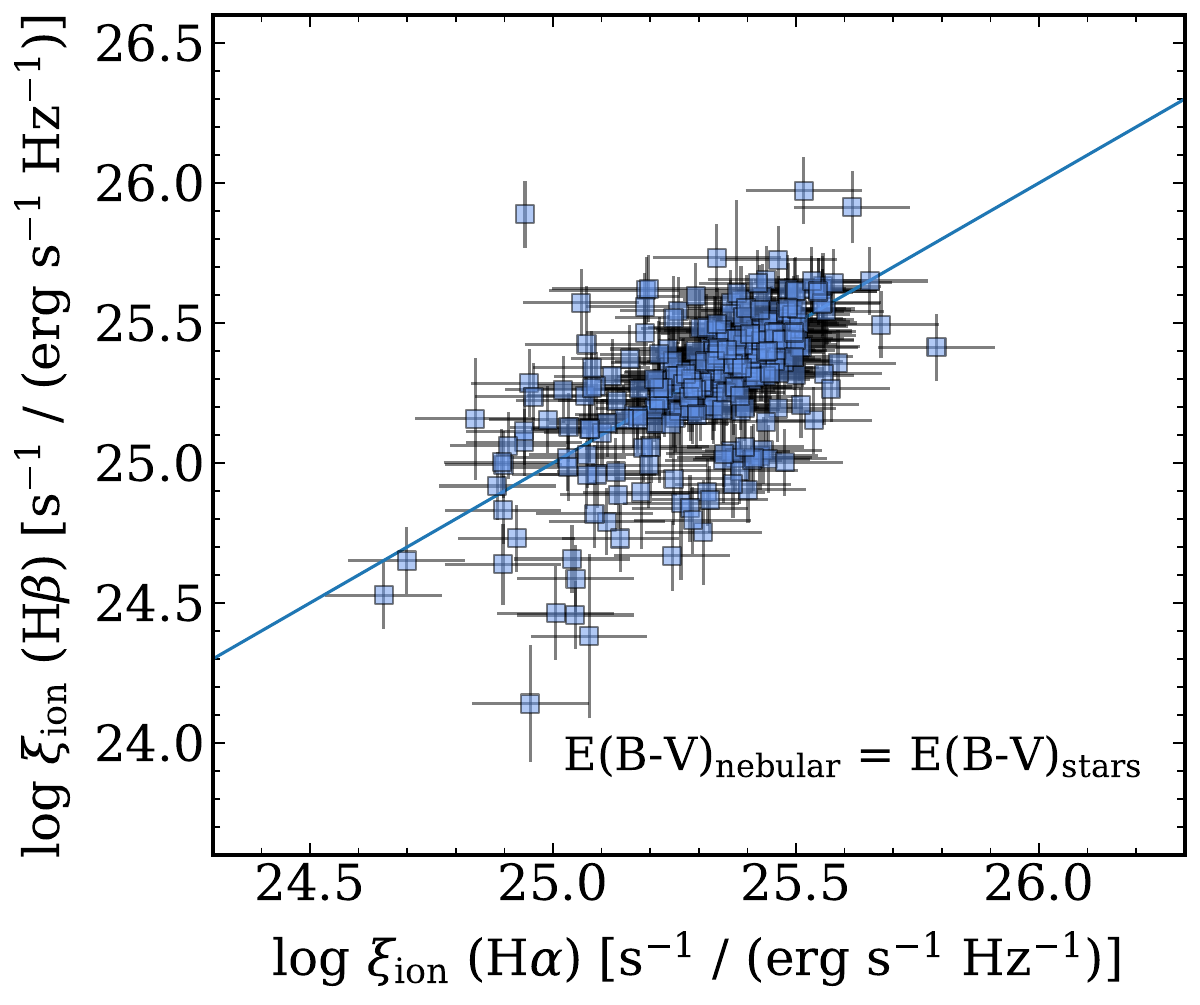}
\end{center}
\vspace{-12pt}
\caption{Comparison of \xiion\ derived from the \ha\ to that from the \hb\ lines.  The \xiion\ values are derived based on each Balmer emission line. 
 In both cases we assume the dust attenuation of the stellar continua is equal to that of the nebular emission, $E(B-V)_\mathrm{nebular}=E(B-V)_\mathrm{stars}$.  The solid line shows the unity relation.   These results show \hb\ is a reasonable surrogate for \ha\ when deriving \xiion\
  for our datasets.  }\label{fig:xi_xiHa}
\end{figure}

As introduced above (Section~\ref{section:introduction}), the ionizing production efficiency, \xiion, is defined as
%
%
$\xiion \equiv Q(\mathrm{H}_0) / L_\mathrm{UV}$,
where $Q(\mathrm{H}_0)$ is the hydrogen-ionizing production rate in units of s$^{-1}$.  $L_\mathrm{UV} = F_{1500} \times (4\pi D_L^2) (1+z)^{-1}$, is the rest-frame UV continuum luminosity, using the redshift ($z$) and rest-frame UV flux density ($F_{1500}$) derived in  Section~\ref{section:bagpipes}.
We define the ionizing production rate, $Q$, assuming Case-B recombination \citep{Leitherer_1995,Izotov_2021,Hayes_2025}, 
\begin{equation}
    Q(\mathrm{H}_0) = 2.12 \times 10^{12}~\mathrm{s^{-1}}~ \left( L(\hb) /\mathrm{erg~s^{-1}} \right), 
\end{equation}
where $L(\hb)$ is the dust corrected \hb\ luminosity.    In cases where we compare to $L(\ha)$, we assume $L(\ha)=2.86\ L(\hb)$ \citep{Osterbrock_1989}.  The definition of \xiion\ above implicitly assumes $\fesc = 0$, and we discuss this further in the sections below.

For galaxies in the redshift range $4.5 < z < 6.8$ we can compare \xiion\ derived from \hb\ to that derived independently from \ha.   Figure~\ref{fig:xi_xiHa} shows that these values agree in the average.   Interestingly, the mean (median) ratio is $\xiion(\hb)/\xiion(\ha) = 1.1$ (1.0), implying the values from $\hb$ are 10\% higher in the mean compared to those from \ha, although this is smaller than the scatter ($\pm 0.4$ dex).   
%
%
There are some objects that show  \xiion\ from \ha\ is higher than that from \hb, and outside the scatter.  These  may be a result of the dust attenuation law, where we have assumed $E(B-V)_\mathrm{stars} = E(B-V)_\mathrm{nebular}$ for all galaxies, which is appropriate for young star-forming galaxies \citep{Lecroq_2024}.  However, some studies find higher levels of attenuation in the nebular gas compared to the stellar continuum \citep[e.g.,][]{Calzetti_2001,Kreckel_2013,Reddy_2015,Salmon_2016,Robertson_2024}, particularly for dustier galaxies \citep{Reddy_2015}.  In the latter case, this would \textit{increase} the dust-corrected \hb\ luminosity and \xiion. It is also possible that these galaxies have higher ratios of \nii/\ha, inflating \xiion(\ha).  It could also be a result of the assumption of Case B recombination \citep[e.g.,][]{Mendez-Delgado_2024,Scarlata_2024}.  Nevertheless, the effect of dust attenuation is low because most of our galaxies have low color excess, $E(B-V) \lesssim 0.3$ (see Appendix~\ref{appendix:lines}).  We therefore adopt the results using \xiion\ from \hb\  as this line is available for all of our galaxies, but these results may include a systematic uncertainty of $\sim$10--20\%.


  \section{Results}\label{section:results}

\subsection{The Ionizing Production Efficiency of EoR Galaxies}\label{section:results_xiion}

\begin{figure*}[t]
\begin{center}
    \includegraphics[width=0.9\textwidth]{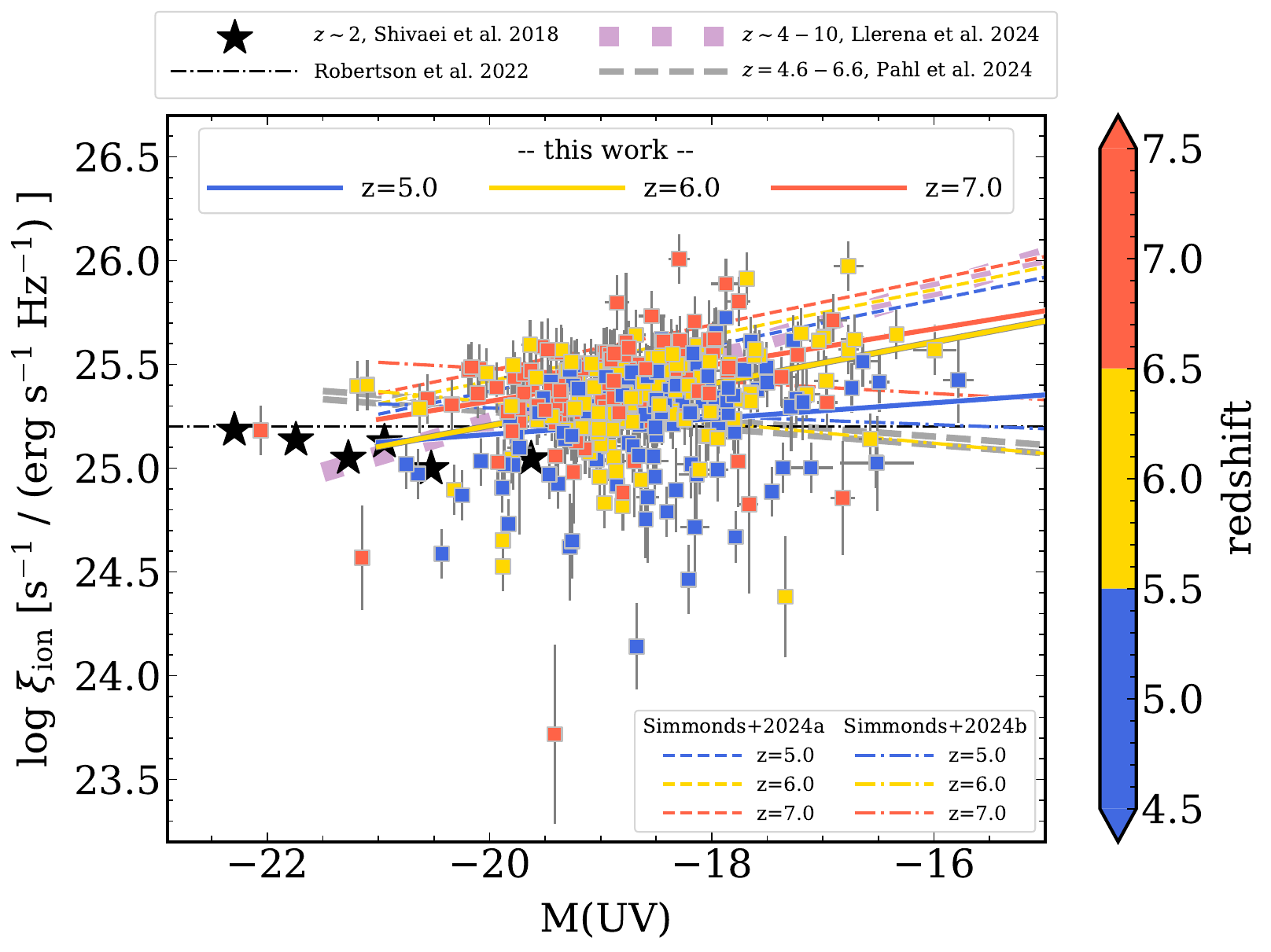}
\end{center}
\vspace{-12pt}
\caption{UV absolute magnitude, \muv, versus the ionizing photon production efficiency, \xiion.  The data points show the results derived from the CEERS and JADES datasets, color-coded by redshift.  The solid lines show linear fits to the relation in bins of redshift, as labeled.  These are consistent with results derived based on the analysis of photometric broad-bands \citep[dashed lines,][]{Simmonds_2024} and other spectroscopic analyses of $4 < z < 10$ \citep{llerena_2024}, with noticeable differences (e.g., \citealt{Pahl_2024}, see text). At the bright end,  $\muv \lesssim -20$, our results are similar to the measurements at $z\sim 2$ from \citet{Shivaei_2018} and the canonical value for stellar populations, $\xiion = 25.2$, argued by \citep{Robertson_2022}.}\label{fig:muv_xi_ion}
\end{figure*}

Our first finding is that fainter galaxies are more efficient at producing ionizing radiation, and this efficiency increases with increasing redshift.  Moreover, the scatter in this trend is significant.   Figure~\ref{fig:muv_xi_ion} shows the distribution of \xiion\ as a function of UV absolute magnitude (\muv) and color-coded by redshift.   We quantify trends by fitting a linear relation between \xiion\ and \muv\ using \texttt{LINMIX} \citep{Kelly_2007} to the data in three bins of redshift, $4.5 < z < 5.5$, $5.5 < z < 6.5$, and $6.5 < z < 9.0$. \texttt{LINMIX}  uses Gaussian Mixture Models to estimate the linear fit incorporating uncertainties on the dependent and independent variables, and allowing intrinsic scatter in the sample.  We fit the data with a linear relation,
\begin{equation}\label{eqn:xi_muv_fit}
\log \xiion = A(z) (\muv + 18)  + \log \xi_\mathrm{ion}^{-18}(z).
\end{equation}
The results are shown in Figure~\ref{fig:muv_xi_ion} and tabulated in Table~\ref{table:xi_muv_fit}.   The slopes ($A$) of the relations show \editone{slight evidence of evolution with redshift, with $A(z) \simeq 0.04\pm 0.3$ at $4.5 < z < 5.5$ to $A(z) = (0.09-0.10) \pm 0.02$ at $6.5 < z < 9.0$. This steepening of the slope may also be indicative of sample incompleteness (which we discuss further below).}  The normalization, $\log \xi_\mathrm{ion}^{-18}$  (defined to be $\log \xiion$ at $\muv = -18$~mag) increases by a \editone{factor of two from $\log \xi_\mathrm{ion}^{-18} = 25.2$ at $4.5 < z < 5.5$, to $25.5$ at $6.5 < z < 9.0$, where the statistical uncertainties are $0.02-0.03$~dex.}  

\begin{deluxetable}{c@{\hskip 12pt} c@{\hskip 12pt} c@{\hskip 12pt}}
\tablecolumns{3}
\tablewidth{0pt}
\tablecaption{Parameters\tablenotemark{$\mathrm{\dag}$} for Linear Fit between \xiion\ and \muv\ as a function of redshift\label{table:xi_muv_fit}}
\tablehead{\colhead{$z$} & \colhead{$A(z)$ } & \colhead{$\log \xi_\mathrm{ion}^{-18}$}}
\startdata  
$4.5 < z < 5.5$ & \editone{0.04} $\pm$ \editone{0.03} & \editone{25.24} $\pm$ 0.03 \\
$5.5 < z < 6.5$ & \editone{0.10} $\pm$ 0.02 & \editone{25.41} $\pm$ 0.02 \\ 
$6.5 < z < 9.0$ & \editone{0.09} $\pm$ \editone{0.02} & \editone{25.50} $\pm$ \editone{0.03} \\
\enddata
\tablenotetext{\dag}{Parameters are defined in  Equation~\ref{eqn:xi_muv_fit}.}
\end{deluxetable}

\begin{figure*}[t]
\begin{center}
    \includegraphics[width=0.9\textwidth]{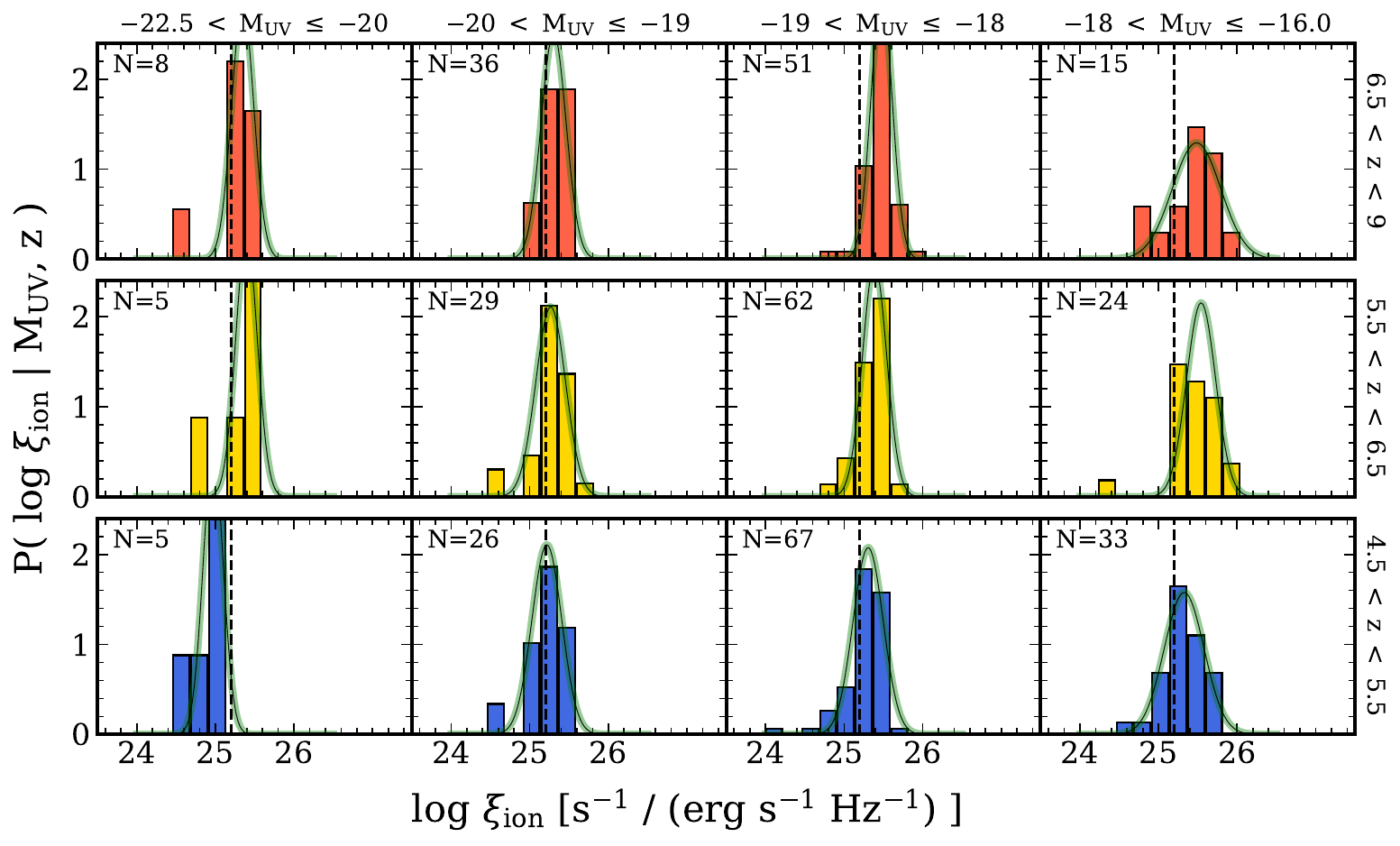}
\end{center}
\vspace{-12pt}
\caption{Distribution functions for $\log$ \xiion\ as a function of redshift, $z$, and UV magnitude, \muv, $P(\log \xiion | \muv, z)$. 
Each row shows the results binned as a function of redshift (as labeled on the right-hand axis).  
 Each column shows the distribution of \xiion\ in bins of \muv, as labeled along the top axis.  The inset labels provide the number of galaxies in the distribution of each panel.  The green curve in each panel shows a Gaussian fit to the distribution with results in Table~\ref{table:xi_muv_hist}.  The vertical dashed lines show the canonical $\log \xiion=25.2$ value of \citet{Robertson_2022}.  
  }\label{fig:xi_muv_hist}
\end{figure*}

The relations between $\log \xiion$ and \muv\ we measure are consistent with previous studies, with some notable differences.    At $4.5 < z < 5.5$ this normalization is consistent with the canonical value expected for star-forming stellar populations, $\log \xiion = 25.2$ \citep{Robertson_2022} and studies of star-forming galaxies at $z\sim 2$ \citep{Shivaei_2018}.  In the \jwst\ era, studies at higher redshift, $z > 5.5$, find a similar increase in \xiion\ compared to what we observe \citep[e.g.,][]{Atek_2024, llerena_2024,Pahl_2024, Simmonds_2024,Simmonds_2024b,Hayes_2025}.  
Interestingly, the relation between $\log \xiion$ and \muv\ may not be linear.  At bright magnitudes, e.g., $\muv \lesssim -19.5$~mag, all galaxies with $4.5 < z < 5.5$ have $\log \xiion$ below the linear fit, and these may connect to the galaxy populations seen at $z\sim 2$ \citep{Shivaei_2018}.  This is important and may motivate exploring non-linear distributions and incorporating the scatter when describing \xiion\ at a given \muv\ and redshift.

We quantify the average and scatter of the ionizing production efficiency by fitting a Gaussian to the $\log \xiion$ distribution in bins of \muv\ and redshift, with results shown in Figure~\ref{fig:xi_muv_hist} and Table~\ref{table:xi_muv_hist}.  In all bins the scatter is typically $\sigma(\log \xiion) \simeq$ 0.20-0.30~dex \editone{consistent with other recent work \citep[e.g.,][]{Pahl_2024,Begley_2025}}. It is notable that a similar amount of scatter is observed in the \xiion\ distributions of $z\sim 2$ galaxies \citep{Shivaei_2018}.  This scatter is important as it tells us that galaxies at these redshifts produce different rates of ionizing photons at a fixed UV luminosity.  This may be a symptom of increased ``burstiness'' in the star-formation histories of galaxies as has been suggested from photometric studies \editone{\citep{Endsley_2024,Simmonds_2024b,Cole_2025}}.  For the mode (peak) of the $\log \xiion$ distributions, these increase with decreasing \muv\ and increasing redshift.  This says that on average fainter galaxies are more efficient at producing ionizing photons, and this efficiency increases with redshift.  
%

\editone{Our sample may be biased by the fact that it requires spectroscopic confirmation.  Recent studies using photometric samples demonstrate this bias \citep{Simmonds_2024b,Begley_2025}.  Figure~\ref{fig:muv_xi_ion} shows the $\xi$--\muv\ relation from \citet{Simmonds_2024b}, which found a flat or decreasing relation between $\xi$\ and \muv.  \citet{Endsley_2024} used photometric samples and found that \xiion\ declines with UV luminosity, opposite to the trend in our data, which they interpret as evidence that ``post burst'' galaxies would have lower \xiion, weaker emission lines, and fainter UV magnitudes.  However, as stated above, these photometric samples could be biased in that galaxies with weaker emissions lines are more difficult to interpret.  While these are consistent with our distributions $4.5 < z < 5.5$, they deviate at higher redshifts.  To test this potential bias will require deeper spectroscopy of fainter $\muv$ galaxies at these redshifts.} 

\begin{deluxetable*}{c@{\hskip 12pt}|@{\hskip 12pt}ccc@{\hskip 12pt}|@{\hskip 12pt}ccc@{\hskip 12pt}|@{\hskip 12pt}ccc@{\hskip 12pt}|@{\hskip 12pt}ccc }
\tablecolumns{13}
\tablewidth{0pt}
\tablecaption{Mean and scatter in $\log \xiion$\tablenotemark{$\mathrm{\dag}$} as a function of \muv\ and redshift\label{table:xi_muv_hist}}
\tablehead{ \colhead{} & \multicolumn{3}{c}{$-22 < \muv < -20~~~~~~$} & \multicolumn{3}{c}{$-20 < \muv < -19~~~~~~$} & \multicolumn{3}{c}{$-19 < \muv < -18~~~~~~$} & \multicolumn{3}{c}{$-18 < \muv < -16~~~~~~$}  \\
  \colhead{$z$} & \colhead{$N$} & \colhead{$\mu$} & \multicolumn{1}{c}{$\sigma~~~~~$}
  & \colhead{$N$} & \colhead{$\mu$} & \colhead{$\sigma~~~~~$}
  & \colhead{$N$} & \colhead{$\mu$} & \colhead{$\sigma~~~~~$}
   & \colhead{$N$} & \colhead{$\mu$} & \colhead{$\sigma$} }
\startdata  
$4.5 < z < 5.5$ & \editone{5} & \editone{24.97} & \editone{0.13} &  \editone{26} & \editone{25.22} & \editone{0.19} & \editone{67} & \editone{25.31} & \editone{0.19} & \editone{33}& \editone{25.33} & \editone{0.25} \\ 
$5.5 < z < 6.5$ & \editone{5} & \editone{25.40} & \editone{0.14} & \editone{29} & \editone{25.27} & \editone{0.19} & \editone{62} & \editone{25.39} & \editone{0.15} & \editone{24} & \editone{25.54} & \editone{0.19}  \\
$6.5 < z < 9.0$ & \editone{8} & \editone{25.35} & \editone{0.14} & \editone{36} & \editone{25.30} & \editone{0.16} &  \editone{51} & \editone{25.48} & \editone{0.14} & \editone{15} & \editone{25.48} & \editone{0.31}  \\
\enddata
\tablenotetext{\dag}{The values in the table show the number of galaxies in each bin, $N$, and the Gaussian fits with mean $\mu$ and variance $\sigma^2$ of $\log \xiion$ in each bin.  These fits are illustrated in Figure~\ref{fig:xi_muv_hist}. }
\end{deluxetable*}


\subsection{The Inferred LyC Escape Fractions of EoR Galaxies}\label{section:results_fesc}

\begin{figure*}
    \begin{center}
        \includegraphics[width=0.8\textwidth]{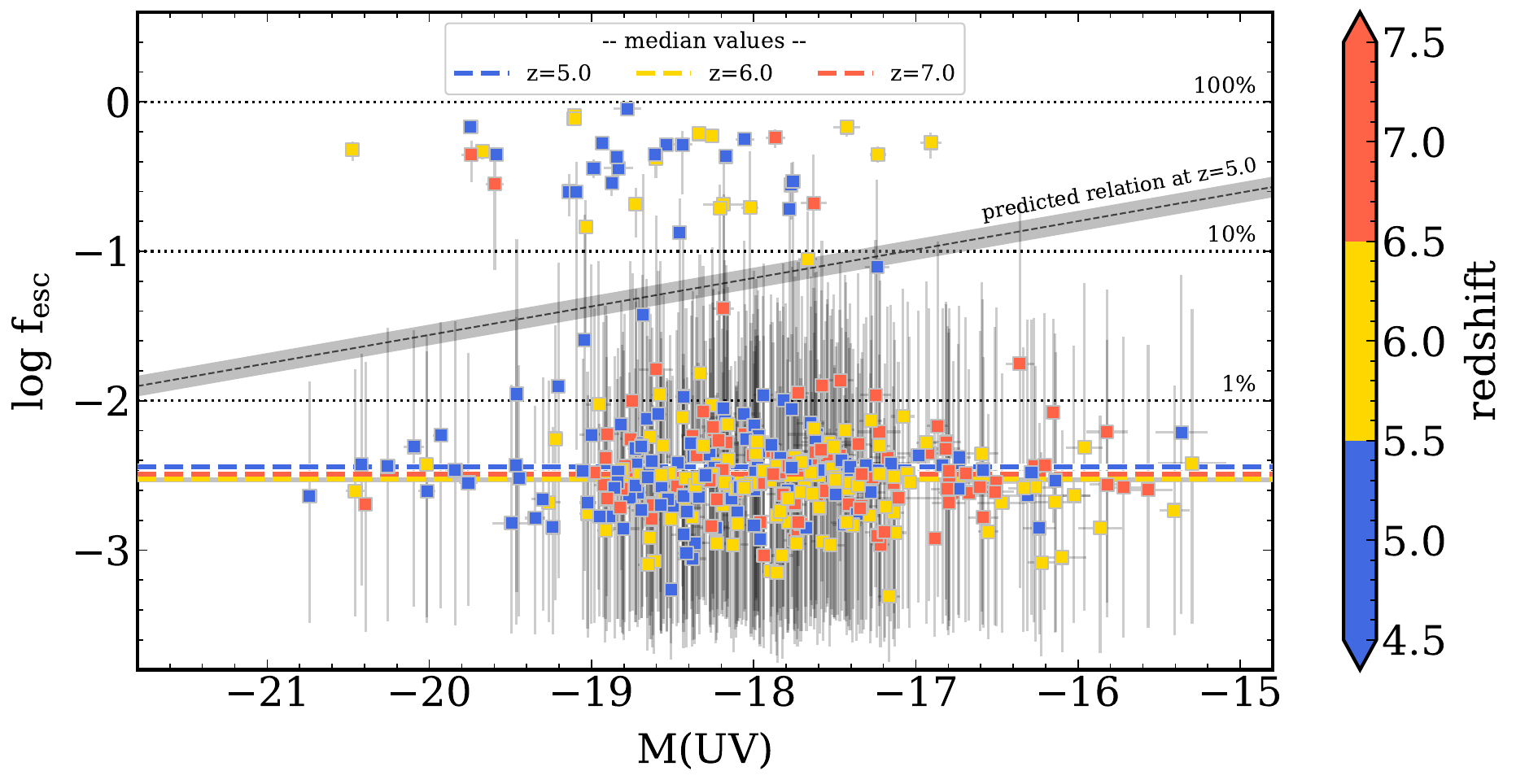}
    \end{center}
    \vspace{-12pt}
    \caption{UV magnitude, \muv, versus the inferred LyC escape fraction, \fesc.  The data points show the median values for the galaxies in the JADES and CEERS samples, color-coded by redshift.  The error bars show the 16--84\% range for each galaxy.   The horizontal dotted lines denote escape fractions of 1, 10, and 100\%. \editone{The thicker horizontal lines show medians derived from the samples in bins of redshift.  These do not include uncertainties on the medians, which are 0.2--0.3~dex (see text).}
    \editone{Less than 20\% of galaxies at $z > 4$ have inferred \fesc\ values greater than a few percent}.   This contrasts with predictions inferred from the UV-spectral slopes and UV magnitudes for low-redshift galaxies, where the diagonal line shows the empirical predictions for $z=5$ derived from LzLCS at $z\sim 0.3$ \citep{Chisholm_2019}. These predictions shift to higher \fesc\ for higher redshift galaxies.  }\label{fig:fesc_muv}
\end{figure*}


\begin{deluxetable}{c@{\hskip 12pt} c@{\hskip 12pt} c@{\hskip 12pt}}
\tablecolumns{3}
\tablewidth{0pt}
\tablecaption{\editone{Medians and $\sigma_\mathrm{NMAD}$ of the $\log$ \fesc\ values in bins of redshift}\label{table:fesc_muv_fit}}
\tablehead{\colhead{$z$} & \colhead{median $\log\fesc$ } & \colhead{$\sigma_\mathrm{NMAD} \fesc$}}
\startdata  
$4.5 < z < 5.5$ & \editone{$-$2.44} & \editone{0.31} \\
$5.5 < z < 6.5$ & \editone{$-$2.52} & \editone{0.32} \\ 
$6.5 < z < 9.0$ & \editone{$-$2.49} & \editone{0.24} \\
\enddata
\end{deluxetable}

Another of our main findings is that the inferred \fesc\ values for $4.5 < z < 9.0$ galaxies are on average low, with no evidence of evolution with redshift or \muv.  
 Figure~\ref{fig:fesc_muv} shows the distribution of \fesc\ as a function of \muv\ for the galaxies in our CEERS and JADES samples.  
 \editone{The lines in Figure~\ref{fig:fesc_muv} show the median values of fesc\ in the three bins of redshift.}
%
%
%
\editone{We list the values of the medians and the scatter, derived from the normalized median absolute deviation, $\sigma_\mathrm{NMAD}$, in Table~\ref{table:fesc_muv_fit}. }

The \editone{median $f_\mathrm{esc}$ values are similar in all redshift bins, with med($\log \fesc$) = $-2.5$ to $-2.4$ with a scatter of $\sigma_\mathrm{NMAD}=0.2-0.3$}.  This implies a median $\fesc$ of less than 1\% (but this is misleading when considering only the medians, see Section~\ref{section:discussion_fesc}).  \editone{Comparing the numbers, 89\% of galaxies in our sample have $\log \fesc < -1$ and 82\% have $\log \fesc < -2$.   This implies that the majority of the galaxies in our sample favor low \fesc.  This is consistent with a recent finding by \citet{Giovanazzo_2025} based on an independent analysis using only \jwst\ spectroscopy that the majority of galaxies, $\gtrsim95$\%, have no statistical evidence for high $\fesc$, though those authors find a much higher fraction of galaxies with median $\fesc > 10$\%, which we disfavor from our analysis. }
%

We reiterate the important caveat that we are constraining the \textit{inferred} LyC escape fractions from the SED modeling.   This is the best we can do given that direct measurements of \fesc\ are not possible given the density of \ion{H}{1} absorbers makes the IGM optically thick to LyC emission at $z \gtrsim 4$ \citep{Inoue_2014}.   Our results say that there is no evidence that galaxies have high \fesc, as nearly all of the LyC photons produced need to be reprocessed to produce the strength of the nebular emission lines (see also Appendix~\ref{appendix:fesc_sed}).   It is therefore prudent to consider what features in the modeling are sensitive to the escape fraction, and what aspects of the data are driving the fits to favor low \fesc.

\begin{figure*}[t]
  \begin{centering}
     \gridline{
    \fig{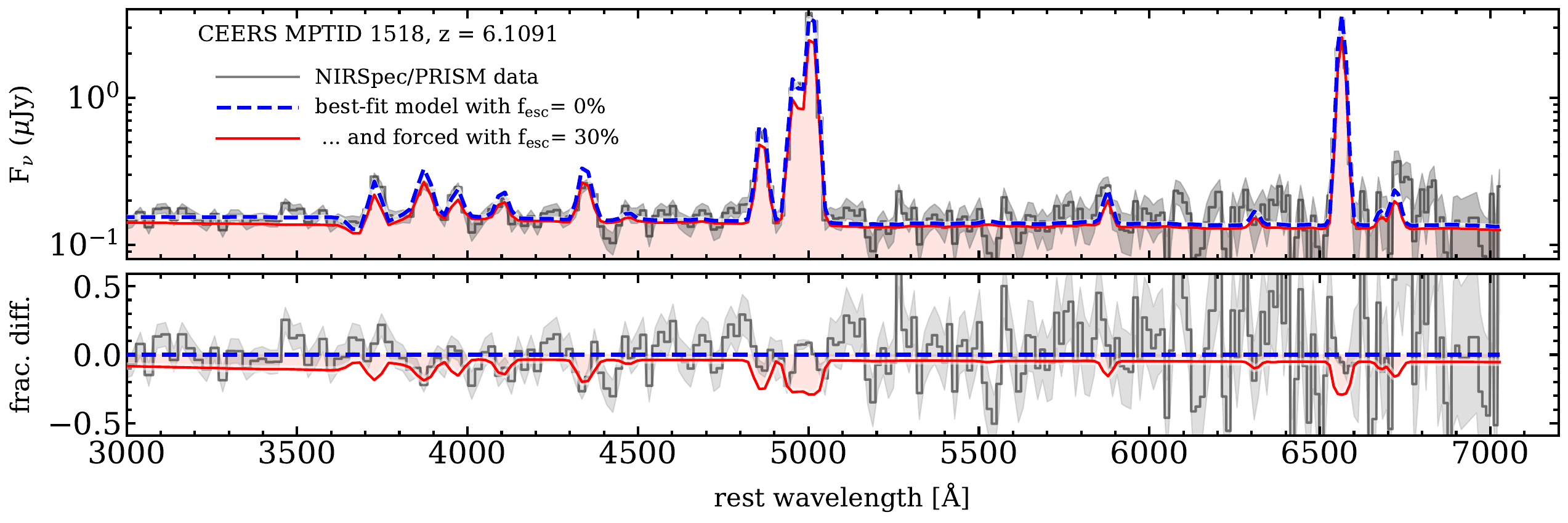}  {0.7\textwidth}{}  \raisebox{0pt}{\fig{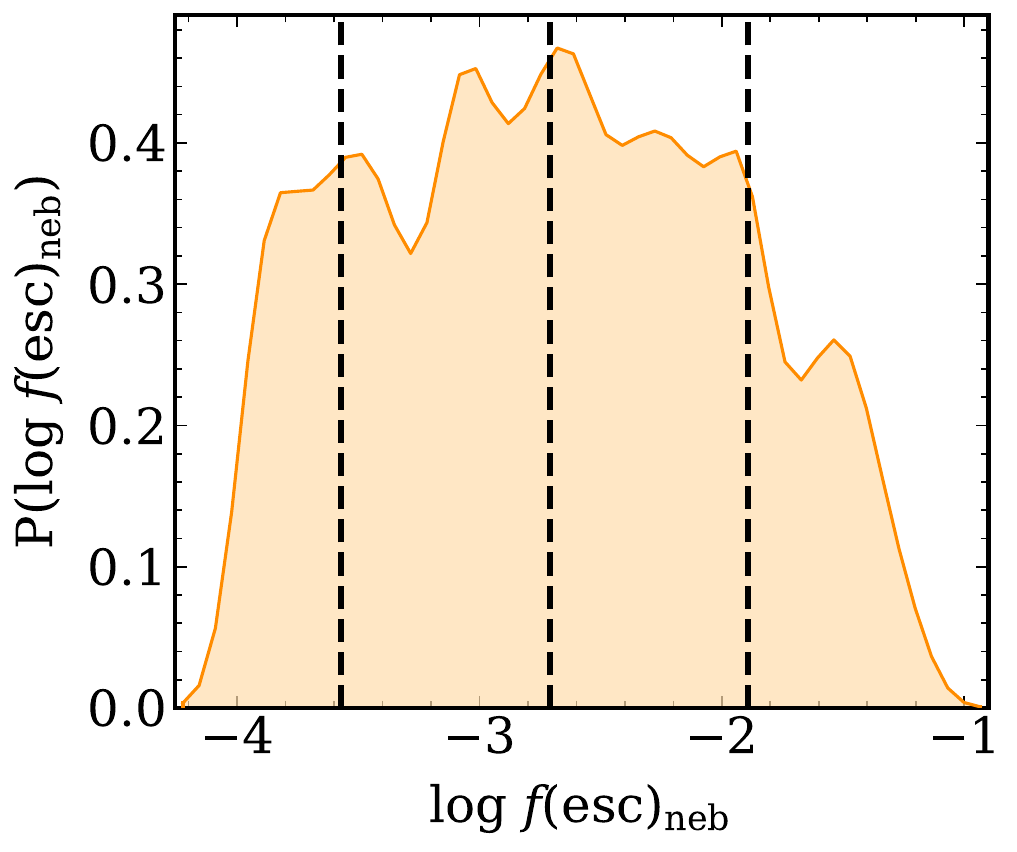}{0.27\textwidth}{}}
    }\vspace{-12pt}
    \gridline{\fig{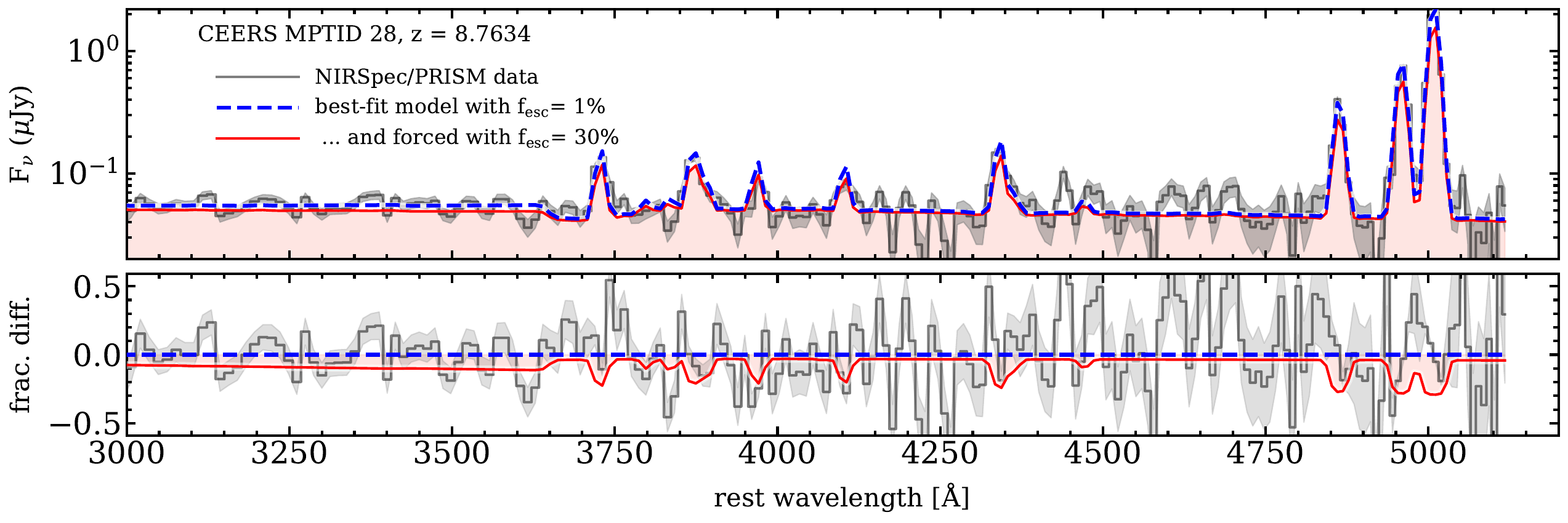}{0.7\textwidth}{}
    \raisebox{0pt}{\fig{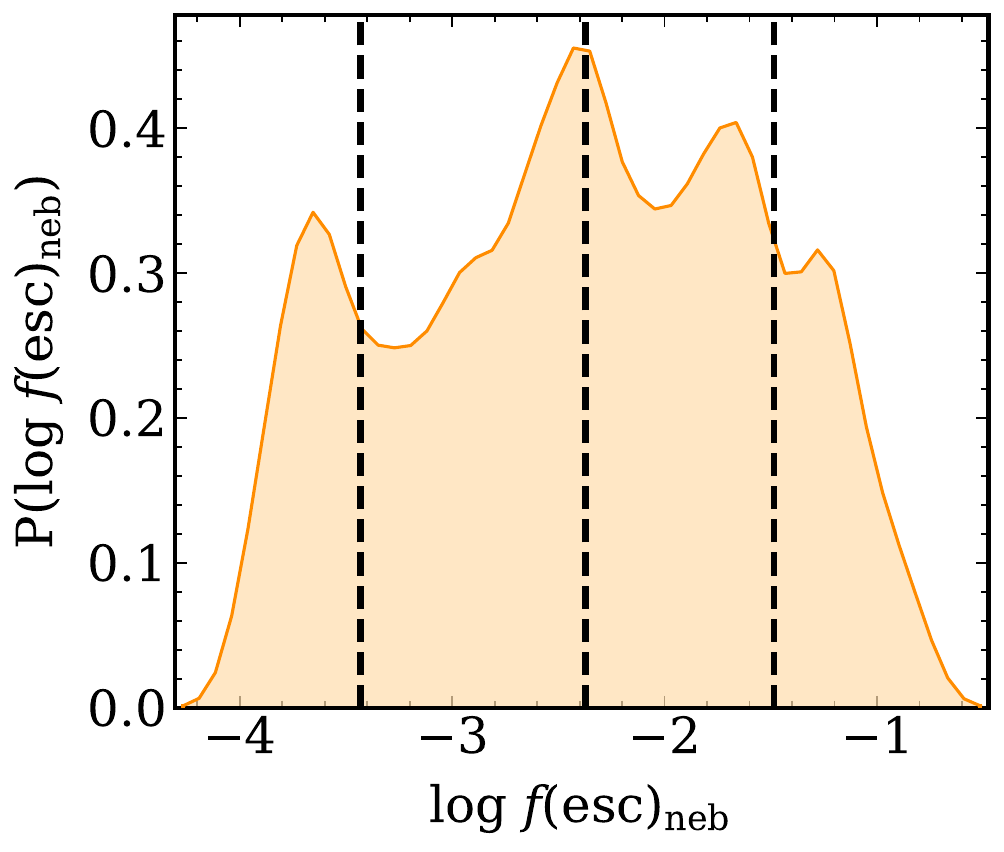}{0.27\textwidth}{}}
    }\vspace{-12pt}
\end{centering}
\caption{Example of how the \fesc\ impacts the SEDs for two galaxies (top and bottom rows, respectively) from Figure~\ref{fig:sedfits_ceers} with low inferred \fesc\ from the SED fitting.   In each galaxy, the left panels show the NIRSpec/PRISM data along with the best-fit from the SED fitting (solid, blue line), which has $\fesc \ll 1$\%.   The model illustrated by the red-dashed line shows the model if we force a high escape fraction, $\fesc = 30$\%.   The bottom panels show the relative flux difference compared to the best-fit model, $\Delta F_\nu / F_\nu$.   In the case of the $\fesc=30$\% models, there are clear differences seen, particularly in the emission lines, but also in the nebular continuum.   These features are not as strong as expected given the ionizing continuum from the stellar population.  The right hand panels show the one-dimensional distributions for $\log \fesc$ for each galaxy.  The vertical, dashed lines show the 16, 50, and 84 percentiles.}\label{fig:spec_force_fesc}
\end{figure*}
   
Figure~\ref{fig:spec_force_fesc} shows SED fits to the spectra for two of the galaxies from Figure~\ref{fig:sedfits_ceers}).   These both have  solutions that favor $\fesc \lesssim 1$\%.  \editone{CEERS 1518 has $\log \fesc = -2.79^{+0.88}_{-0.87}$, with only 3.5\% of the likelihood having \fesc\ $>$ 3\% (and no likelihood of \fesc\ $> 10$\%)}.  \editone{CEERS DDT 28 has $\log \fesc = -2.22^{+0.76}_{-1.17}$, with 17\% of the likelihood with \fesc\ $>$ 1\% (but only 2\% of the likelihood above \fesc\ $>$ 10\%)}.   These are representative for the majority of galaxies in the samples as evidenced by the error bars on \fesc\ on the data points in Figure~\ref{fig:fesc_muv}. 

The models in Figure~\ref{fig:spec_force_fesc} include both a ``best-fit'' model (a model selected to have near-maximum likelihood, blue, solid lines) and a model where we forced a higher \fesc\ = 30\% (red dashed lines).    In the latter case, it is evident that the model underpredicts the strength of the emission lines, particularly at \hb+\oiii\ (and \ha\ in the case of CEERS 1518).  This model also underpredicts the rest-UV ($\sim 1500-3000$~\AA) continuum, because it has a lower contribution from the nebular continuum \citep{Katz_2024}.   These factors lead the SED fitting to disfavor models with higher \fesc\ values. 
%

Motivated by these examples, in Appendix~\ref{appendix:fesc_sed}, we consider how well the SED modeling can recover \fesc\ for simulated galaxies.  As expected, as \fesc\ increases, the  strength of the emission lines decreases, compared to the continuum.  As the continuum drives the number of produced ionizing photons, in order to fit the weaker emission lines requires higher \fesc.   We show that for \editone{model galaxies with properties like those in our sample, if we simulate an increase in \fesc}, the SED fitting is able to recover these values (see Appendix~\ref{appendix:fesc_sed}).  %
%

  
 \subsection{Constraints on the Cosmic Ionizing Rate Density}\label{section:ndot_ion}

Our measurements of the ionizing production efficiency, \xiion, and the constraints on the LyC escape fraction, \fesc, allow us to compute the cosmic ionization production rate density (also called the emissivity), $\dot n_\mathrm{ion}$, which is the rate of ionizing photons produced per cubic (comoving) Mpc.   This quantity is derived by integrating the UVLF, $\phi(\muv)$, multiplied by the product of $(\fesc \times \xiion\ \times L_\mathrm{UV})$, that is
 \begin{equation}\label{eqn:ndot_ion}
\dot n_\mathrm{ion} = \int\ \phi(M_\mathrm{UV}) L_\mathrm{UV} \xiion f_\mathrm{esc}\ dM_\mathrm{UV}.
\end{equation}

To compute $\dot n_\mathrm{ion}$ we use a Monte Carlo simulation, which allows us to sample the range of uncertainties in parameters of the UVLF, and to sample the full distributions that we have measured for \xiion\ and \fesc\ as a function of \muv\ and redshift.  In each iteration of the Monte Carlo, we first draw a sample of galaxies with \muv\ selected to reproduce the UVLF of \citet{Finkelstein_2022a} at different redshifts, with $z \in (5,6,7,8)$, \editone{where use bins in redshift that match those of the UVLF}.  We also extend these simulations to higher redshift using the UVLF of \citet{Donnan_2024} in order to consider the redshift evolution of the cosmic ionized-hydrogen fraction, $x_e$ (see Section~\ref{section:discussion_xe}). In each instance of the simulation, we vary the parameters of the UVLF within the values and covariances.  In this way the total number of galaxies, and the number of galaxies per unit magnitude include the full range of uncertainties in the UVLF.

Second, for each galaxy in the simulated sample we \editone{draw a random value for \xiion\ }from our measured distribution, $P(\xiion|\muv, z)$, using \editone{the galaxy's} $\muv$ and $z$, using our empirically measured distributions (Table~\ref{table:xi_muv_hist}).   \editone{For \fesc, we select a value of \fesc\ by taking at random a (real) galaxy from our sample at similar \muv\ and $z$ as the simulated galaxy, and drawing a value from its $P(\log \fesc | \muv, z)$ distribution}.  In this way we each simulated galaxy receives \xiion\ and \fesc\ drawn from the measured galaxy distribution functions.   

\begin{figure*}[t]
\begin{center}
    \includegraphics[width=0.8\textwidth]{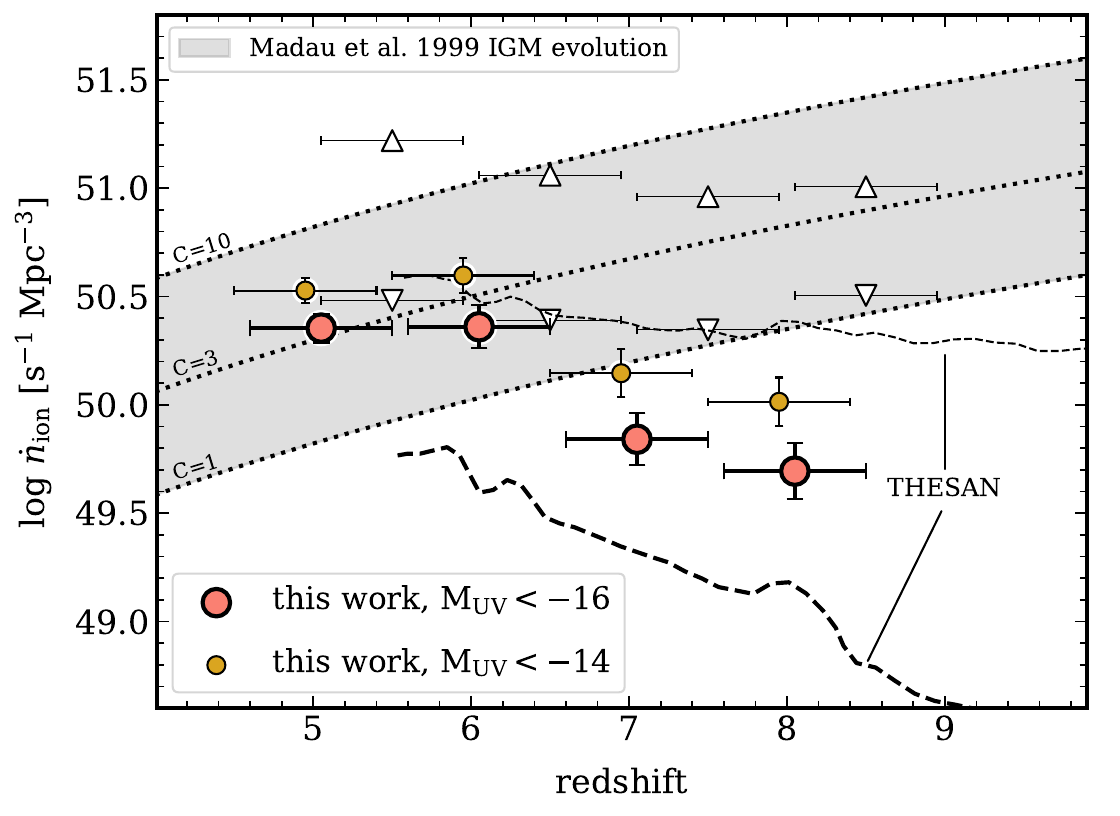}
\end{center}
\caption{Evolution of \ndotion\ as a function of redshift.  The large circles show the results from our analysis of the JADES and CEERS samples, where we integrate the UV luminosity function to $\muv < -16$ (large circles) and $\muv < -14$ (small circles). The vertical error bars show the 16--84\% range, and the horizontal error bars show the bounds of the redshift bin.  The triangles show results from the analysis of \jwst\ photometric data with different assumptions of the LyC escape fractions from \citet{Simmonds_2024} \editone{(upward triangles) and \citet{Simmonds_2024b} (downward triangles).}
 The shaded region shows the \ndotion\ required to reionize the IGM at a given redshift for different assumptions of the \ion{H}{1} clumping factor, $C$ \citep{Madau_1999}. 
 The dashed lines show predictions from the \textsc{Thesan} simulation for the global galaxy population (light line) and galaxies with $\log M_\ast / M_\odot = 8-9$ (thick line, \citealt{Yeh_2023}).}\label{fig:ndotion}     
\end{figure*}

\begin{deluxetable}{c@{\hskip 24pt} c@{\hskip 24pt} c@{\hskip 12pt}}
\tablecolumns{3}
\tablecaption{Measured Ionizing Photon Production Rate Density}\label{table:ndotion}
\tablehead{\colhead{} & \multicolumn{2}{c}{$\log( \dot{n}$ / s$^{-1}$ Mpc$^{-3}$)} \\ \colhead{$z$~~~~~~~~~~~~~} & \multicolumn{1}{c}{~~($\muv < -16$)~~~~~~~~~~~~~~~}  & \multicolumn{1}{c}{~~~~~($\muv < -14$)~~~~~~~~~~ } }
\startdata  
    5.0 & \editone{50.52 $\pm$ 0.08} & \editone{50.72 $\pm$ 0.07} \\
    6.0 & \editone{50.59 $\pm$ 0.09} & \editone{50.85 $\pm$ 0.08} \\ 
    7.0 & \editone{50.29 $\pm$ 0.11} & \editone{50.59 $\pm$ 0.10} \\
    8.0 & \editone{50.10 $\pm$ 0.13} & \editone{50.49 $\pm$ 0.11} \\
\enddata
\end{deluxetable}

 We repeated the Monte Carlo simulation 4000 times for each of two cases:  one assuming a faint-end cut off of the UVLF at $\muv_\mathrm{cut} = -16$, which is the observational limit of our dataset, and one assuming $\muv_\mathrm{cut} = -14$, near the lowest value to which the UVLF has been measured \citep{Atek_2024}.  In the latter case, we must extrapolate our relations, so we use the \xiion\ and \fesc\ distributions for galaxies with $\muv = -16$ for galaxies fainter than this value.  We then calculate $\dot n_\mathrm{ion}$ as a function of redshift using Equation~\ref{eqn:ndot_ion} for each simulation.  We take the median $\dot n_\mathrm{ion}$ of the distribution as the quoted value with the 16-to-84th percentile as the error. 
 
The results of the Monte Carlo simulation are shown in Figure~\ref{fig:ndotion} and Table~\ref{table:ndotion}.   The figure compares the $\dot{n}_\mathrm{ion}$ values to the rate needed to sustain an ionized universe at a given redshift from \citet{Madau_1999}.   One of the remaining uncertainties is the ``clumping factor'' of the gas in the IGM, $C$, where we must rely on predictions from simulations \citep[e.g.,][]{Madau_1999,Finlator_2011,Pawlik_2015,Davies_2024}.  Figure~\ref{fig:ndotion} shows that the production rate of ionizing photons needed to reionization increases with this clumpiness factor and that if we include galaxies with $\muv < -16$ ($-14$) we do not reach this level until $z\sim 5$ ($z\sim 6$) for $C\sim 3$.  



\section{Discussion }\label{section:discussion}

\subsection{Why are the ionizing production efficiencies of \\EoR galaxies high?}\label{section:discussion_xiion}

In Section~\ref{section:results_xiion} we showed that the ionizing production efficiency for our samples increase for fainter galaxies and increase toward higher redshift.   This echoes trends observed from other \jwst\ studies that find high values,  $\log \xiion = 25-26$, for galaxies at $z\simeq 7-8$.   This includes both studies that have used SED fitting of \jwst\ imaging data to infer \xiion\ \editone{\citep[e.g.,][]{Prieto-Lyon_2023,Endsley_2024,Simmonds_2024,Simmonds_2024b,Begley_2025}}, and spectroscopic studies with \jwst\ find similar trends of an increase in \xiion\ for EoR galaxies \citep[e.g.,][]{Atek_2024,Calabro_2024,llerena_2024,Mascia_2024,Pahl_2024,Saxena_2024,Hayes_2025}.  Ours is the first to estimate \xiion\ by combining the photometry and spectroscopic data with a consistent treatment. 

The finding that $\xiion$ is higher at high redshifts seems robust as most studies have employed different methods, assumptions, and samples.   For example, \citet{Hayes_2025} considered the properties of stacked (average) NIRSpec/PRISM spectra for galaxies at $4 < z < 10$, finding $\log \xiion \simeq 25.2-25.7$ with a strong correlation between \xiion\ and many signatures of LyC escape.  Using individual galaxy spectra, \citet{llerena_2024} measured \xiion\ from Balmer-emission line measurements from JWST/NIRSpec prism and $R\sim 1000$ data, combined with photometric modeling to obtain $L_\mathrm{UV}$ for galaxies at $4 < z < 10$. Their results yield results similar to those here, at least in the mean (see Figure~\ref{fig:muv_xi_ion}).    

\editone{Not all analyses conclude that \xiion\ increases with decreasing UV luminosity.  This could be related to systematics or to selection effects.  \citet{Pahl_2024} measured} the evolution of \xiion\ using JWST spectroscopy for galaxies at $1.06 < z < 6.71$, finding that \xiion\ is higher for UV-luminous galaxies and declines with decreasing UV luminosity.  This may be related to the sample selection as \citet{Pahl_2024} required $>3\sigma$ detections in both the \hb\ and \ha\ lines, or because of differences in the SED modeling and spectroscopic slitloss corrections.  \editone{Studies using multiband imaging data including medium-band photometry also find that UV-fainter galaxies show lower \xiion\ \citep{Endsley_2024,Simmonds_2024b}.  One explanation for this is that some fainter galaxies are ``post-burst'', which would have lower \xiion\ and fainter UV magnitudes \citep{Endsley_2024}.  These could be missed by spectroscopic surveys, but these galaxies also suffer larger photometric redshift uncertainties.} This does highlight the potential for selection effects in spectroscopic samples. Testing these selection effects will require deeper \jwst/NIRSpec observations 
%
%
\citep[e.g.,][]{Dickinson_2024}.  

\begin{figure*}[t]
    \begin{center}
    \gridline{
        \includegraphics[height=0.38\textwidth, trim={0 0 110pt 0}, clip=true]{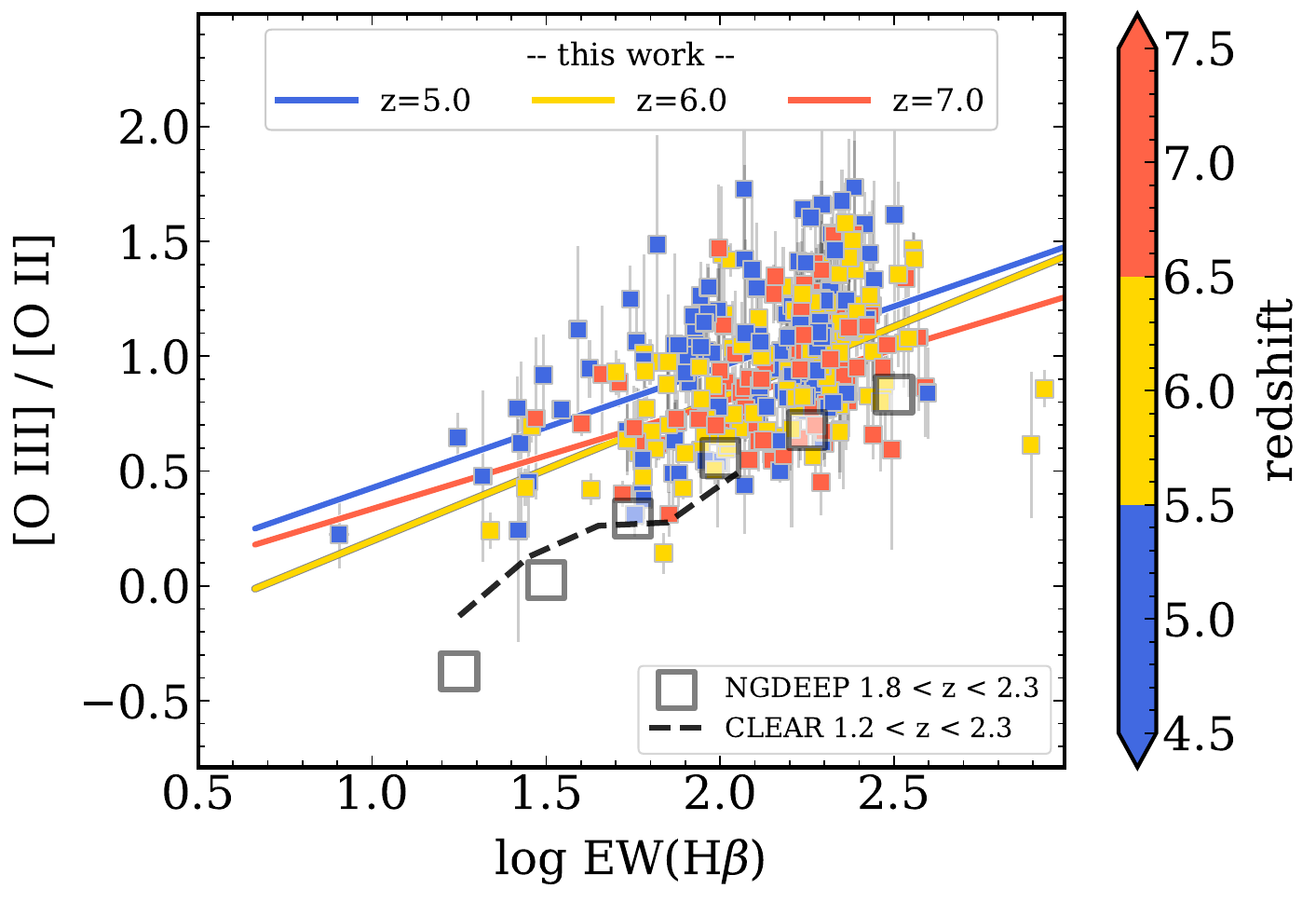}
        \includegraphics[height=0.38\textwidth]{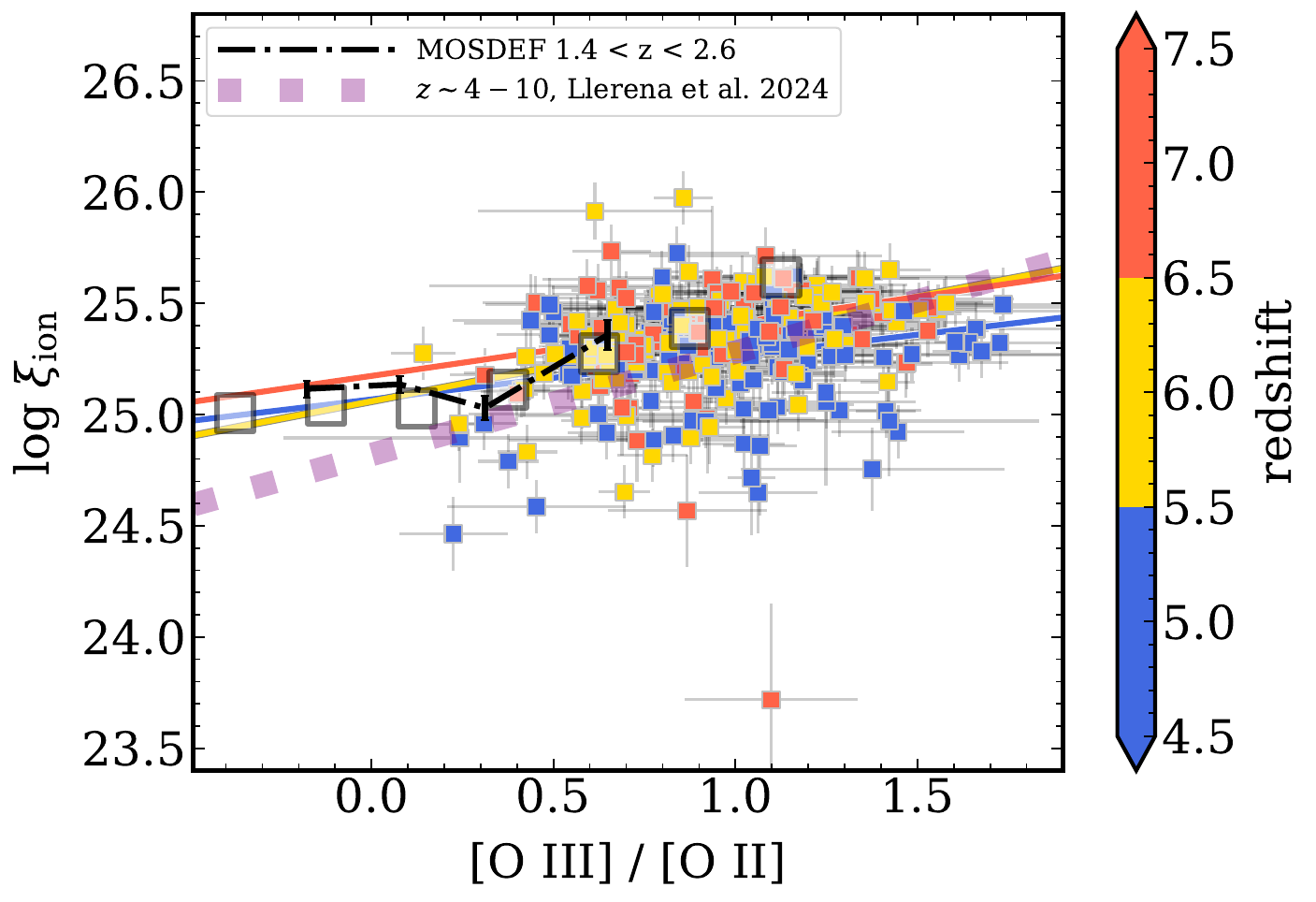}
    }\vspace{-12pt}
    \gridline{
            \includegraphics[height=0.38\textwidth, trim={0 0 110pt 0}, clip=true]{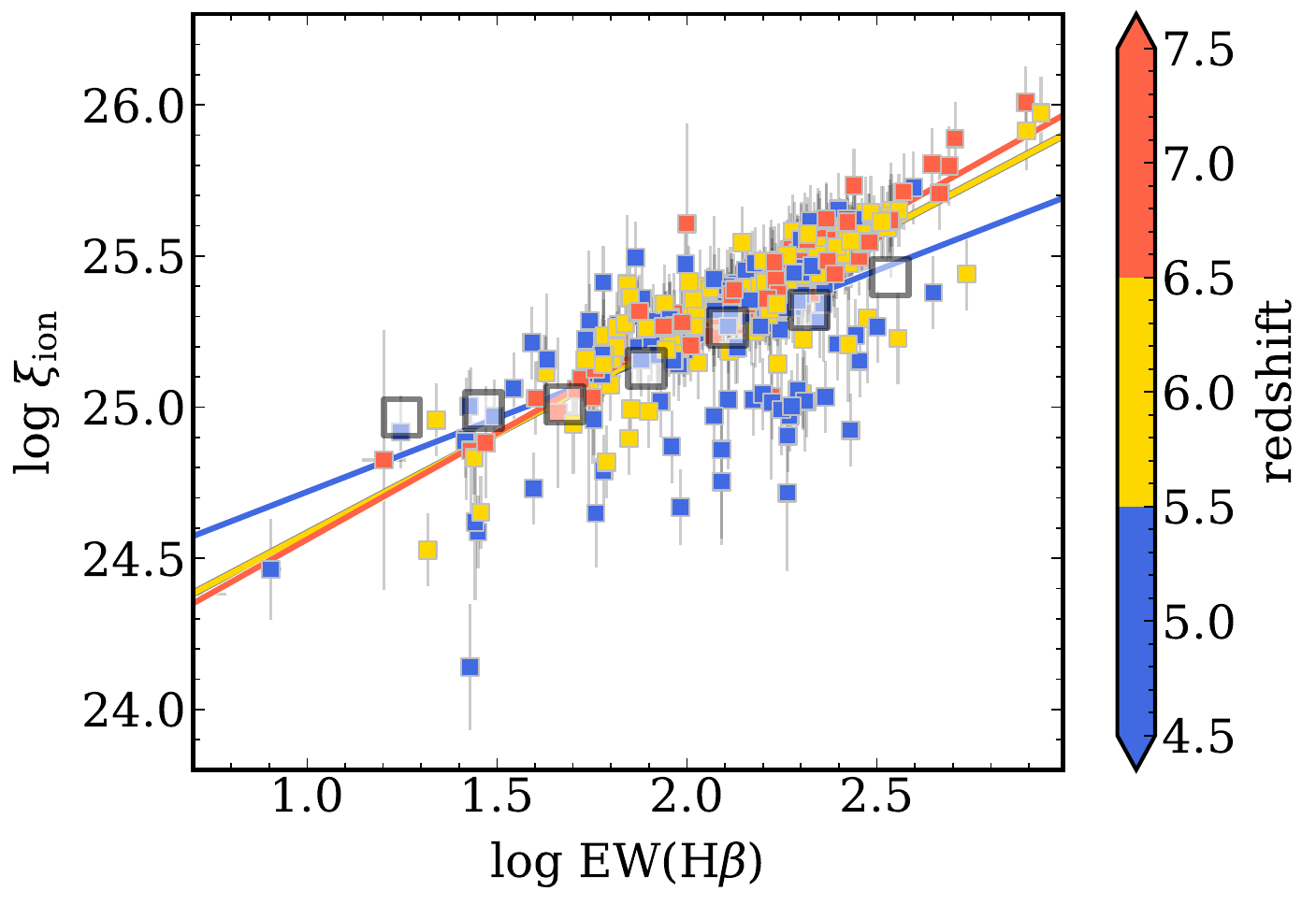}
        \includegraphics[height=0.38\textwidth]{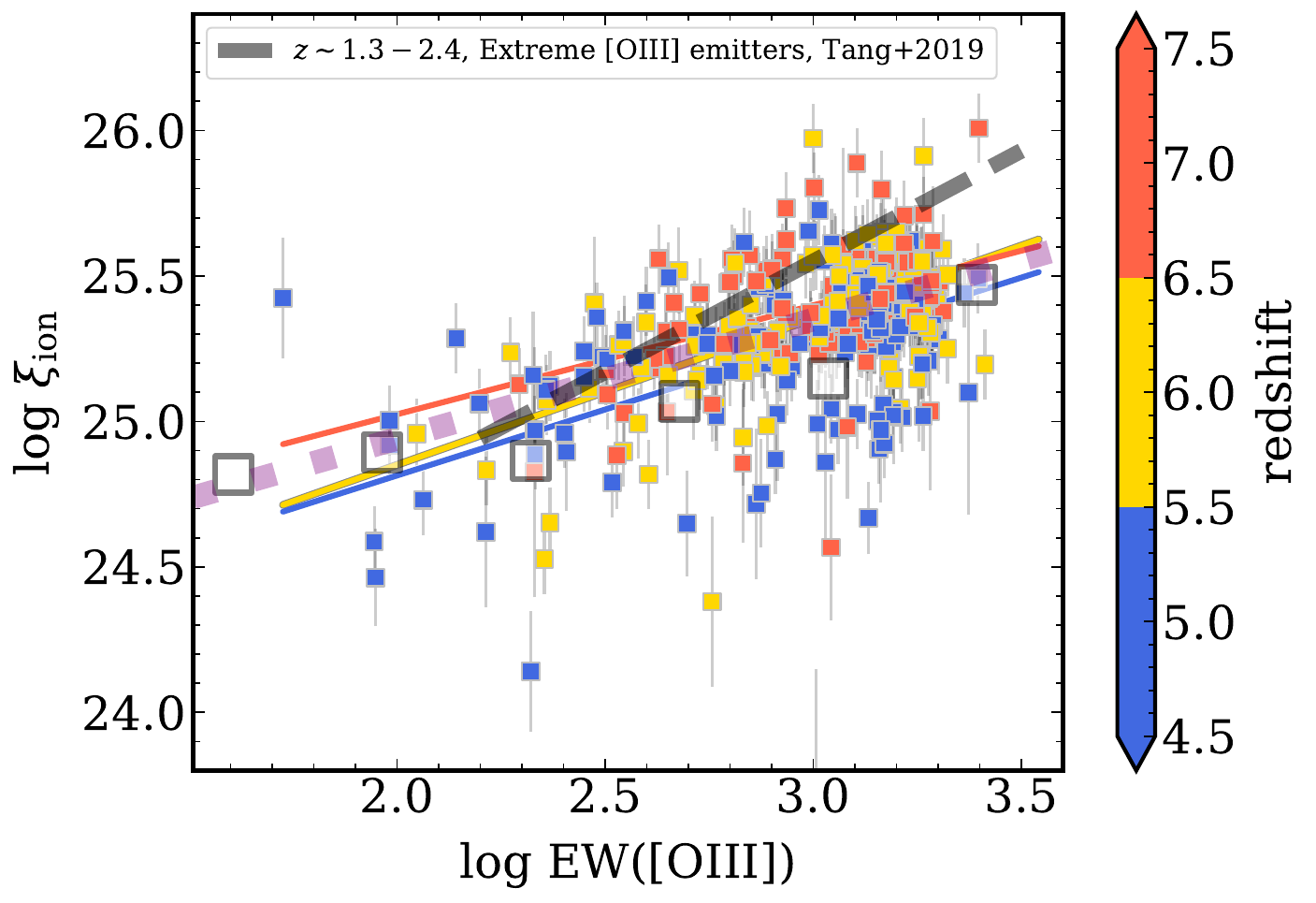}
    }
    \end{center}
    \vspace{-24pt}
    \caption{\textit{Top Left}: Equivalent width (EW) of \hb\ versus the \oiii/\oii\ ratio.  The data points show the  JADES and CEERS galaxies, color coded by redshift.   \textit{Top Right}: \oiii/\oii\ emission-line ratio versus the ionizing photon production efficiency, \xiion.  \textit{Bottom Left}: EW(\hb) versus the ionizing photon production efficiency.  \textit{Bottom Right:} EW(\oiii) versus the ionizing photon production efficiency.   In all plots, the data points show the  JADES and CEERS galaxies, color coded by redshift.  The solid lines show fits to the different redshift samples (as labeled).  \editone{The thick, purple-dashed line shows the measured relations at $4 < z < 10$ by \citet{llerena_2024}, which are consistent with the measurements here.}   \editone{The black, dot-dashed line shows the relation from MOSDEF at $1.4 < z < 2.6$ \citep{Shivaei_2018}, and the heavy, gray-dashed line shows the relation for extreme \oiii-emitters at $1.3 < z < 2.4$ \citep{Tang_2019}.} \editone{The open, large squares show measurements from NGDEEP \jwst\ slitless spectroscopy at $1.2 < z < 2.3$ \citep{Shen_2024} and the black, dashed line shows the relation from CLEAR \hst\ slitless spectroscopy at $1.2 < z <2.3$ \citep{Papovich_2022}.}
}\label{fig:O32_comparisons}
\end{figure*}

An increase in $\xiion$ requires a greater ionizing production rate, $Q(\mathrm{H_0})$, per unit far-UV continuum luminosity, $L_\mathrm{UV}$.   This occurs when galaxies either have a higher fraction of O-type stars, or when these stars have hotter effective temperatures, or both.  \citet{Shivaei_2018} measured empirical trends for galaxies at $z\sim 0-2$, showing that \xiion\ increases with lower dust content, stellar population age and metallicity.  Higher effective temperatures are predicted as a consequence of lower metallicity stars, which lowers the opacity in stellar photospheres, leading to higher LyC production \citep[see, e.g.,][]{Yung_2020a,Mascia_2024}. Studies using \jwst\ observations of the nebular emission find that the gas-phase metallicity is declining with increasing redshift \citep[e.g.,][]{Trump_2022,Curti_2023}.  Assuming this extends to lower stellar metallicities, this could provide one physical reason for the galaxies' \xiion\ to increase as the metallicity declines with redshift \citep{llerena_2024}.

A related consideration is that galaxies in our samples may be dominated by young, $\lesssim 5$~Myr, stellar populations, which would increase the relative number of O-type stars.  This could be a consequence of ``burstier'' star-formation histories in these galaxies, which have been reported in some studies using \jwst\ \citep[e.g.,][]{Endsley_2024,Looser_2023,Simmonds_2024b,Cole_2025}.  Indeed, \citet{Yung_2020a} show that young stellar populations, with ages $\lesssim$ 10~Myr, are required to keep $\log \xiion > 25.5$, without invoking changes to the IMF.   In nearby galaxies, the age of \ion{H}{2} regions correlates inversely with \ha/UV and EW(\ha) \citep{Scheuermann_2023}.  If most of our galaxies are experiencing bursts, and if the frequency of bursts increases with redshift \citep[as argued by, e.g.,][]{Cole_2025}, then this provides a physical explanation for the trends in our sample.   It may be that galaxies experiencing lulls in their star formation have lower \xiion, which would potentially be underrepresented in spectroscopic samples such as ours \citep[see, e.g.,][]{Looser_2024}.  \citet{Simmonds_2024b} considered this by using a mass-complete photometric sample that potentially includes objects with weaker emission lines.  They found an increase in the scatter of \xiion, including a tail of objects with lower \xiion, primarily at $z \lesssim 7$, but that \xiion\ remains high, $\log \xiion \simeq 25.4$ at $\muv = -18$. \editone{\citet{Begley_2025} obtain similar results.}  

It is interesting then to consider what properties of galaxies correlate with \xiion\ at higher redshifts.   
%
%
Figure~\ref{fig:O32_comparisons} shows the relation between the \oiii/\oii\ ratio, EW of \hb\ and \oiii, and \xiion\ for the galaxies in our samples.   The \xiion\ of our galaxies increases with (1) increasing \oiii/\oii\ ratio, (2) increasing EW(\hb), and (3) increasing EW(\oiii).  Comparing our sample to galaxies at redshifts $1 < z < 3$ \citep{Shivaei_2018,Shen_2024} the relation between \xiion\ and \oiii/\oii\ appears nearly unchanged with redshift, consistent with other \jwst\ studies \citep[e.g.,][]{llerena_2024}. \editone{The relation between EW(\oiii) and \xiion\ for our samples is also shallower than those of ``extreme'' line emitters at lower redshift \citep{Tang_2019} indicating EoR galaxies may not fall in this category.}.   \oiii/\oii\ correlates with the ionization parameter with a secondary dependence on metallicity, such that lower metallicity galaxies will have higher \oiii/\oii\ ratio at fixed ionization parameter \citep{Strom_2018,Papovich_2022}.  This is able to explain the fact that at fixed \oiii/\oii\ ratio, galaxies at higher redshift have 
higher \xiion, as they are expected to have lower metallicity. 
Interestingly, the fact that we see some galaxies have higher \oiii/\oii\ at fixed EW(\hb) compared to lower-redshift samples \citep{Shen_2024}, this could indicate evidence for radiation-bounded \ion{H}{2} regions \citep{Plat_2019}. 



\subsection{Why are the inferred escape fractions of \\EoR galaxies low?}\label{section:discussion_fesc} 

\begin{figure*}[t]
    \begin{center}
    \gridline{
        \includegraphics[height=0.42\textwidth]{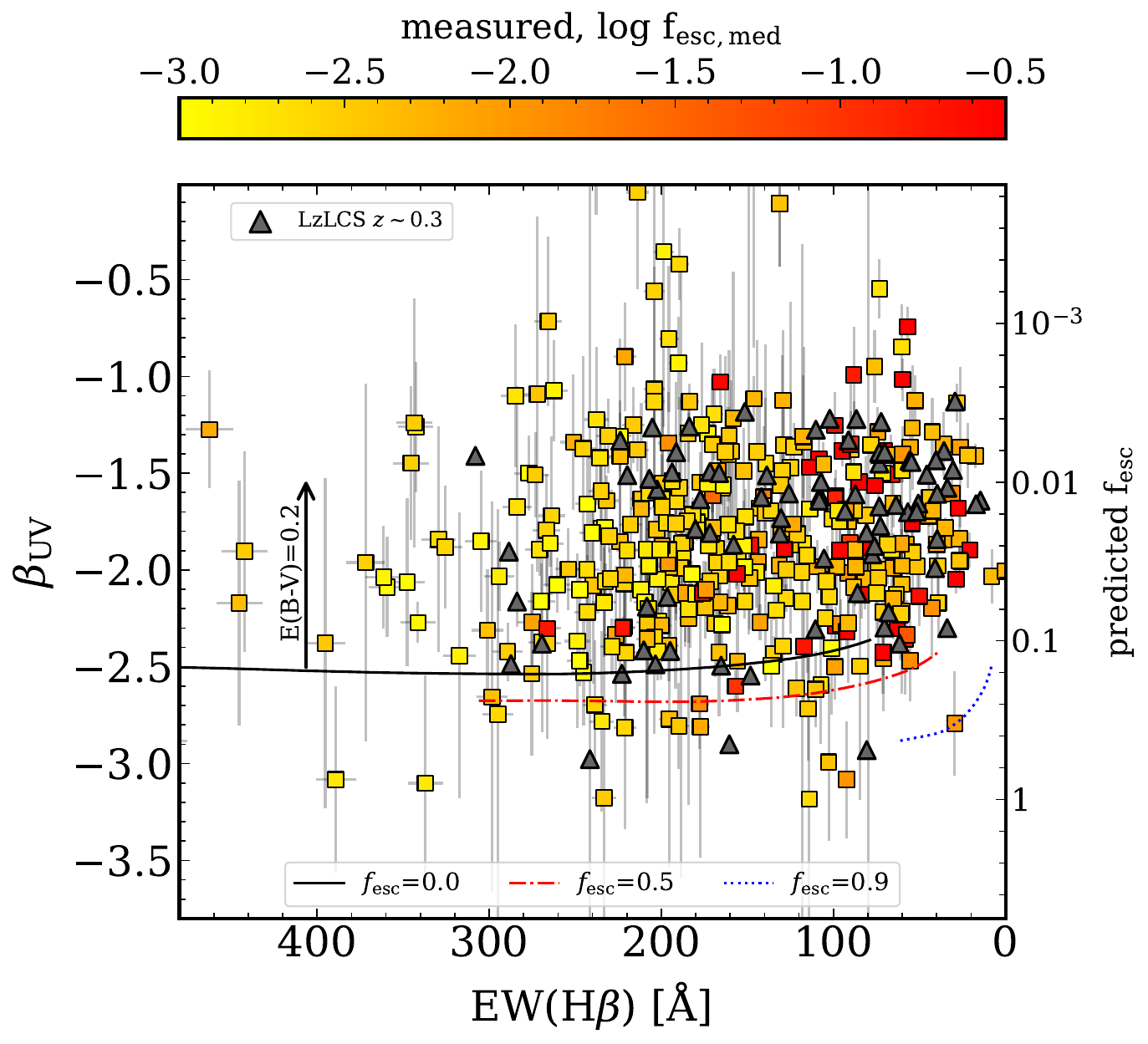}
        \includegraphics[height=0.42\textwidth]{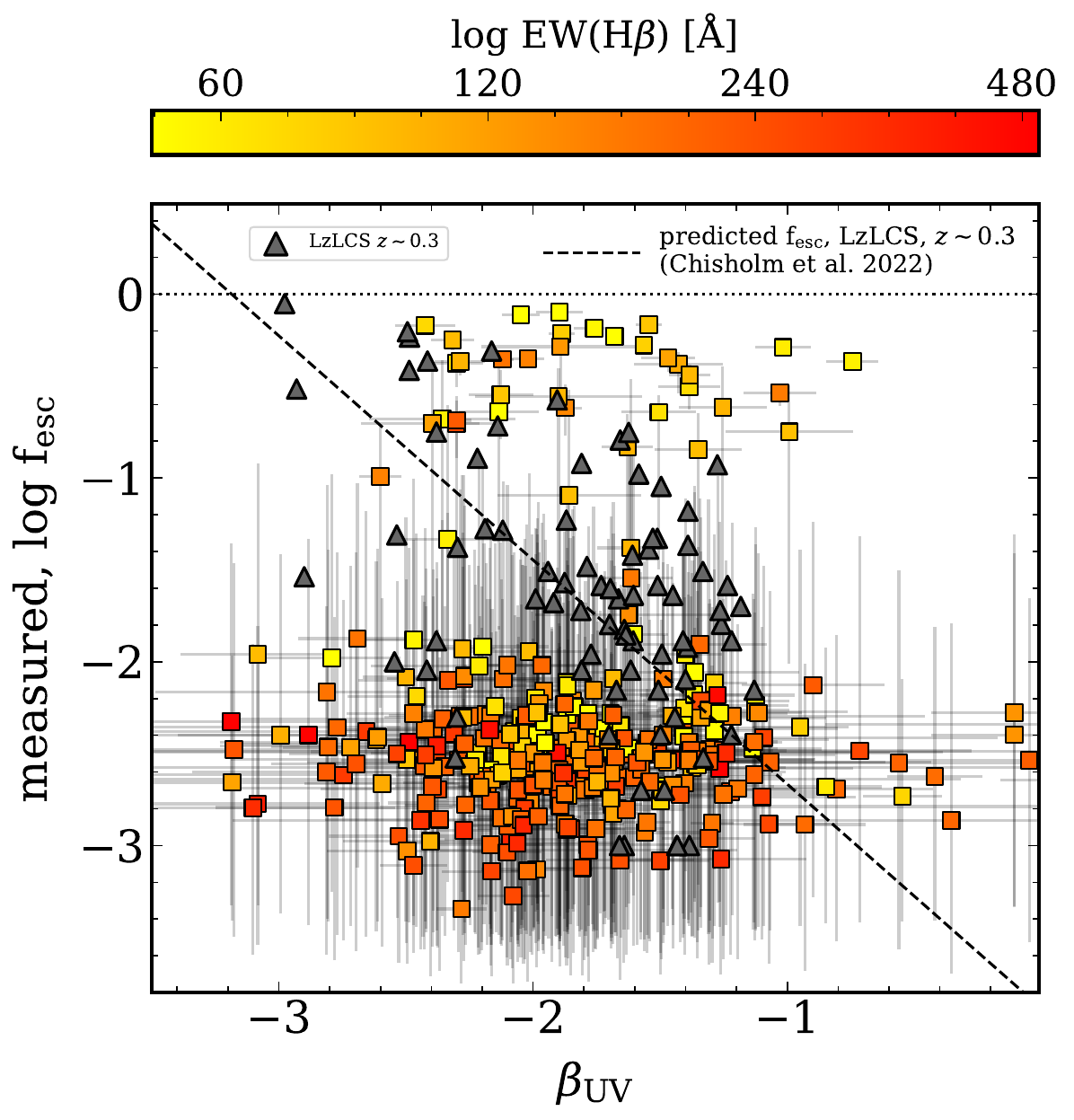}
    }\vspace{-3pt}
    \gridline{
            \includegraphics[height=0.42\textwidth]{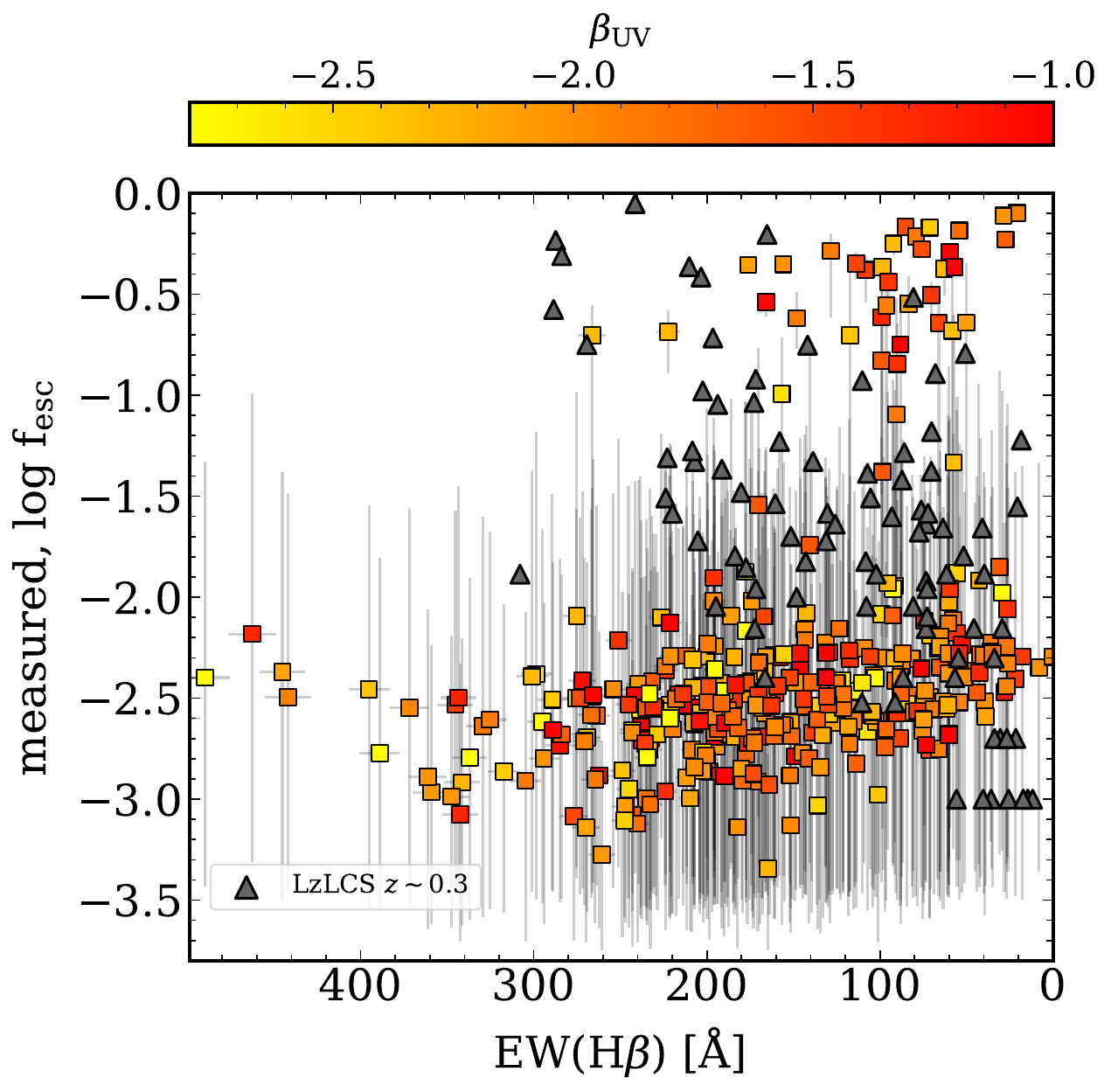}
        \includegraphics[height=0.42\textwidth]{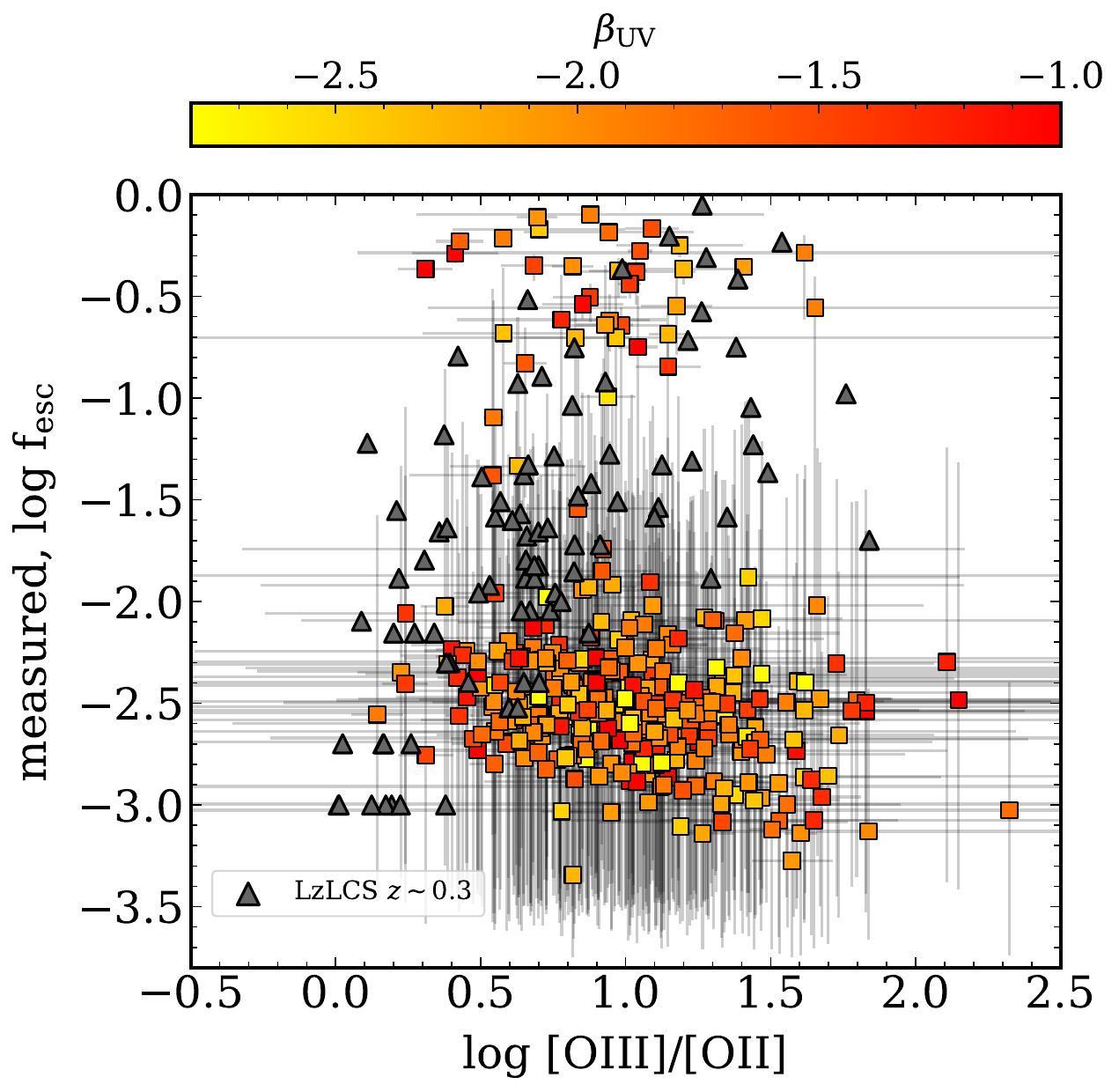}
    }
    \end{center}
    \vspace{-24pt}
    \caption{Relation between the UV spectral slope, $\beta_\mathrm{UV}$, \hb\ EW, \oiii/\oii ratio, and the measured \fesc\ for galaxies in our sample.  Each panel shows different projections of these quantities. The large data points show the samples here, and are color coded as labeled.  The \editone{triangles} show measurements from LzLCS at $z\sim 0.3$ \citep{Flurry_2022b}.   In the top-right panel, the right-hand axis shows the \textit{predicted} \fesc\ using the UV spectral slope based on LzLCS from \citet{Chisholm_2022}.  Unlike the prediction, there is no relation between the measured \fesc\ and UV spectral slope in our samples. 
    In the top left panel, the curves show the expected relation between $\beta_\mathrm{UV}$ and EW(\hb) for a range of \fesc\ (as labeled). Each model assumes $Z=0.05$~$Z_\odot$ and a constant star-formation history.  These roughly bound the lower envelope of $\beta_\mathrm{UV}$ and EW(\hb), where the arrow indicates the effects of dust attenuation with $E(B-V) = 0.2$.
    %
    }
    \label{fig:uvbeta_fesc}
\end{figure*}

In Section~\ref{section:results_fesc}, we showed that the inferred escape fractions of $4.5 < z < 9.0$ galaxies are low, a few percent or less on average for most galaxies.  Our \fesc\ values are based on the SED fitting, and implies that the strength of the emission lines requires nearly all of the ionizing photons produced by the stellar population to be absorbed in the \ion{H}{2} region and reprocessed into the nebular emission.  The observable factors that drive this relation are therefore the relative shape of the rest-far-UV continuum to the rest-optical continuum, and the strength of the emission lines themselves.  As we show above (Section~\ref{section:results_fesc}), changes in \fesc\ impact the relative strength of these components (see also Appendix~\ref{appendix:fesc_sed})
%
Nevertheless, an underlying uncertainty in this analysis is that \fesc\ requires that the stellar populations and nebular emission resemble reality.  We discuss this further below (Section~\ref{section:fesc_challenges}), but ultimately to test the details of these models will require analyses of a multitude of emission lines for individual galaxies at these redshifts.  This is beyond the scope of our present work, but may be possible with forthcoming datasets from \jwst\ \citep{Hutchison_2024,Papovich_2024}.  

We also show in Section~\ref{section:results_fesc} that there is no \editone{strong} evidence for evolution in the \fesc\ values with either redshift nor \muv\ (see Section~\ref{section:results_fesc}, Figure~\ref{fig:fesc_muv}). 
This is striking given the expectations based on low-redshift studies. Figure~\ref{fig:fesc_muv} shows the predicted relation for $z=5$ galaxies from  \citet{Chisholm_2022} based on correlations of \fesc\ and \muv\ and the UV-spectral slope, $\beta$ from LzLCS.   When applied to galaxies at $z > 5$ these relations predict a high \fesc, which increases with redshift, exceeding 5\% (10\%) for galaxies at $z > 5$ with $\muv$ fainter than $-19$ ($-17.5$)~mag, consistent with findings from studies that apply these relations to EoR galaxies from \jwst\ data \citep{Topping_2022,Cullen_2023,Mascia_2024,Hayes_2025}. These predictions favor even higher \fesc\ values for galaxies into the EoR, $z > 6-7$ \editone{\citep[see also,][and discussion below]{Choustikov_2024}.}

At low redshift, $z\sim 0.3$, there are measured correlations between various galaxy properties and \fesc\ \citep[e.g.,][]{Chisholm_2022,Flurry_2022b}.  Figure~\ref{fig:uvbeta_fesc} shows the relation between $\beta_\mathrm{UV}$ and EW(\hb), \fesc\ and $\beta_\mathrm{UV}$, \fesc\ and EW(\hb), and \fesc\ and $\log \oiii/\oii$ for our sample and for LzLCS galaxies at $z\sim 0.3$.   In general, there is no trend between bluer $\beta_\mathrm{UV}$ and higher \fesc\ seen in the galaxies in our samples.   Interestingly, a small fraction of our sample appears to follow the trend see in LzLCS between increasing \fesc\ and with decreasing (bluer) $\beta_\mathrm{UV}$, implying some galaxies may follow the \editone{relation seen in low-redshift galaxies}.    Also of interest is that  in our samples the EoR galaxies with the highest EW(\hb) also have the lowest \fesc\ values.  This contrasts somewhat with the $z\sim 0.3$ samples, which show a trend between increasing EW(\hb) and \fesc\ \citep{Flurry_2022b} (also illustrated in Figure~\ref{fig:uvbeta_fesc}).   The main reason for this is that in the SED modeling strong EW(\hb) requires most of the LyC photons to be absorbed (and therefore not to escape).    We also observe no correlation between ionizing parameter, tracked by \oiii/\oii, \editone{\hb}, and \fesc, in contrast to the measurements at $z\sim 0.3$ \citep{Flurry_2022b}.   This may indicate there is a difference in the physical conditions between galaxies in the EoR and ionizing galaxies at $z \sim 0.3$.  \editone{Other predictive relations between \fesc\ and multivariate galaxy properties such as $\beta_\mathrm{UV}$, $L(\hb)$, O32, or (\oiii+\oii)/\hb\ \citep{Choustikov_2024,Jaskot_2024}, show no correlation with the values we measure \fesc. Interestingly, however, applying these to our sample, both multivariate relations derived empirically \citep{Jaskot_2024} and from simulations \citep{Choustikov_2024} favor low \textit{average} values of $\langle \fesc \rangle \simeq 4$\% for our sample, equal to the mean value from from our analysis.  However, these do \textit{not agree} on which galaxies have high or low \fesc\ as the distributions show no correlation (we obtain similar values using the medians, where we obtain 0.0032 for our analysis, compared to 0.0031 and 0.0082 for \citealt{Jaskot_2024} and \citealt{Choustikov_2024}, respectively).  Why this is the case is not clear.   Therefore, the multivariate measures predict low average \fesc\ for the population, but they disagree on the details.  We plan to study these differences in a future work.}


\subsection{On the evolution of the volume-averaged escape fraction}\label{section:average_fesc}

In Section Section~\ref{section:ndot_ion} we calculated $\dot n_\mathrm{ion}$ using a Monte Carlo simulation.  This modeling included the fact that for most galaxies the \fesc\ distribution functions, $P(\log \fesc)$, are broad.    This is apparent by looking at the 16th-to-84th percentile range for individual galaxies in Figure~\ref{fig:fesc_muv}.  While the distance from the 16th--50th percentile and 50th-84th percentile is similar in logarithmic spacing, this means the distance between the median to the 84th percentile is substantially larger in \textit{linear} spacing compared to the distance from the median to the 16th percentile.  This means that a distribution drawn from $P(\log \fesc)$ randomly for each galaxy will, \textit{on average}, have an expectation value that will be higher than the median from the $P(\log \fesc)$ likelihood. 

\begin{figure}[t]
    \begin{center}
    \includegraphics[width=0.48\textwidth]{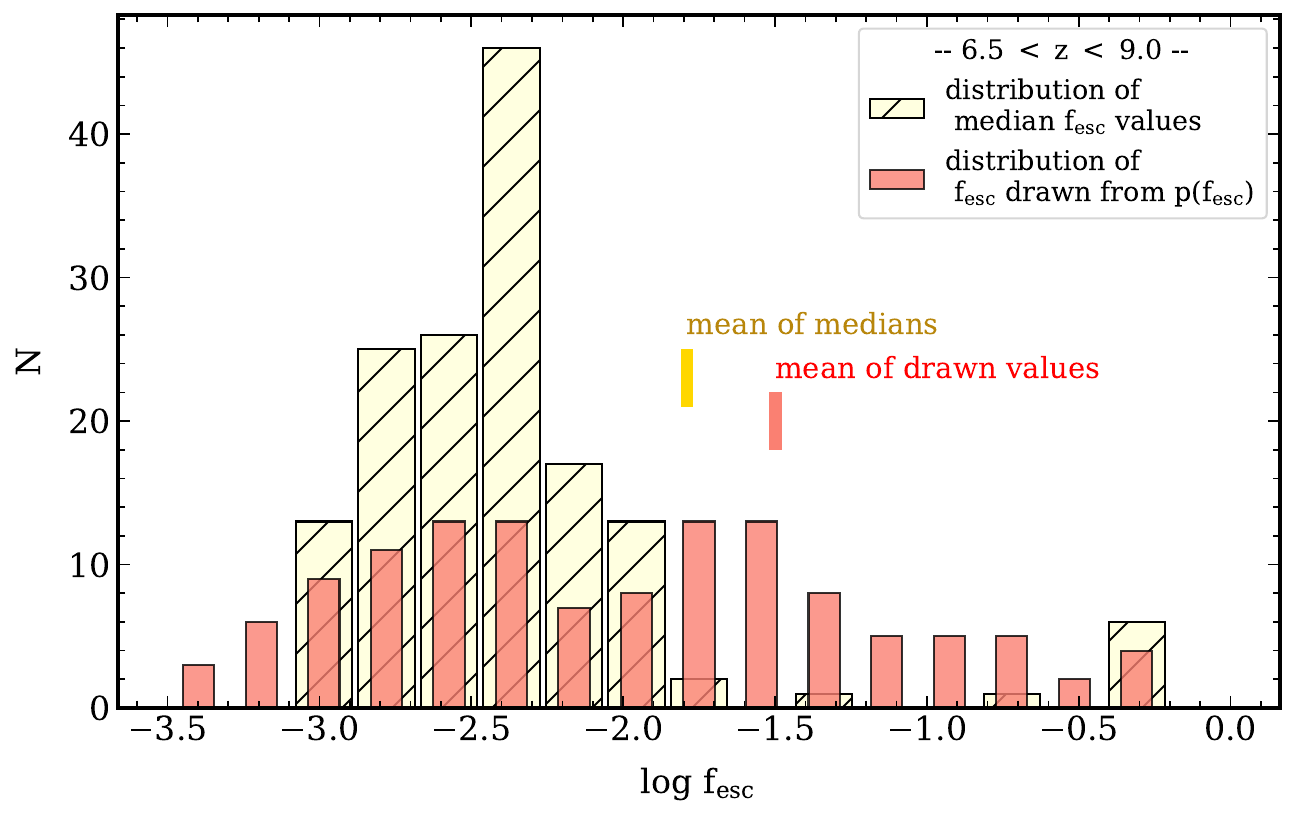}
    \end{center}
    \vspace{-12pt}
    \caption{Distribution of the inferred LyC escape fraction, \fesc, for CEERS and JADES galaxies at $6.5 < z < 9.0$.  The yellow, hash-filled histogram shows the distribution of \textit{median} values for each galaxy.   These show that the mean \fesc\ from the galaxy medians is $\simeq 1.5$\%.  However, the red, shaded histogram shows the distribution when we draw a random value from the $P(\fesc)$ for each galaxy. In this case, the mean \fesc\ is $\simeq 3$\%, roughly twice as high as for the mean of the \textit{medians}, and there exists a small tail of objects with $\fesc \gtrsim 10$\%.  This is a result of the $P(\fesc)$ for most galaxies having a large tail extending to higher \fesc\ (see Figure~\ref{fig:fesc_muv}). We obtain similar results for galaxies at $4.5 < z < 6.5$. }
    \label{fig:fesc_distribution}
\end{figure}

Figure~\ref{fig:fesc_distribution} illustrates this effect.   The figure shows the distribution of the median $\log \fesc$ values for galaxies in our sample at $6.5 < z < 9.0$ (yellow--shaded histogram). The mean of these medians is $\langle \fesc \rangle$ = 0.015.   However, this is \textit{not} to be confused with the expectation value for \fesc\ of the galaxies.  For the latter, we can then recompute the distribution if we randomly draw a $\log \fesc$ value from the $P(\log \fesc)$ of each galaxy, which we show as the red-shaded histogram in the figure.  Because the tail of possible values skews toward higher \fesc, the resulting is similarly skewed.   The mean of this resampled distribution is $\langle \fesc \rangle = 0.03$, a factor of two higher.   This highlights the importance of considering the full range of the \fesc\ probability distribution functions, $P(\log \fesc)$, from the posteriors of the SED fitting when considering the global escape fraction.  

Following \citet{Munoz_2024}, we calculate the \textit{volume-averaged} escape fraction as the ratio of $\dot n_\mathrm{ion}$ using the measured distribution of \fesc\ to the value assuming all of the ionizing radiation escapes, $\fesc = 1$, that is, 
\begin{equation}\label{equation:average_fesc}
\langle f_\mathrm{esc} \rangle = \frac{ \dot n_\mathrm{ion}(f_\mathrm{esc}) }{\dot n_\mathrm{ion}(f_\mathrm{esc}=1)}
\end{equation}
  This makes $\langle \fesc \rangle$ the average weighted by the ionizing emissivity.   We  compute the result from Equation~\ref{equation:average_fesc} for each iteration of the Monte Carlo simulation, and take the median as the quoted value with the 16th--to--84th percentile as the uncertainty.  These values for $\langle \fesc \rangle$ are listed in Table~\ref{table:average_fesc}.

\begin{deluxetable}{c@{\hskip 12pt} c@{\hskip 12pt} c@{\hskip 12pt}}
\tablecolumns{3}
\tablewidth{0pt}
\tablecaption{Inferred Volume-averaged Escape Fractions\label{table:average_fesc}}
\tablehead{\colhead{} & \multicolumn{2}{c}{$\langle \fesc \rangle$} \\ \colhead{$z$~~~~~~} & \colhead{~~~~~~($\muv < -16$)~~~~~~~~~~~} & \multicolumn{1}{c}{~~~~~~~($\muv < -14$)~~~~~~~~~~~~~~} }
\startdata 
5.0 & \editone{0.065 $\pm$ 0.022} & \editone{0.058 $\pm$ 0.024} \\
6.0 & \editone{0.055 $\pm$ 0.023} & \editone{0.058 $\pm$} 0.026 \\ 
7.0 & \editone{0.027 $\pm$ 0.015} & \editone{0.024 $\pm$ 0.024} \\
8.0 & \editone{0.026 $\pm$ 0.014} & \editone{0.025 $\pm$ 0.017} \\
\enddata
\end{deluxetable}

\begin{figure}[t]
    \begin{center}
    \includegraphics[width=0.48\textwidth]{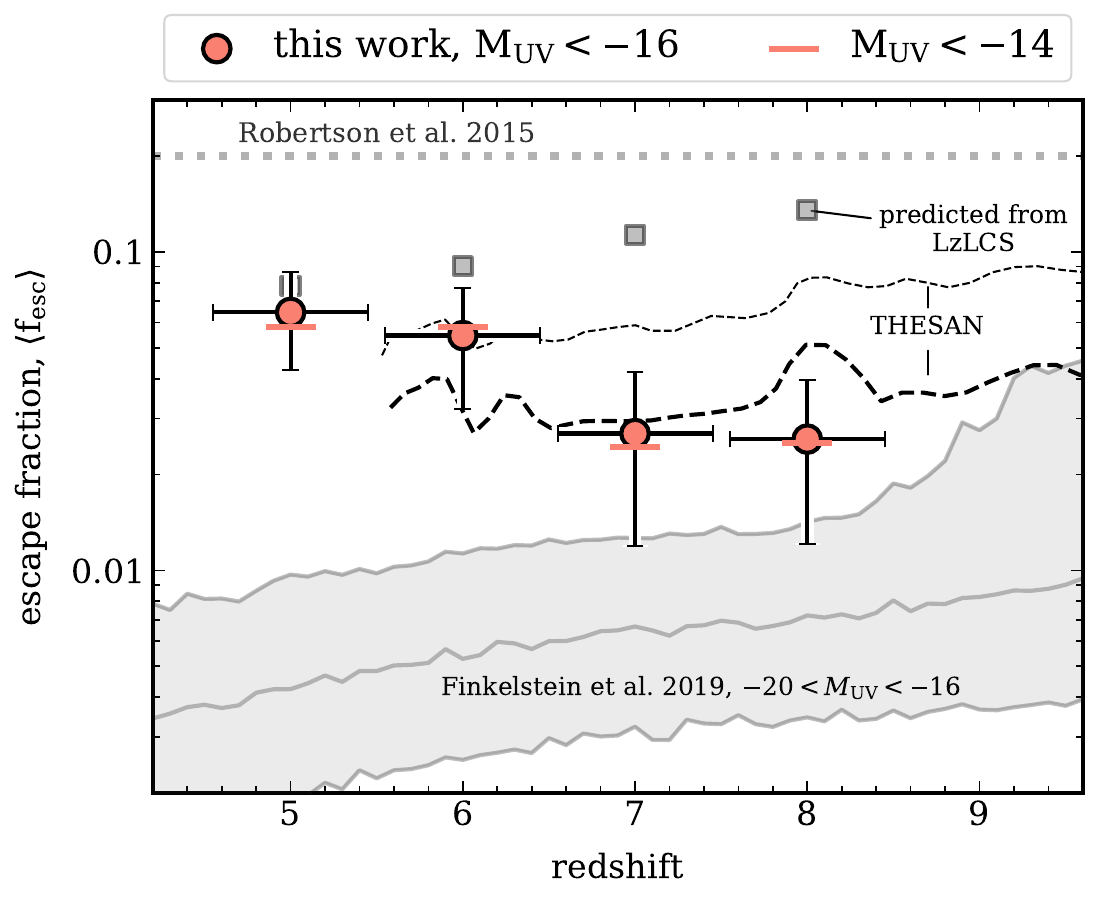}
    \end{center}
    \vspace{-12pt}
    \caption{The volume-averaged inferred LyC escape fraction, $\langle \fesc \rangle$ for galaxies as a function of redshift.  For each datum, we randomly select a value of \fesc\ for each galaxy in the bin from its $P(\fesc)$, and compute the volume-averaged value from Equation~\ref{equation:average_fesc}, integrating the UV luminosity function to $\muv < -16$.  The error bars show the 16--84\% range derived from the Monte Carlo.  The red-horizontal dashes show the results if we integrate to $\muv < -14$.   The small gray boxes show the results if we assumed \fesc\ values based on the relation from LzLCS at $z\sim 0.3$ \citep{Chisholm_2019}. The dotted horizontal line shows the canonical value of \fesc=20\% \citep{Robertson_2015}.  The dashed lines show predictions from the \textsc{Thesan} simulation for the global galaxy population (light line) and galaxies with $\log M_\ast / M_\odot = 8-9$ (thick line, \citealt{Yeh_2023}).
The shaded region shows simulation-based \fesc\ values from \citet{Finkelstein_2019}, which were required to match reionization using a variable $\xiion$ and $M_{lim}$ values. }
    \label{fig:fesc_volume_averaged}
\end{figure}

Figure~\ref{fig:fesc_volume_averaged} shows the results of this Monte Carlo for the volume-averaged \fesc.     The average inferred escape fractions show no evidence for evolution in redshift from $z\sim 5$ to 8, where the \editone{median values are $\langle \fesc \rangle \simeq 3-7$\%.  The 16th--to--84th percentile range varies a bit with redshift, but falls within $\simeq 1-9$\%}.   This conclusion does not change appreciably if we integrate the UVLF to $\muv = -14$~mag (indicated by the short, vertical red lines in Figure~\ref{fig:fesc_volume_averaged}).  This is because we have extended our relations to the fainter \muv\ galaxies, assuming they follow the same distributions as those for galaxies at $\muv = -16$.  However, if $\muv > -16$ galaxies have higher escape fractions, they will certainly increase the volume averaged \fesc\ values \citep[see, e.g.,][]{Yeh_2023}.  

\editone{The cosmic-averaged values of \fesc\ we derive at $4.5 < z < 9.0$ are consistent with measurements of the \fesc\ derived from galaxies at $z\sim 3-3.5$ through direct measurements.  These studies derive average \fesc\ by stacking spectroscopic or photometric data, and find $\fesc \simeq 2-7$\% \citep[and references therein]{Grazian_2017,Steidel_2018,Pahl_2021,Begley_2022} averaged over the population.  Similar results are found by \citet{Saldana-Lopez_2023} using relations between ISM absorption lines and other galaxy properties.  While these are not directly comparable to the averages we derive (which are weighted by the UVLF), it is additional evidence that  the average \fesc\ at high redshifts is relatively low.}

  Our finding of low average \fesc\ runs contrary to most pre-JWST predictions, and predictions made using observations of local galaxies.  \citet[and references therein]{Robertson_2015} argue \fesc\ needs to be 20\% for reionization to occur at a redshift consistent with the CMB polarization.  This is in part because that these past studies assumed EoR galaxies will have relatively low ionizing production efficiencies, $\log \xiion = 25.2$, and the higher \fesc\ values are then required.   As we show below, high \fesc\ values are unnecessary when the ionizing production efficiencies are high \citep[see also][]{Munoz_2024, Simmonds_2024b}.   Interestingly, our results exceed the simulation-based \fesc\ values from \citet{Finkelstein_2019}, who used these to match reionization timing constraints using variable $\xiion$ and $M_{lim}$ values.  However, our results are consistent with predictions from the \textsc{Thesan} simulation for galaxies in our stellar-mass range \citep{Yeh_2023}. 
  
  The volume averaged values we derived would be much higher if we instead used \fesc\ values based on correlations from LzLCS at $z\sim 0.3$.   Figure~\ref{fig:fesc_volume_averaged} includes values for $\langle \fesc \rangle$ when we repeat the Monte Carlo simulation using correlations between \fesc\ and \muv\ and the UV spectral slope ($\beta_\mathrm{UV}$) predicted from LzLCS \citep{Chisholm_2019}. 
  %
  %
  In this case we would have expected the average \fesc\ to increase with redshift over our redshift range $4.5 < z < 9$, rising to $\langle \fesc \rangle > 0.1$ at $z \gtrsim 7$.   This is strongly \editone{disfavored by $>5\sigma$ from our results.}      


\subsection{Challenges in interpreting the escape fraction of \\EoR galaxies}\label{section:fesc_challenges}

One  challenge in interpreting our inferred \fesc\ values is that they depend on the underlying stellar population models used in the SED fitting.   The BPASS models used here have been demonstrated to reproduce the properties of young, metal--poor stellar populations dominated by short-lived massive stars under a variety of conditions \citep[see, e.g.,][]{Eldridge_2022}, but it is conceivable that galaxies during the EoR host stellar populations with different properties \citep[e.g.,][]{Zackrisson_2011}. Currently some of the best evidence supporting the use of BPASS models comes from observations that these models reproduce the colors and emission-line properties of high-redshift galaxies \citep[e.g.,][]{Strom_2017,Steidel_2018,Olivier_2022,Larson_2023}.   However, photoionization models using BPASS models fail in some galaxies to produce extreme emission properties, including galaxies with strong \ion{He}{2}, \ion{C}{4}, \oiii/\oii,  and other high ionization emission lines \citep[e.g.,][]{Lecroq_2024}.  Several of these studies have invoked stellar populations that include stars with effective temperatures of $T_\mathrm{eff} \sim 80-100$~kK to explain the high-ionization emission lines and nebular continuum of galaxies \citep{Olivier_2022,Cameron_2024,Katz_2024}.  However, even if these situations exist in the galaxies in our samples, this would \textit{increase} the ionizing photon rate per unit UV luminosity, increasing \xiion.  The existence of Pop~III or a top-heavy IMF would have a similar impact.  If \xiion\ is higher, then to match the emission line strength of the data would require increasing \fesc\ such as to not overproduce the strength of the emission lines of the galaxies in our sample.  That is, the product of $\xiion \times \fesc$ would increase, leading to an increase in the cosmic ionization rate, $\dot n_\mathrm{ion}$ leading to an earlier reionization in contrast to the CMB observations and studies of the IGM opacity from QSO studies.   Therefore this scenario seems disfavored.  

Another challenge is that the inferred escape fractions of the galaxies at $4.5 < z < 9.0$ contrast strongly with expectations from scaling relations observed in low-redshift samples \citep[e.g., LzLCS,][]{Flurry_2022b}.  The \hb\ and \oiii\ emission line EWs, \oiii/\oii\ line ratios, and UV spectral slopes of the galaxies in our sample are similar in many ways to those in the LzLCS at $z\sim 0.3$.  However, the escape fractions are very different (see Figure~\ref{fig:uvbeta_fesc}).  Why this is the case is unclear.   It may be that there are spatial disparities in the galaxies such that the regions of strong emission lines are physically distinct from regions of higher \fesc, or that \fesc\ is not emitted isotropically such that that average \fesc\ is much lower when averaged of $4\pi$ steradians\editone{ \citep[see, e.g.,][]{Martin_2024}.  This would impact direct measurements of the LyC emission but not values such as ours that are inferred from the reprocessed nebular emission. This possibility may be} supported by some studies of the kinematics of emission lines in $z\sim 0.3$ galaxies with LyC detections that show they can produce outflows that can clear such channels to promote higher \fesc\ \citep[e.g.,][]{Amorin_2024,Flury_2024}.  Therefore, some of the important missing ingredients in interpreting the escape fractions may result from variations in the spatial distribution of star formation, the gas-covering fractions, or both, all of which may be very different between low-redshift and high-redshift galaxies.

Yet another challenge is that theoretically the escape fraction is difficult to calculate because it depends heavily on the connection between the effects of star formation and stellar feedback on the geometry, density, and ionization structure of the ISM \citep{Ferrara_2013,Xu_2016}.  The radiation pressure and feedback of massive stars can create channels of ionized gas, through which ionized photons can escape, but in other sight lines, there may be significant clumps of neutral gas that absorb the ionizing radiation \citep{Clarke_2002}.   The escape fraction of each galaxy will have a temporal, radial, and angular dependence.  In addition, many simulations may not have sufficient spatial resolution to resolve the details of the escape fraction on scales of \ion{H}{2} regions \citep[see, e.g.,][]{Mascia_2024}.  

Nevertheless theory provides guidance as many simulations predict low \fesc\ for galaxies with masses like those in our sample. Simulations generally predict that galaxies in the lowest-mass halos have the highest escape fractions, with $\fesc \gtrsim 25$\% \citep[e.g.,][]{Wise_2014,Paardekooper_2015,Kostyuk_2023,Yeh_2023}.  
%
%
However, these galaxies typically have $\log M_\ast/M_\odot < 7$, and are expected to contribute less than 10\% of the total ionizing radiation \citep{Lewis_2020}.  \editone{Other work argues that there is a ``peak mass'' at which galaxies produce maximal \fesc\ \citep{Ma_2020}.  There may also be changes in the production of dust, as ``attenuation-free-models'' favor \fesc $<$ 10\% at $5\lesssim z \lesssim 9$ \citep{Ferrara_2025}}.   
Many studies predict that higher mass galaxies, $\log M_\ast \gtrsim 8$, like those in our sample, have $\fesc \sim 1-5$\% \citep{Xu_2016,Kostyuk_2023,Yeh_2023}, though other work finds higher values, $\fesc \sim 5-15$\% \citep{Xu_2016}.  This highlights that theory has not yet converged, and that it is important to constrain the escape fraction empirically.
  
Simulations also predict a large variance in \fesc\ among galaxies.  \citet{Kimm_2014} showed the galaxies may experience periods of high \fesc, with a delay following a period of intense star formation. \citet{Rosdahl_2018} used the \textsc{Sphinx} simulation to show that galaxies with high escape fractions, $\fesc > 10$\% occur only after a starburst, as feedback from the burst expels gas, temporarily halting star formation and increasing \fesc\ \citep[see also,][]{Barrow_2020,Ma_2020, Rosdahl_2022,Kostyuk_2023,Katz_2023}.    
%
%
Therefore, it may be that the periods of highest specific SFR and \xiion\ production occur with low \fesc, and vice versa, such that the product $(1-\fesc)\xiion$ is roughly constant \citep[e.g.,][]{Kimm_2014}.  If so, then at any time only a small fraction of galaxies contribute to the leakage of LyC emission into the IGM \citep{Paardekooper_2015}.  This scenario may also help explain the timing of reionization \citep[whose low volume-averaged \fesc\ values are illustrated in Figure~\ref{fig:fesc_volume_averaged}]{Finkelstein_2019}, and the strong \oiii/\oii\ emission in \lya-selected galaxies \citep{Secunda_2020,Naidu_2022}.    
%

%
%
%
%

To summarize the challenges from theory, most cosmological simulations predict that \fesc\ is low in most star-forming galaxies in the mass range like those in our sample because they have high gas densities, with few clear channels through which LyC radiation may escape.   This is a very different physical situation than observed in starburst galaxies at $z\sim 0.3$ with high LyC escape, and may explain the differences between the observed properties and \fesc\ values of in our sample (see, e.g., Figure~\ref{fig:uvbeta_fesc}).   This may be related to the fact that higher redshift galaxies appear to have higher variance in their star-formation histories \citep[e.g.,][]{Cole_2025}, where theory predicts that the strong feedback associated with star formation may temporarily expel the gas, pausing star formation and allowing more LyC emission to escape \citep[e.g.,][]{Kostyuk_2023}. 
%
%
The net effect is that the low values of \fesc\ we infer appear consistent with models, and this has implications for reionization.  
%




\begin{figure*}
    \begin{center}
        \includegraphics[width=0.49\textwidth]{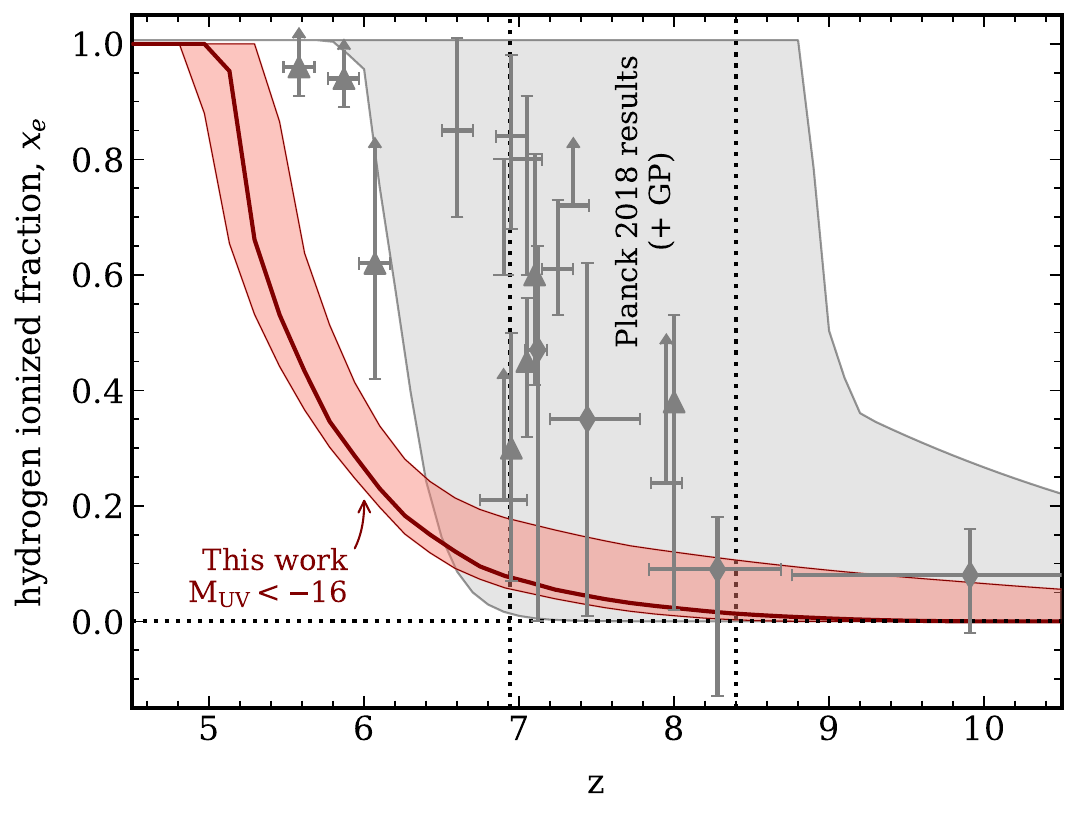}
        \includegraphics[width=0.49\textwidth]{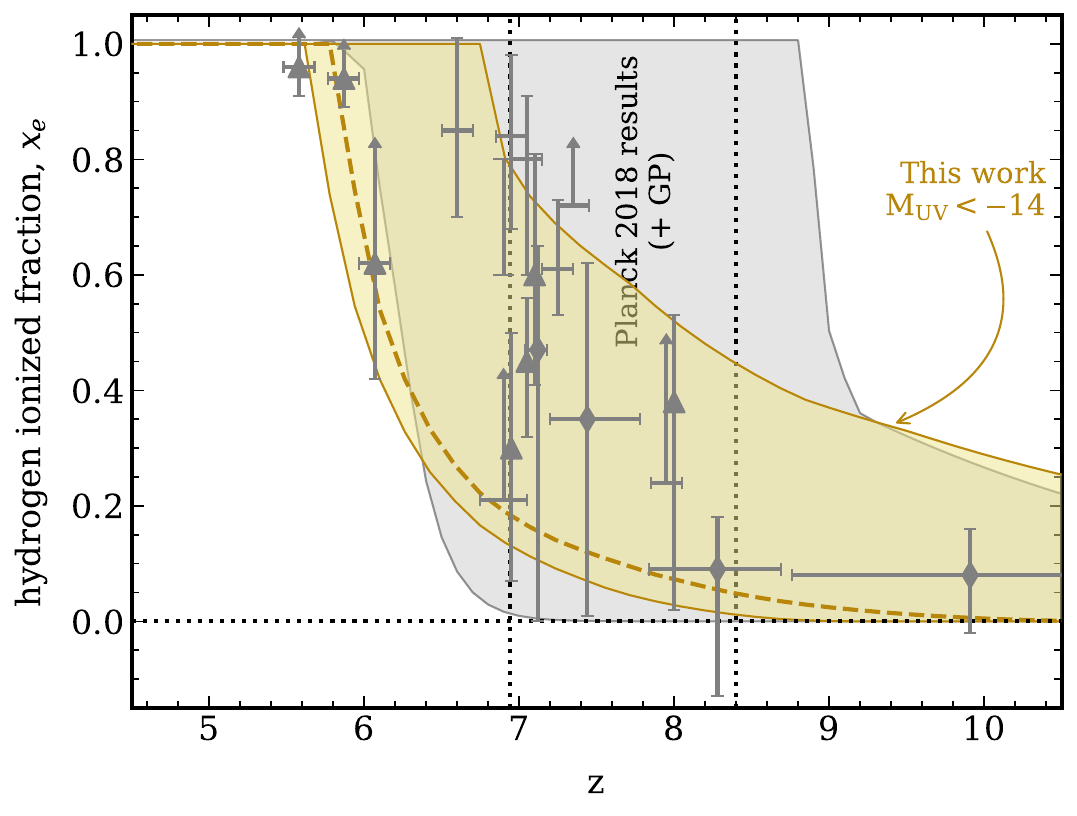}
    \end{center}
    \vspace{-12pt}
    \caption{Evolution of the ionized fraction of the IGM, $x_e$.  The curves and shaded regions show the results of our analysis based on integrating the UV luminosity function to $\muv < -16$ (left panel, red-shaded region) and to $\muv < -14$ (right panel, yellow-shaded region).  In each case, the width of the shading denotes the 16--84\% range derived from the Monte Carlo.  
    In both panels the gray-shaded region shows constraints from an analysis of Planck data with a prior using the measurements of the Gunn-Peterson (GP) trough, where $x_e = 0.5$ corresponds to $z_\mathrm{re}=7.67\pm 0.73$ \citep{Planck_2020}, indicated by the vertical dotted lines.   The data points show estimates from the optical depth of absorbers in QSO sightlines,  \editone{the evolution of \lya-emission in galaxies  \citep{McGreer_2015,Greig_2017,Hu_2019,Mason_2019,Wang_2020,Whitler_2020,Goto_2021,Nakane_2024}, and measurements from damped \lya\ profiles from the analysis of \jwst\ spectra \citep{Umeda_2024}}.   Our analysis favors a scenario where EoR galaxies produce sufficient ionizing photons to reionize the universe at times consistent with the CMB and QSO sightline data.  }\label{fig:Qvz} 
\end{figure*}

\subsection{The impact of high ionizing photon production efficiency and low escape fraction on reionization}\label{section:discussion_xe} 
 
The evolution of the cosmic ionizing rate density, $\dot n_\mathrm{ion}$, from Section~\ref{section:ndot_ion} and Figure~\ref{fig:ndotion}, allows us to estimate the evolution in the hydrogen-ionized fraction of the IGM, $x_e$.   For each of our Monte Carlo simulations we solve the differential equation for this evolution.  Following \citet[and references therein]{Mason_2015}, we define,
\begin{equation}\label{eqn:xe}
\dot{x}_e = \frac{\dot n_\mathrm{ion}}{\langle n_\mathrm{HI} \rangle} - \frac{x_e}{t_\mathrm{rec}}.
\end{equation}
Here, $\langle n_\mathrm{HI}\rangle$ is the average IGM neutral hydrogen density, $\langle n_\mathrm{HI}\rangle = (1-Q) n_\mathrm{H,0}$, and $n_\mathrm{H,0} = (1-Y_p) (\Omega_{b,0}/m_p) \rho_{c,0}$ for a primordial helium mass-fraction, $Y_p = 0.2454$, baryon density parameter, $\Omega_{b,0}=0.049$, and critical density, $\rho_{c,0}$, defined by $H_0 =67.4$ km s$^{-1}$ Mpc$^{-3}$  \citep{Planck_2020}.   The recombination time for \ion{H}{2} is  $t_\mathrm{rec}$, where we follow \citet[and references therein]{Robertson_2015} and take
$t_\mathrm{rec} = [ C \alpha_B(T) n_e(1+z)^3]^{-1}$.  The hydrogen recombination coefficient, $\alpha_B(T)$, depends on the temperature, and we adopt the functional fit of \citet{Hui_1997}.  The electron density is $n_e = (1+Y_p/4 X_p)\langle n_H \rangle$, where $X_p$ is the primordial hydrogen abundance. We further adopt the evolution of the clumpiness factor, $C$, from the simulations of \citet{Pawlik_2015}, who find that this evolves from $C=4.8$ at $z=6$ to $C=1.5$ at $z=14$ \citep[see also][although see \citealt{Davies_2024}, who argue for higher $C$ values]{Madau_1999,Finlator_2011,Finkelstein_2019}.  We then solve the differential equation (Equation~\ref{eqn:xe}) for each iteration of the Monte Carlo (Section~\ref{section:ndot_ion}) using the \texttt{ODEINT} routine in the \texttt{SciPy} package.    

Figure~\ref{fig:Qvz} shows the evolution of the hydrogen-ionized fraction of the IGM, $x_e$, from our analysis.  The thick solid line shows the median evolution from the Monte Carlo simulations, and the  color-shaded regions show the 16--84 percentile range of our results integrated to $\muv < -16$ (left panel) and $\muv < -14$ (right panel).    
As a comparison, the gray-shaded region in both panels in the figure show the $1\sigma$ constraints on reionization from an analysis of the CMB Planck large-scale polarization measurements along small-scale measurements of the kinematic Sunyaev–Zeldovich (kSZ) effect \citep{Raghunathan_2024} with a prior on $z_{\rm end} = 6$ using the measurements of the Gunn-Peterson (GP) trough \citep{Fan_2006}. Excluding kSZ data \citep{Raghunathan_2024} reduces the constraining power on the high-$z$ end by a few percent. 
  This shows that reionization is expected to begin by $z=8.4$ and end shortly after $z=7$ (vertical dotted lines,  \citealt{Planck_2020}).   Considering only galaxies brighter than $\muv < -16$, our results find that the hydrogen ionized fraction reached $x_e=0.5$ at \editone{(median) $z_\mathrm{re} = 6.3$, with a range of $5.3 - 5.8$ (68\% confidence)}.  Including fainter galaxies down to $\muv < -14$, and again assuming they have \fesc\ and \xiion\ distributions similar to galaxies at $\muv = -16$, our results find the \editone{hydrogen ionizing fraction of $x_e=0.5$ is reached earlier, with a median $z_\mathrm{re} = 7.7$, and with a range of $6.0-8.1$  (68\% confidence).} 

Our results provide evidence that galaxies in the EoR have high ionizing photon production efficiencies, $\xiion$, and low LyC escape fractions, $\fesc$.  The combination of these with the current constraints on the UVLF yield a cosmic photon ionization rate, $\dot n_\mathrm{ion}$, that causes reionization at times consistent with other observations.  This averts the photon crisis \citep{Munoz_2024}, but only if our measurements reflect the \xiion\ and \fesc\ distributions of the full galaxy population. 
%
%
While we find low \fesc, this depends on the balancing between the stellar population synthesis models we have assumed and the strength of the emission lines.   
It is possible that \jwst\ spectroscopic samples like ours are biased toward galaxies with strong nebular emission, with higher \xiion\ and lower \fesc.  \editone{As mentioned above in Section~\ref{section:discussion_xiion}, studies using photometric samples of galaxies and some spectroscopic studies \citep{Pahl_2024} find a weaker relation between \muv\ and \xiion.  If our sample is biased against galaxies with lower \xiion, particularly at faint UV magnitude, then the cosmic ionizing rate density, $\dot{n}$, will be lower, leading to reionization that occurs at later times.  Candidate galaxies for this case have been found in stellar-mass-limited photometric studies \citep{Simmonds_2024b}. Confirming their \xiion\ values, and attempting to infer \fesc, will be paramount to understand these biases.  This }will require deeper spectroscopy with JWST.


These results depend on some remaining uncertainties.   One uncertainty is the clumping factor, $C$, of the IGM, which we have assumed to be $C \simeq 2-5$ (see above).  If this is higher, as recently argued by \citet{Davies_2024}, then the universe would require more ionizing photons leaking into the IGM at earlier times. 

Another uncertainty is how fainter galaxies, $\muv > -16$, impact reionization.  
%
%
This depends both on the galaxy properties, and on the behavior of the UVLF, specifically the faint-end slope and if there is some minimum UV luminosity a galaxy in the EoR can have.   For the faint galaxies, some recent \jwst\ studies have measured high \xiion\ values for faint galaxies in the EoR \citep{Atek_2024,Naidu_2024}.  However, it is unclear if these are a biased subset of galaxies for which it is easiest to detect their emission lines.  Regarding the faint end of the UVLF, \citet{Atek_2024} recently measured constraints from \jwst, finding the steep slope extends at least to $\muv = -13$.  If these fainter galaxies are plentiful, if they have high \xiion, and especially if have higher \fesc, as predicted from models, then they may contribute significantly to reionization, reigniting the photon budget crisis \citep{Munoz_2024}.    Regardless, our findings are that \editone{brighter galaxies $\muv < -16$ have $\fesc \simeq 2-6$\%,} which mitigates this crisis, and that fainter galaxies in the EoR with $\muv > -16$ must contribute for reionization to occur at times consistent with current constraints.

\section{Conclusions}\label{section:summary}

 We have used new modeling of \hst and \jwst\ imaging and \jwst/NIRSpec prism spectroscopic data to study the ionizing production efficiency and LyC escape fractions of galaxies at $4.5 < z < 9.0$.  We use data from the CEERS and JADES surveys which provide imaging and NIRSpec/PRISM data for \editone{412} galaxies.  We fit these data for each galaxy simultaneously with stellar population and nebular emission models with \bagpipes, including parameters to correct for slit losses, and we include a parameter to account for the fraction of ionizing photons that escape the galaxy, \fesc. 
 
 We measure the ionization production efficiency, $\xiion = Q(\mathrm{H_0})/ \luv$, using measurements of the Balmer emission line strength taken from the NIRSpec data, combined with measures of the UV continuum luminosity from the SED modeling. (Section~\ref{section:results_xiion}). 
 In this way, the ionizing rate, $Q$, and UV luminosity, $L_\mathrm{UV}$, are taken from data that have the same flux calibration applied.   We find that the ionizing production efficiency increases with increasing redshift and decreasing UV luminosity.  There is a scatter in the ionization production efficiency at fixed UV magnitude and redshift, $\sigma( \log \xiion) \simeq 0.3$ dex.  This means galaxies of fixed UV magnitude produce a range of ionizing photon rate.
 
We also constrain the escape fractions, which are driven by the strength of the nebular emission lines and the number of ionizing photons predicted from the stellar population models.  The inferred escape fractions of the galaxies at $4.5 < z < 9.0$ are typically low, with medians $\fesc\lesssim 1-3$\% (Section~\ref{section:results_fesc}).  We furthermore see no evidence for evolution in redshift nor in \muv\ over the range of our samples.  
 
We use a Monte Carlo simulation to compute the cosmic ionizing rate density, $\dot n_\mathrm{ion}$ (Section~\ref{section:ndot_ion}).  This simulation includes uncertainties in the UVLF, and including our empirical distributions of \xiion\ and \fesc.   We find that considering galaxies down to our observational limit, $\muv < -16$~mag, the \editone{cosmic ionizing rate density is $\log \dot{n} = 50.1\pm 0.1$ at $z=8$ }and rises to $50.5\pm 0.1$ at $z=5$.   \editone{At $z > 6.5$, this is too low} compared to the the amount needed to reionization the universe as predicted by \citet{Madau_1999}.   Extrapolating to fainter galaxies, $\muv < -14$~mag, increases the cosmic ionizing rate to $\log \dot n = 50.5\pm 0.1$ at $z=8$ and $\log \dot n = 50.7 \pm 0.1$ at $z=5$, \editone{producing the needed radiation for reionization by $z\sim 6$.}  
 
The relations between \xiion\ and galaxy properties in our $4.5 < z < 9$ samples are similar to those at lower redshifts, $1 < z < 3$ (Section~\ref{section:discussion_xiion}). The high \xiion\ values for galaxies in our sample correlate with increased EW(\hb) and EW(\oiii), and \oiii/\oii\ ratio.  This implies that the elevated \xiion\ is related to a relative increase in the number of O-type stars, and may be suggestive of burstier star-formation histories.  

The relations between \fesc\ in our $4.5 < z < 9$ samples differ from predictions from correlations between \fesc\ and galaxy properties observed at $z\sim 0.3$ (Section~\ref{section:discussion_fesc}).   We find no correlation between the inferred \fesc\ the UV spectral slope, $\beta_\mathrm{UV}$.  We also find no strong correlations between \fesc\ and EW(\hb), nor \fesc\ and \oiii/\oii.   This may be related to different physical conditions in the EoR galaxies compared to local starburst galaxies. 

We use our Monte Carlo simulation of the ionizing rate density to estimate the volume-averaged escape fraction, $\langle \fesc \rangle$ (Section~\ref{section:average_fesc}).  For galaxies brighter than $\muv < -16$, \editone{we find $\langle \fesc\rangle$ ranges from $3\pm 1$\%  at $z\sim 8$ to $7\pm 2$\% at $z\sim 5$, with weak or no evolution over $4.5 < z < 9.0$.} There is only a slight change when we include fainter galaxies, $\muv < -14$, but this is likely a result of our assumption that galaxies with $\muv > -16$ have similar properties as those with $\muv = -16$.   Our measurement of $\langle \fesc \rangle$ contrasts strongly with pre-\jwst inferences from $z\sim 0.3$ galaxies.  However, our results are consistent with predictions from simulations (Section~\ref{section:fesc_challenges}), where high gas column densities reduce \fesc\ except in cases where strong feedback from star formation expels the gas, and clears channels through which LyC emission can escape.  However, these events typically halt star formation, leading to lower \xiion.  Therefore, it may mean that galaxies in the EoR conspire to keep the product of $\xiion \times \fesc$ roughly the same.  This will need to be tested using future data sets. 

 We estimate the impact of our measurements of \xiion\ and \fesc\ on reionization (Section~\ref{section:discussion_xe}), using the Monte Carlo simulations and then solving for the evolution of the hydrogen-ionized fraction, $x_e$, as a function of redshift.  Including galaxies brighter than $\muv < -16$ mag, the universe would not reach a hydrogen-ionized fraction of $x_e = 0.5$ until $5.3 < z < 5.8$ (68\% confidence), consistent with the lower-end of reionization based on QSO sight lines.   Including an estimate of the contribution from fainter galaxies, $\muv < -14$ mag, reionization occurs sooner, with $x_e = 0.5$ obtained by $6.0 < z < 8.1$,  broadly consistent with the CMB constraints.  However, this assumes that galaxies with $-16 < \muv < -14$ have ionizing production efficiencies and escape fractions similar to galaxies with $\muv = -16$~mag.   The implication is that while EoR galaxies produce plenty of ionizing photons, this emission does not efficiently escape.  We argue that this is a result of high gas fractions combined bursty star-formation histories which lead to short times in which the galaxies have sufficiently clear channels through the gas through which ionizing photons can escape. 

Our results have several caveats.  Among these is the assumption that we are able to constrain the \fesc\ in galaxies through the SED fitting.  Our tests with simulated galaxy SEDs shows this is reasonable (Appendix~\ref{appendix:fesc_sed}), but additional tests will require extending our analysis to galaxies at lower redshifts with similar data, and direct measurements of the LyC escape fractions.  We plan to pursue this in a future study.   Our results also depend on selection effects, where our samples may be biased toward objects with stronger nebular emission, favoring galaxies in a ``burst'' stage of star formation. \editone{However, even in this case, the implied \xiion\ values of galaxies in post-burst stages would be lower, reducing the number of ionizing photons, unless these galaxies also have higher \fesc.}   To test this will require more complete samples of galaxies in the EoR with \jwst\ imaging and prism spectroscopy.  This will become possible with forthcoming datasets that include larger samples of fainter galaxies\citep[e.g.,][]{Dickinson_2024}.

\begin{acknowledgements}

We wish to 
thank our colleagues in the CEERS collaboration for their
hard work and valuable contributions on this project.  \editone{We extend our sincerest thanks to the anonymous referee whose critical and constructive report improved the quality of this manuscript.}  We also thank the JADES team for providing an excellent dataset for science.   We with to thank colleagues for valuable discussions, feedback, and suggestions, including John Chisholm, Kevin Huffenberger, Jessica Meh, Julian Mu\~noz, Irene Shivaei, Justin Spilker, Aaron Smith, and Romain Teyssier.       

Portions of this 
research were conducted with the advanced computing resources provided
by Texas A\&M High Performance Research Computing (HPRC,
\url{http://hprc.tamu.edu}).  

This work benefited from support from
the George P. and Cynthia Woods Mitchell Institute for Fundamental
Physics and Astronomy at Texas A\&M University. CP thanks Marsha and Ralph
Schilling for generous support of this research.  This work was partially support by the Future Investigators in NASA Earth and Space Science and Technology (FINESST) program Grant No.\ 80NSSC23K1487.  RA acknowledges support of Grant PID2023-147386NB-I00 funded by MICIU/AEI/10.13039/501100011033 and by ERDF/EU, and the Severo Ochoa grant CEX2021-001131-S funded by MCIN/AEI/10.13039/50110001103. ACC acknowledges support from a UKRI Frontier Research Guarantee Grant (PI Carnall; grant reference EP/Y037065/1)
 This work
acknowledges support from the NASA/ESA/CSA James Webb Space Telescope
through the Space Telescope Science Institute, which is operated by
the Association of Universities for Research in Astronomy,
Incorporated, under NASA contract NAS5-03127. Support for program JWST-ERS-01345.009-A, JWST-GO-02079.013-A, JWST-GO-06368.011-A, and JWST-GO-01837.030-A, was provided by NASA through a grant from the Space Telescope Science Institute, which is operated by the Association of Universities for Research in Astronomy, Inc., under NASA contract NAS 5-03127.  

This work made use of v2.2 of the Binary Population and Spectral Synthesis (BPASS) models as described in \citet{Stanway_2018}.  

\editone{All the data used this work are available on MAST: JADES, \dataset[doi: 10.17909/8tdj-8n28]{\doi{10.17909/8tdj-8n28}}; CEERS, \dataset[doi: 10.17909/z7p0-8481]{\doi{10.17909/z7p0-8481}}.}

\end{acknowledgements}

\appendix

\section{Comparison of Nebular Emission Line Fluxes from Models and Direct Measurements.}\label{appendix:lines}

In our SED fits, \bagpipes\ models the nebular emission using the photoionization rate, $Q$, from the stellar populations, and converts it into a nebular spectrum.  The nebular spectrum depends on the metallicity of the gas (taken to be equal to the metallicity of the stellar populations) and the ionization parameter, $U = n_\gamma / n_{H}$, the ratio of the density of ionizing photons to the gas density.  The models assume a gas density, $n_e = 100$~cm$^{-3}$, and that the models are ``radiation bounded'' such that all of the ionizing photons are absorbed by the gas and converted into nebular emission \citep{Carnall_2018}.    All of these assumptions are reasonable in general, but may have deviations in practice.

\begin{figure*}[h]
    \gridline{ \includegraphics[width=0.48\textwidth]{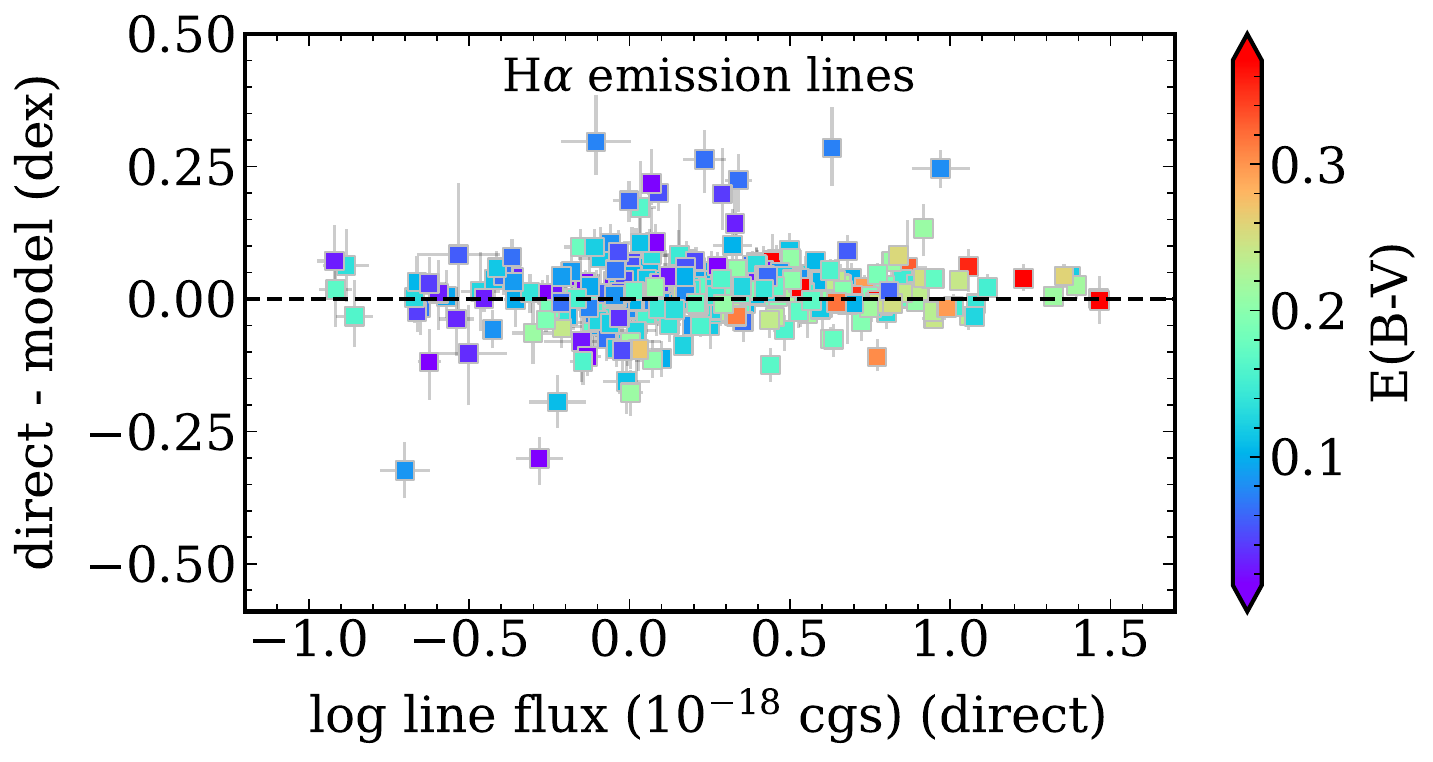}
    \includegraphics[width=0.48\textwidth]{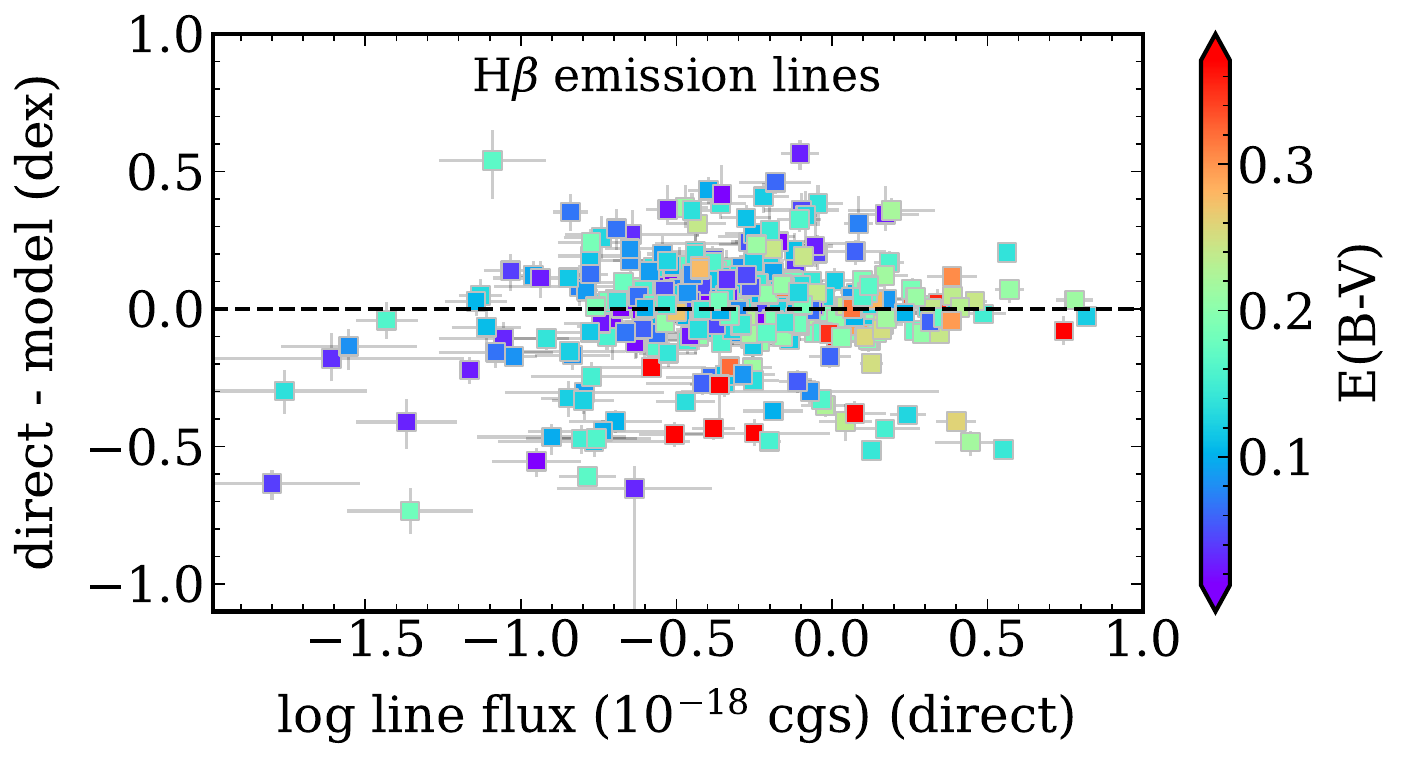}
    }\vspace{-12pt}
    \gridline{ \includegraphics[width=0.48\textwidth]{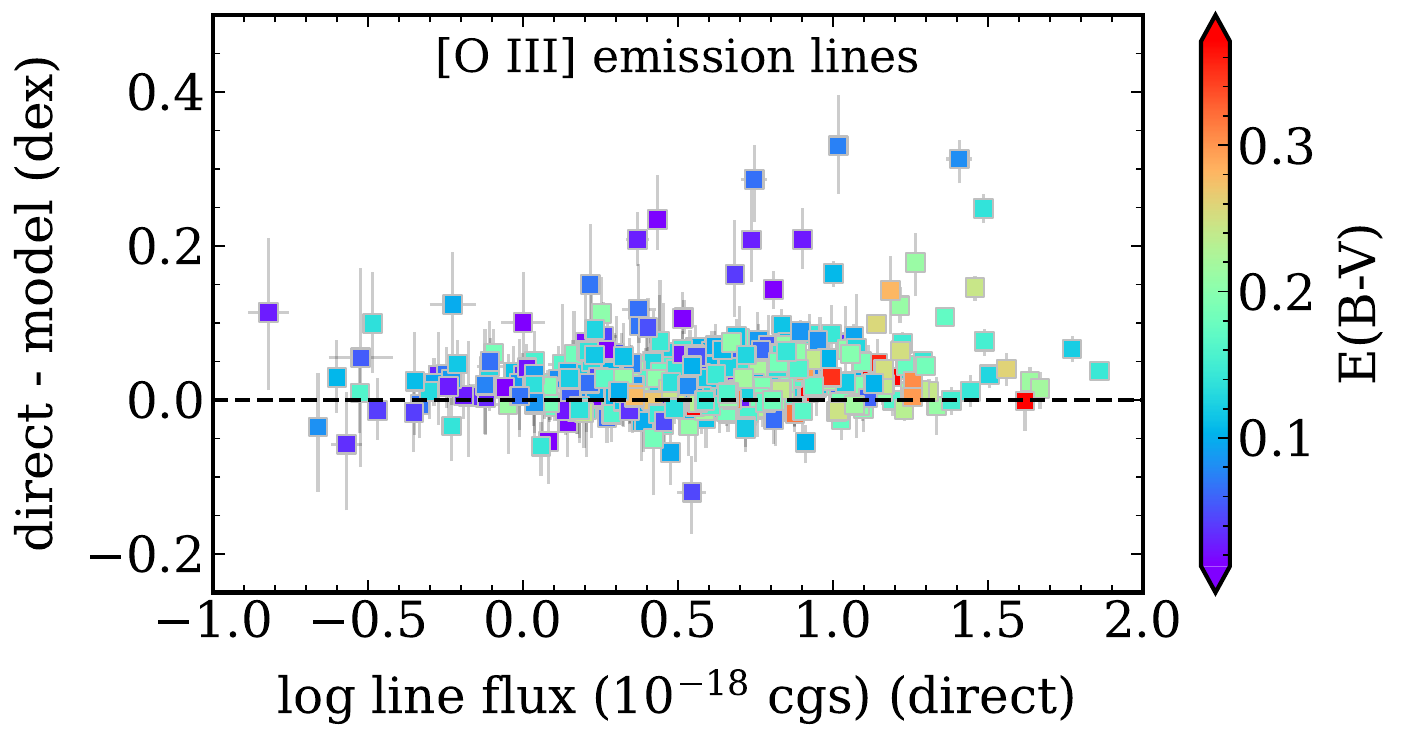}
    \includegraphics[width=0.48\textwidth]{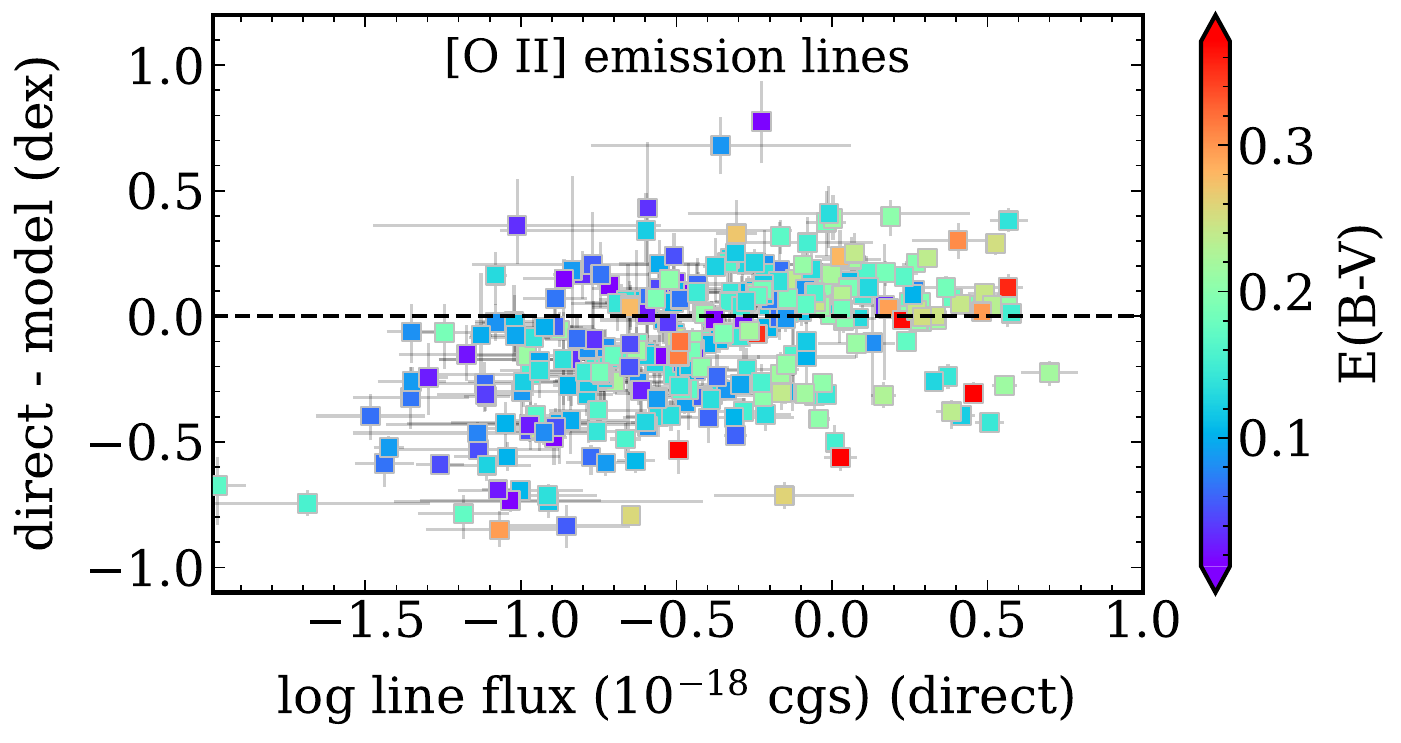}
    }\vspace{-12pt}
    \caption{Comparison of emission line fluxes for the sample of galaxies at $4.5 < z < 9.0$ considered here.  Each panel shows the comparison for strong nebular lines, \ha, \hb, \oiii, and \oii, as labeled, in units of $10^{-18}$ erg s$^{-1}$ cm$^{-2}$. The abscissa in each panel shows the emission line fluxes measured directly from the NIRSpec/PRISM data for each galaxy, as described in Section~\ref{section:lines}.  The ordinate in each panel shows the difference between the directly measured flux and the emission line flux from the \bagpipes\ models for each galaxy.  The data points are color-coded by the median color excess derived from the SED modeling.  }\label{fig:line_comparison}
\end{figure*}

One way to test the validity of the nebular models is to compare the emission line fluxes from the model fits to those measured directly from the spectra.  We described our method to measure directly the emission line fluxes in Section~\ref{section:lines}.  We also repeated that process on the \bagpipes\ model stellar populations that included the emission lines.    Figure~\ref{fig:line_comparison} compares these emission line measurements for \ha, \hb, \oiii, and \oii.   The \ha\ measurements are most consistent in the comparison, where the difference in the direct and model measurements have a median is 0.01 dex, and the scatter (derived from the normalized absolute deviation) is 0.04~dex.  Given that \ha\ has a strong connection to the number of ionizing photons, it is encouraging that the models match the direct measures.     The \hb\ emission is also very consistent, where the difference has a median is 0.0~dex, and the scatter is 0.12~dex, consistent with the systematic uncertainty we estimated in Section~\ref{section:xiion}.   The higher scatter for \hb\ compared to \ha\ is driven in part by the fact that the line is weaker than \ha.  However, there is also a population of galaxies where the models favor higher \hb\ compared to the direct measurements.  Many of these galaxies show higher dust attenuation (and these lie exclusively at $z < 6$), and may indicate a departure in the dust--attenuation law from \citep{Calzetti_2001}, which we have assumed here.

The comparison of the \oiii\ and \oii\ emission lines is instructive.  For \oiii, the difference in the directly measured line flux and model flux is relatively small, with a median of 0.02~dex and a scatter of 0.03~dex.   For \oii, however, there exists much larger discrepancy in the differences, with a median of $-0.08$~dex and a scatter of $0.30$~dex.   This means that \textit{on average} the \oii\ emission predicted by the models is stronger than what we measure by 20\%, but can extend to a factor of $2-3$.  There is no indication that this is caused by uncertainties in the dust modeling.    Interestingly, the largest offsets occur for galaxies with \textit{lower} overall measured \oii\ emission.  This may be a result of assumptions about the metallicity, density, temperature, and ionization parameter in the models.  Alternatively, this may indicate that the \ion{H}{2} regions are density bounded, such that the \oii\ regions of the nebula are not fully formed, which was considered by \citet[and references therein]{Papovich_2022} to explain high \oiii/\oii\ ratios in \hst\ grism spectra of galaxies at $z\sim 1-2$.  We plan to explore these aspects in a future study.


\section{Measuring the Escape Fraction from SED fitting}\label{appendix:fesc_sed}

\begin{figure}[h]
    \centering
    \includegraphics[height=1.55in]{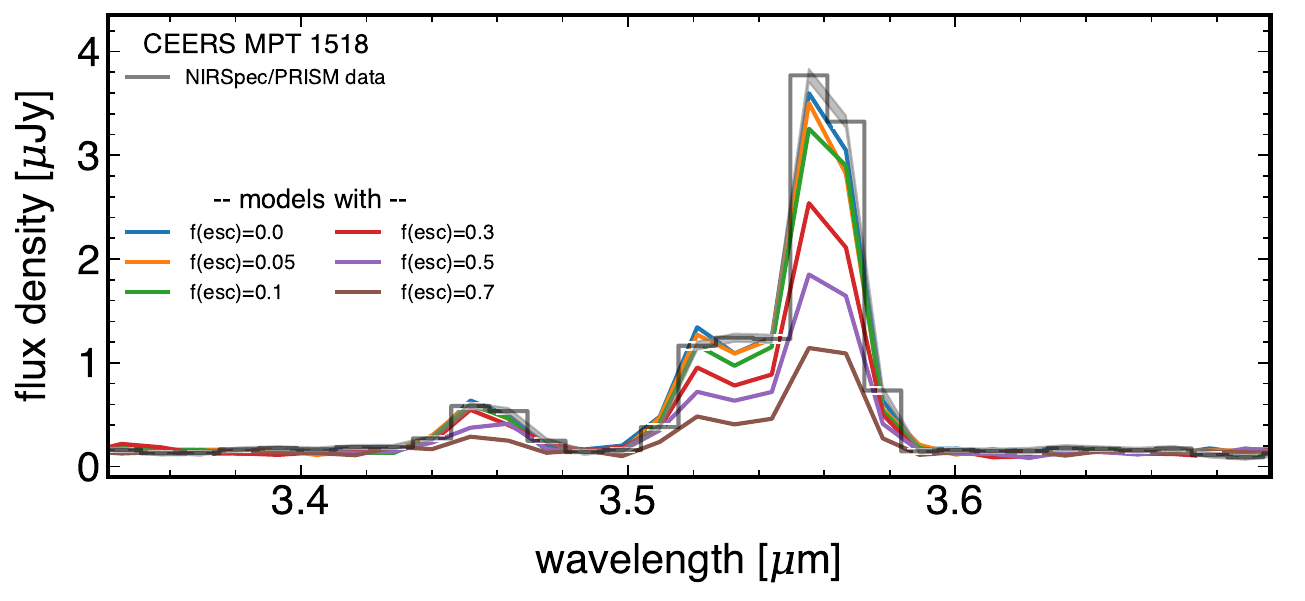}
    \includegraphics[height=1.5in]{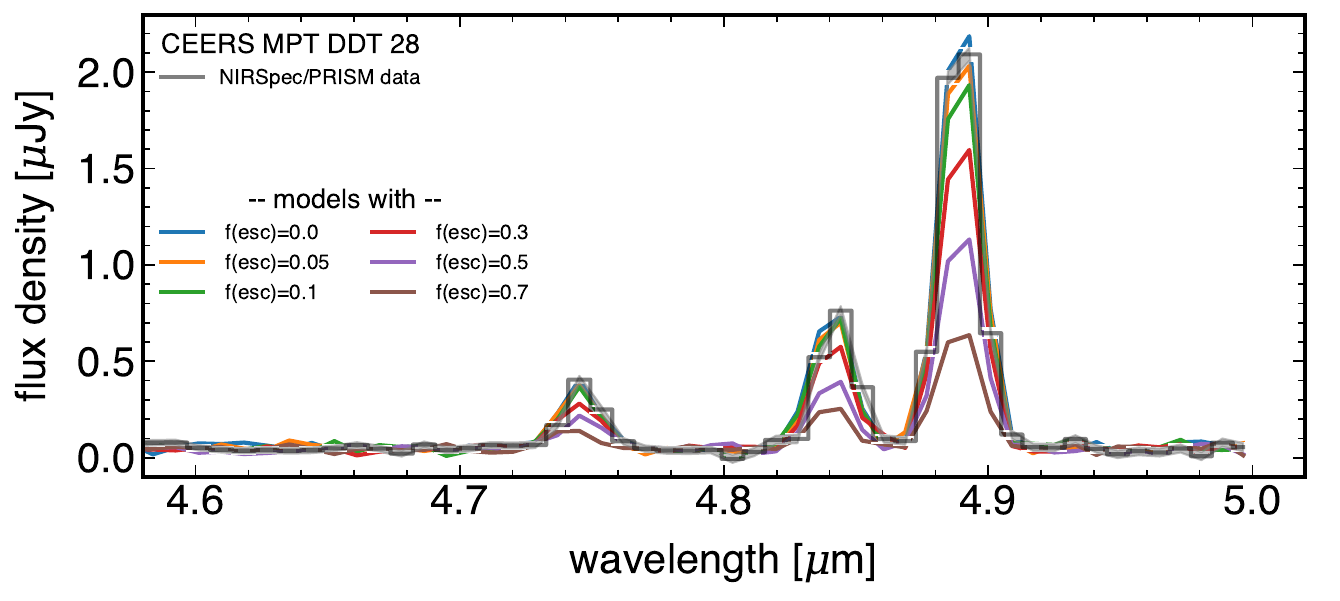}
    \caption{Example of varying \fesc\ for two of the galaxies in our sample, CEERS 1518 and DDT 28.  In each panel, the gray line and shading shows the NIRSpec/prism data and uncertainties in the region around redshifted \hb\ and \oiii.  The colored curves show a best fit stellar population model, but with \fesc\ varying from 0.0 to 0.7 (as labeled).   As \fesc\ declines, the strength of the nebular emission drops as $(1-\fesc)$, which is evident in the relative strength of the emission features.   } 
    \label{fig:fesc_test}
\end{figure}

Here, we test the efficacy of SED fitting to constrain \fesc.   We start with the SED model fit to \editone{ galaxies in the our CEERS sample.  These include CEERS DDT 28 and 1518, depicted in Figures~\ref{fig:sedfits_ceers}, and \ref{fig:spec_force_fesc}, and 30 other galaxies that span the range of redshift of the sample, $4.5 < z < 9.0$, and dust attenuation, $0.02 < A(V)$/mag $< 1.0$.  These galaxies} have relatively high S/N, which allows for relatively good constraints on the galaxies' stellar population parameters. \editone{ The galaxies include redshifts $z > 6.8$ and $z < 6.8$, so that the NIRSpec/PRISM data does and does not include \ha, respectively. }   We then take the best-fitting SED models for each galaxy, \editone{and force \fesc\ to have different values, \fesc\ = (0.005, 0.02, 0.05, 0.1, 0.2, 0.3, 0.5, 0.7). While we hold the other values fixed, the galaxies in this test span the full range of parameter space in $\log U$, $A(V)$, $Z/Z_\odot$, age and star-formation history, so these tests implicitly cover these parameters.}  Figure~\ref{fig:fesc_test} show examples of these models compared to the NIRSpec/PRISM data for CEERS 1518 and DDT 28 in the region around \hb\ and \oiii.   It is evident from these figures that as \fesc\ increases the strength of the nebular emission decreases, as expected.   

\begin{figure}[h]
    \centering
    \includegraphics[width=1\textwidth]{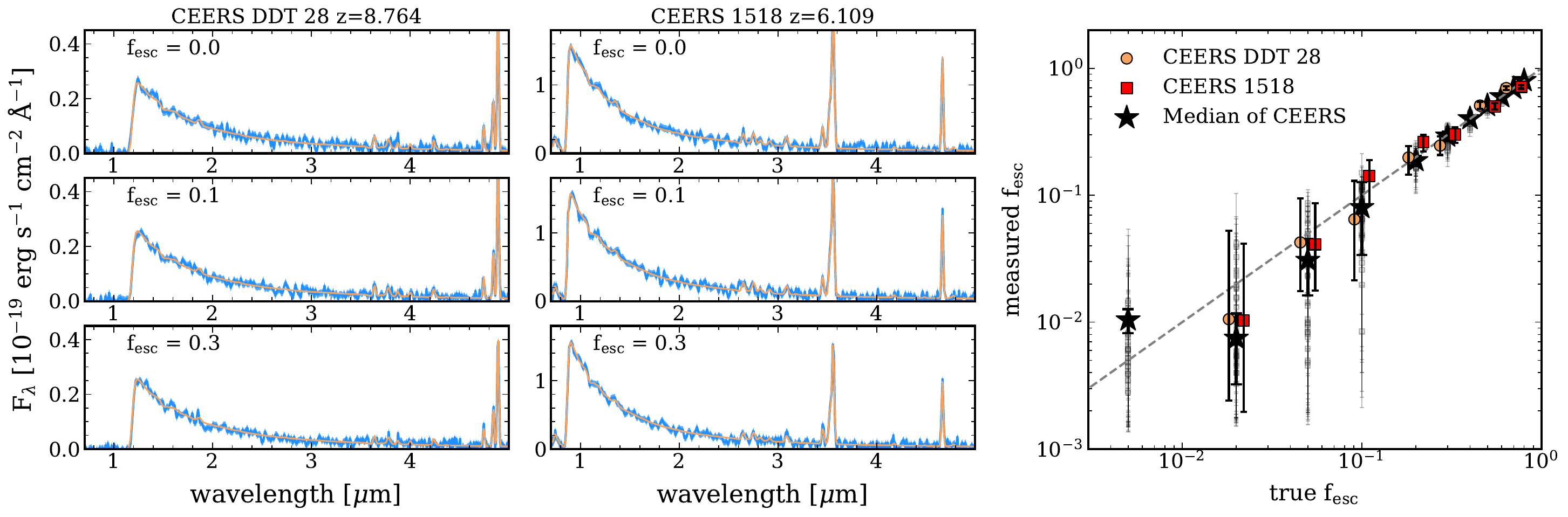}
    \caption{Results of fitting \fesc\ for models with known input values.  In the left and middle panels, the blue curves show the model spectra based on fits to two galaxies in our sample (CEERS DDT 28 at $z=8.764$ and CEERS 1518 at $z=6.109$), which are forced to have S/N $\approx$ 3 in the region around the redshifted \hb\ and \oiii\ lines.   The orange curves show the best-fit SED model.  The right panel shows the recovered \fesc\ values as a function of the (true) input values; values have been shifted slightly on the abscissa for clarity. \editone{The colored symbols show results for CEERS 1518 and DDT 28.  The black stars show the median values for all galaxies in the CEERS sample used here.}  The error bars denote the inner-68\% confidence interval.   The recovered values trace well the input values demonstrating the efficacy of SED fitting to derive \fesc\ values.  \editone{At high \fesc\ the results show good agreement.  At low values, $\fesc \lesssim 0.03$ the results have larger fractional uncertainty, but the measured values span the ``true'' values.} } 
    \label{fig:fesc_test_results}
\end{figure}

We take the best-fit SED model for each galaxy and create a ``mock'' galaxy by simulating its NIRSpec prism spectrum and photometry in the CEERS \jwst\ and \hst\ bandpasses.  We repeat this varying \fesc.  We then add noise to mimic S/N=10 for the photometric bands and S/N = 3 for the spectrum in the region around \hb\ and \oiii.   We then refit the data, deriving a posterior likelihood for \fesc.  

Figure~\ref{fig:fesc_test_results} shows the results of this test.  The panels on the left show examples of the simulated data and the best-fit models to those data \editone{for CEERS 1518 and DDT 28}.  The right panel compares the measured \fesc\ to the input values \editone{for all galaxies used here}.   There is good agreement, which implies the SED fitting is able to recover the \fesc\ values in this test. \editone{At high \fesc\ the results show good agreement.  At low values, $\fesc \lesssim 0.03$ the results have larger fractional uncertainty, but the measured values span the ``true'' values.} This test is somewhat idealized as we are fitting the simulated data with the same suite of models used to create them.   It will be important to test the  \fesc\ values inferred from SED-derived like those here, and \fesc\ values measured directly for galaxies with LyC observations, which we plan in a future study.
%


\bibliographystyle{aasjournalv7}
\bibliography{ceers_xi_ion_fesc}{}

\end{document}

%% file: defs.tex
\def\lesssim{\mathrel{\hbox{\rlap{\hbox{%
 \lower4pt\hbox{$\sim$}}}\hbox{$<$}}}}
\def\gtrsim{\mathrel{\hbox{\rlap{\hbox{%
 \lower4pt\hbox{$\sim$}}}\hbox{$>$}}}}
\def\I.{\kern.2em{\sc i}} 
\def\II.{\kern.2em{\sc ii}} 
\def\III.{\kern.2em{\sc iii}} 
\def\IV.{\kern.2em{\sc iv}} 
\def\arcs{\hbox{$^{\prime\prime}$}}

\newcommand{\hst}{\textit{HST}}
\newcommand{\spitzer}{\textit{Spitzer}}
\newcommand{\chandra}{\textit{Chandra}}
\newcommand{\sirtf}{\textit{Spitzer}}
\newcommand{\cxo}{\textit{CXO}}
\newcommand{\hdf}{HDF--N}
\newcommand{\wfu}{\hbox{$U_{300}$}}
\newcommand{\wfb}{\hbox{$B_{450}$}}
\newcommand{\wfv}{\hbox{$V_{606}$}}
\newcommand{\wfi}{\hbox{$I_{814}$}}
\newcommand{\acsv}{\hbox{$V_{606}$}}
\newcommand{\acsi}{\hbox{$I_{814}$}}
\newcommand{\nicj}{\hbox{$J_{110}$}}
\newcommand{\nich}{\hbox{$H_{160}$}}
\newcommand{\mirifive}{\hbox{$[5.6]$}}
\newcommand{\miriseven}{\hbox{$[7.7]$}}
\newcommand{\mirirten}{\hbox{$[10]$}}
\newcommand{\ab}{{\rm AB}}
\newcommand{\vega}{{\rm Vega}}
\newcommand{\ks}{\hbox{$K_s$}}
\newcommand{\tfit}{\hbox{TFIT}}
\newcommand{\sigmarms}{\sigma_\mathrm{rms}}
\newcommand{\gsim}{\gtrsim}
\newcommand{\lsim}{\lesssim}
\newcommand{\etal}{et al.}
\newcommand{\eg}{e.g.}
\newcommand{\ie}{i.e.}
\newcommand{\hbeta}{\hbox{H$\beta$}}
\newcommand{\he}{\hbox{H$\epsilon$}}
\newcommand{\hd}{\hbox{H$\delta$}}
\newcommand{\hg}{\hbox{H$\gamma$}}
\newcommand{\hc}{\hbox{H$\gamma$}}
\newcommand{\hb}{\hbox{H$\beta$}}
\newcommand{\ha}{\hbox{H$\alpha$}}
\newcommand{\lstar}{\hbox{$L^\ast$}}
\newcommand{\extinctA}{\hbox{$A_{1700}$}}
\newcommand{\mathS}{\hbox{$\mathcal{S}$}}
\newcommand{\mathR}{\hbox{$\mathcal{R}$}}
\newcommand{\NLBG}{33} 
\newcommand{\dchi}{\hbox{$\Delta \chi^2$}}
\newcommand{\ebv}{\hbox{$E(B-V)$}}
\newcommand{\royalsoceity}{Phil.\ Trans.\ R.\ Soc.\ Lond.\ A}

\newcommand{\mstar}{\hbox{$M_\ast$}}
\newcommand{\Mstar}{\hbox{$M_\ast$}}

\newcommand{\zstar}{\hbox{$Z_\ast$}}
\newcommand{\Zstar}{\hbox{$Z_\ast$}}

\newcommand{\Msol}{\hbox{$M_\odot$}}
\newcommand{\Zsol}{\hbox{$Z_\odot$}}
\newcommand{\msol}{\hbox{$M_\odot$}}
\newcommand{\zsol}{\hbox{$Z_\odot$}}
\newcommand{\fion}[2]{\hbox{[\ion#1#2]}}
\newcommand{\infinity}{\hbox{$\infty$}}
\newcommand{\efolding}{$e$--folding}
\newcommand{\lya}{Lyman~$\alpha$}
\newcommand{\qso}{{Q0122+0338}}
\newcommand{\fos}{{FOS}}
\newcommand{\za}{\hbox{$z_\mathrm{a}$}}
\newcommand{\ze}{\hbox{$z_\mathrm{e}$}}
\newcommand{\NH}{\hbox{$N(\mathrm{H})$}}
\newcommand{\phc}{\phm{:}}
\newcommand{\inprep}{\textit{in prep}}
\newcommand{\distf}{\hbox{$f(M,\dot M,t)$}}
\newcommand{\degree}{\hbox{$^\circ$}}
\newcommand{\mydot}{\hbox{$\bullet$}}
\newcommand\myRoman[1]{\@Roman{#1}\relax}%
\newcommand{\fourge}{{\sc FourGE}}
\newcommand{\sfrten}{\hbox{SFR$_\mathrm{10}$}}
\newcommand{\sfrcen}{\hbox{SFR$_\mathrm{100}$}}

\newcommand{\mone}{\hbox{$[3.6]$}}
\newcommand{\mtwo}{\hbox{$[4.5]$}}
\renewcommand{\plotfiddle}[7]{
        \centering 
        \leavevmode
        \vbox to#2{\rule{0pt}{#2}}
        \includegraphics{#1}
}
\newcommand{\zfourge}{\hbox{ZFOURGE}}

\newcommand{\zftwo}{ZFK2}

\newcommand{\herschel}{\textit{Herschel}}
\newcommand{\jwst}{\textit{JWST}}
\newcommand{\kb}{\hbox{$K_b$}}
\newcommand{\kr}{\hbox{$K_r$}}
\newcommand{\jone}{\hbox{$J_1$}}
\newcommand{\jtwo}{\hbox{$J_2$}}
\newcommand{\jthree}{\hbox{$J_3$}}
\newcommand{\hs}{\hbox{$H_s$}}
\newcommand{\hl}{\hbox{$H_l$}}
\newcommand{\lir}{\hbox{$L_\mathrm{IR}$}}

\newcommand{\wfcj}{\hbox{$J_{125}$}}
\newcommand{\wfch}{\hbox{$H_{160}$}}
\newcommand{\OH}{\hbox{$12 + \log(\mathrm{O/H})$}}

\newcommand{\myhref}[1]{\href{#1}{#1}}

\newcommand{\lcdm}{$\Lambda$CDM}
\newcommand{\LCDM}{$\Lambda$CDM}
\renewcommand{\mone}{\hbox{$[3.6]$}}
\renewcommand{\mtwo}{\hbox{$[4.5]$}}
\newcommand{\mthree}{\hbox{$[5.8]$}}
\newcommand{\mfour}{\hbox{$[8.0]$}}
\newcommand{\reff}{\hbox{$r_\mathrm{eff}$}}
\newcommand{\sersic}{S\'ersic}
\newcommand{\um}{\hbox{$\mu$m}}
\newcommand{\pstart}[1]{\noindent \textbf{#1}}
\definecolor{aggiemaroon}{HTML}{500000}

\newcommand{\neiii}{\hbox{[Ne\,{\sc iii}]}}
\newcommand{\oii}{\hbox{[O\,{\sc ii}]}}
\newcommand{\oiii}{\hbox{[O\,{\sc iii}]}}
\newcommand{\nii}{\hbox{[N\,{\sc ii}]}}
\newcommand{\sii}{\hbox{[S\,{\sc ii}]}}
\newcommand{\siiv}{\hbox{Si\,{\sc iv}}}
\newcommand{\hii}{\hbox{H\,{\sc ii}}}
\newcommand{\heii}{\hbox{He\,{\sc ii}}}
\newcommand{\llambda}{\lambda\lambda}
\newcommand{\myshrink}{\vspace{-7pt}}
\newcommand{\mylinebreak}{\vspace{8pt}}

\newcommand{\grizli}{\hbox{\texttt{grizli}}}
\newcommand{\linmixi}{\hbox{\texttt{LINMIX}}}
\newcommand{\izi}{\hbox{\texttt{IZI}}}
\newcommand{\ppxf}{\hbox{\texttt{pPXF}}}
\newcommand{\bagpipes}{\hbox{\texttt{BAGPIPES}}}
\newcommand{\sextractor}{\hbox{\texttt{SE}}}

\newcommand{\fesc}{\hbox{$f_\mathrm{esc}$}}
\newcommand{\xiion}{\hbox{$\xi_\mathrm{ion}$}}
\newcommand{\luv}{\hbox{$L_\mathrm{UV}$}}
\newcommand{\muv}{\hbox{$M_\mathrm{UV}$}}
\newcommand{\ndotion}{\hbox{$\dot n_\mathrm{ion}$}}

\definecolor{aggiemaroon}{HTML}{500000}

\def\vshiftfig#1#2#3#4{\hfill 
\vbox{\parskip=0pt\hsize=#2
\vskip{#4} \includegraphics[width=#2]{#1}\vskip2pt
\vtop{\centering
\footnotesize
\hsize=#2
#3\vskip1pt
}}\hfill}